\documentclass[a4paper,11pt]{article}
\pdfoutput=1 % if your are submitting a pdflatex (i.e. if you have
\usepackage{jheppub} % for details on the use of the package, please
\usepackage[utf8]{inputenc}
\usepackage[english]{babel}

\usepackage[T1]{fontenc} % if needed
\usepackage{slashed}
\usepackage[table]{xcolor}% http://ctan.org/pkg/xcolor

\title{Combinatorial Superstatistics for Soft QCD}

\author{Mikael Mieskolainen}
\affiliation{Department of Physics, Division of Particle Physics and Astrophysics, University of Helsinki, P.O. Box 64, FI-00014 Helsinki, Finland}

\emailAdd{mikael.mieskolainen@helsinki.fi}
\abstract{High energy diffraction and soft QCD span exciting final state topologies and fluctuations which have not yet been measured or characterized in a fully exhaustive way. In this work, we go beyond the standard measures and formulate a new framework to generalize rapidity gap counting with emphasis on abstract structure, fiducial observables, experimental limitations and factorizing model dependence. The construction is based on a higher dimensional Bernoulli vector statistics with a combinatorial incidence algebra structure, independent of the underlying field theory. This is the first experimentally feasible framework designed to probe the seemingly arcane sign alternating scattering amplitude cutting rules of style Abramovski-Gribov-Kancheli, rules being of interest in Regge theory, perturbative QCD and even in stringy black hole calculus. As an additional novel show case, we pose, construct and solve a highly related combinatorial stochastic superposition Poisson inverse problem using the M\"obius inversion theorem.}

\usepackage{amsmath}     % Math
\usepackage{mathtools}   % Math
\usepackage{algorithm}   % Algorithm boxes
\usepackage{algorithmic}
\usepackage{graphicx}
\usepackage{bm}
\usepackage{float}       % Allows command [H] ~ here for figures
\usepackage{subcaption}  % multiple images
\usepackage{feynmp}      % Feynman diagrams
\graphicspath{{./figs/}}

\def\multiset#1#2{\ensuremath{\left(\kern-.3em\left(\genfrac{}{}{0pt}{}{#1}{#2}\right)\kern-.3em\right)}}

\def\stirling#1#2{\ensuremath{\left\{\kern-.3em\genfrac{}{}{0pt}{}{#1}{#2}\kern-.3em\right\}}}

\DeclareMathOperator*{\argmin}{arg\,min}

\usepackage{natbib}
\begin{document}
\noindent 
\textsc{} % Needed for empty
% date

% Only sections in the table of contents
\setcounter{tocdepth}{1}

\maketitle

\newpage

%\flushbottom % CAN create gaps in the text

% FEYNMAN GRAPHS
%\begin{fmffile}{fgraphs} % Feynman diagrams
%\unitlength = 1mm        % Feynman diagrams

%Mellin-transform, Poisson-Mellin-Newton cycle, N\"orlund–Rice integral. Partitions in number theory. The number of positive divisors, the Euler phi function, the M\"obius function, the Zeta function, the Partition function.

\section{Introduction}

The physics signatures used in searching and analysing interactions at high energy colliders are often based on specific final state `topologies'. Examples of these span from the soft diffractive QCD processes, where different event classes are identified by detector responses in the pseudorapidity space to new physics searches spanning complex event topologies with jets, leptons and missing transverse energy. In this work, we represent the final state topologies using binary vector spaces. This allows us to use powerful mathematical tools of combinatorics, namely, the incidence algebras introduced by G.C. Rota \cite{rota1964foundations}.

The combinatorial problems in high energy physics are numerous. To name a few, they span from combinatorial symmetry factors and Feynman diagram enumeration in QFT, Wick's theorem (Isserlis' theorem), Gribov-Glauber cancellations in heavy ion physics, Abramovski-Gribov-Kancheli cutting rules in Regge and field theory, numerous experimental combinatorial puzzles in track fitting and primary vertex association and in the resonance search analysis of final states with combinatorial background. Here we take a slightly different angle on this topic than what has been done previously. From the experimental point of view, the framework here can be utilized for the extraction of diffractive cross sections and for completely novel fiducial measurements, and can be considered as a complete generalization of the double diffractive measurement done by TOTEM experiment \cite{antchev2013double}. The approach developed in this work is the first explicit utilization of the M\"obius inversion theorem in high energy physics, as far as we know. Perhaps the most well known M\"obius inversion paper in physics is by Chen \cite{chen1990modified}. The connection to supersymmetric mathematical physics models was first developed in papers by Spector \cite{spector1990supersymmetry} and Julia \cite{julia1990statistical}, and later by Onofri, Veneziano and Wosiek, which they called `supercombinatorics' \cite{Onofri2007}.

In soft QCD, different minimum bias processes have very different final state configurations over rapidity. However, when the luminosity conditions grow high, nearly all capabilities to study diffractive processes are lost due to the multiple proton-proton interactions per bunch cross -- the pileup. This can be both on-time and off-time pileup, that is, it can propagate from the same or previous bunch cross due to the short 25 ns spacing at the LHC and long detector integration time windows. As an illustrative example, when the average number of inelastic interactions is $\mu \sim 10$, the so-called large rapidity gap events vanish almost completely when we talk about low-$p_T$ events, and we are left only with uniform event signatures as our observables over pseudorapidity. Thus, there are naturally some fundamental limitations in how large the luminosity can be even without considering all instrumentation and detector level complications. The optimal running setup depends heavily on the physics signal and background. The pile-up rates, being order of 30-50 in ATLAS and CMS, even higher in future, pose a major challenge for physics analysis, cross section measurements and searches for new physics phenomena. However, some high-$p_T$ channels are still experimentally manageable up to $\mu \sim \mathcal{O}(10^2)$ or more. Thus we point out that one practical application of our work is the inversion of (minimum bias) trigger combinations under pileup, which was treated in work by ATLAS \cite{atlas2010luminosity} and ALICE \cite{oyama2011cross} collaborations with less powerful tools. In our other paper \cite{mieskolainen2019inversion}, we solve a related inverse problem of multiplicity distributions undergoing a compound Poisson autoconvolution process, by fusing techniques of characteristic functions and algorithms.

Section \ref{sec:vectorspaces} starts with multivariate Bernoulli observables together with vector spaces over the finite Galois field GF(2) and sets them in the context of high energy diffraction. In Section \ref{sec:incidencealgebra} we go through the partially ordered sets, incidence algebras and the M\"obius inversion and Section \ref{sec:measurements} is devoted for concrete measurements of observables. The other half of the paper deals with the combinatorial superposition problem in a compound Poisson scenario. Section \ref{sec:PoissonProblem} constructs the compound Poisson problem and a mathematical model is built, which is then solved using the combinatorial techniques from the first half of the paper. Finally in Section \ref{sec:simulations}, we study the inverse problem numerically and summarize in Section \ref{sec:conclusions}. Appendix \ref{sec:appendix} provides supplementary information.

\newpage
\section{Binary vector spaces and Diffraction}
\label{sec:vectorspaces}

For the algebraic construction of Bernoulli observables, we need a vector space $\mathbb{F}_2^N$ over the finite Galois field GF(2) with addition as a component wise Boolean OR $(\vee)$ and multiplication as a component wise Boolean AND $(\wedge)$. As the inner product we can use $\langle \mathbf{a} | \mathbf{b} \rangle = \bigvee_{i = 1}^N a_i \wedge b_i \in \mathbb{F}_2$, where $\mathbf{a},\mathbf{b} \in \mathbb{F}_2^N$. The vector space is spanned by $N$ standard basis vectors as $\text{span} ({\mathbb{F}_2^N}) = \{\mathbf{e}_k\}_{k=1}^N$ and the number of elements in an $N$-dimensional binary vector space is $|\mathbb{F}_2^N| = 2^N$, illustrated in Figure \ref{fig: d6}.  The total number of subspaces is given by a sum over $q$-binomial coefficients, given in Appendix \ref{sec:appendix}.

The most useful property with respect to the physics involved here is that the subspaces form an orthomodular poset (partially ordered set). Another aspect is that Boolean operators which are diagonal in one basis are diagonal in all bases. Projection operators are always diagonal, which is semi-intuitive if one thinks about the $N$-dimensional unit-hypercube structure. In \cite{dai2013multivariate} it was pointed out that the multivariate Bernoulli can be considered as a certain generalization of the Ising model. However, here we apply these to high energy physics.

\subsection{Bernoulli random variables and partial cross sections}

Every component observable of vectors in $\mathbb{F}_2^N$ is treated as a Bernoulli random variable $B_i$ with $1 \leq i \leq N$. The random variables $B_i$ follow the distribution
\begin{equation}
\text{Ber}(B_i|m_i) = P_B(B_i = b| m_i) = m_i^b(1-m_i)^{1-b},
\end{equation}
where $b \in \{0,1\}$. The mean is $m_i$ and variance $\text{Var}[B_i] = m_i(1-m_i)$. Together, these can be encapsulated with a multivariate Bernoulli distribution $P(B_1 = b_1, B_2 = b_2, \dots, B_N = b_N | \theta)$. The full distribution is either described directly with $2^N-1$ parameters ($-1$ from normalization) associated with every $2^N$ elements of the binary vector space, what we may call the \textit{fundamental} or \textit{natural} representation, or with parameters describing the expectation values and multipoint correlations between different Bernoulli variables, what we may call the \textit{correlation} representations. We use the notation and representations constructed by Teugels \cite{teugels1990some}.

Example $N = 3$: A centralized correlation representation is
\begin{align}
m_i &= \langle B_i \rangle, \;\;\; i = 1,2,3 \\
\theta_{ij} &= \text{Cov}[B_i,B_j] = \langle (B_i-m_i)(B_j-m_j) \rangle \\
&= \langle B_i B_j \rangle - \langle B_i \rangle \langle B_j \rangle, \;\;\; (i,j) \in \{ (1,2), (1,3), (2,3)\} \\
\theta_{123} &= \langle (B_1 - m_1)(B_2 - m_2)(B_3 - m_3) \rangle,
\end{align}
where the moments or correlation functions are centralized, that is, the mean is subtracted. The structure is only slightly more complicated for larger values of $N$, and is described in the following chapters. One interesting property is that for the multivariate Bernoulli distribution component variables which are pairwise uncorrelated $\text{Cov}[B_i,B_j] = 0$ are also always independent, that is, the density factorizes $P(B_i,B_j) = P(B_i)P(B_j)$ \cite{dai2013multivariate}. This does not hold in general for multivariate probability distributions.

\subsection{Probabilities and cross sections}

In the fundamental representation, a probability vector $\mathbf{p}$ encapsulates the probabilities of \textit{mutually exclusive} combinations of observables in the binary vector space
\begin{equation}
\mathbf{p} = (P_{\mathbf{b}_0}, P_{\mathbf{b}_1}, \dots, P_{\mathbf{b}_{2^N-1}})^T, \;\; \sum_c p_c = 1
\end{equation}
where indices go through all the elements of the binary vector space. The one-to-one correspondence is given uniquely by the binary expansion\footnote{Here we remark on the difference between left and right ordered bit strings. We use the convention of most significant `bit' (MSB) on \textit{right}. In the most significant `bit' on \textit{left} representation, this expansion would be $c = \sum_{i = 1}^N b_i 2^{d-i}$.}
\begin{equation}
c = \sum_{i = 1}^N b_i2^{i-1} \; \Leftrightarrow \; \mathbf{b}^c = (b_1,b_2,\dots,b_N),
\end{equation}
between the index $0 \leq c \leq 2^N-1$ and the binary vector $\mathbf{b}^c \in \mathbb{F}_2^N$.

Now turning into physical observables, the differential cross section element is given by the usual
\begin{equation}
d\sigma_{n'} = \frac{1}{F} \frac{1}{S_{n'}} \, d\Pi_{n'} |\mathcal{M}_{i \rightarrow {n'} }|^2, 
\end{equation}
where the matrix element squared $|\mathcal{M}_{i \rightarrow {n'}}|^2$ is the probability amplitude for the transition from the initial state to the ${n'}$-particle final state, encapsulating all dynamics. The $S_{n'}$ is a QFT combinatorial symmetry factor for identical final states, $F = 4\sqrt{(p_1 \cdot p_2)^2 - m_1^2m_2^2} \longrightarrow 2s$ (high energy limit) is the incoming invariant M\"oller flux and the standard Lorentz invariant phase space element is
\begin{equation}
d\Pi_{n'} = (2\pi)^4 \prod_{j=1}^{n'} \frac{d^3 k_j}{ (2\pi)^3 2k_j^0}.
\end{equation}

Using the combinatorial incidence algebra notation, now the so-called $2^N$ partial cross sections can be expressed with

\begin{align}
\bm{\sigma} =
\nonumber
\frac{1}{F}
&\sum_{n'} \frac{1}{S_{n'}}\int d\Pi_{n'} \delta^{(4)}\left( p_1+p_2 - \sum_{n'} k_j \right) 
|\mathcal{M}_{2\rightarrow n'}|^2
\begin{pmatrix}
 1  &   -1 \\
 0  &   1 \\
\end{pmatrix}^{\otimes \,N} \\
\nonumber
&\begin{pmatrix}
 1  \\
A \{ \{ k_{j} \}; \Xi_N \} \\
\end{pmatrix} \otimes \\
&\hspace{2em}\begin{pmatrix}
 1  \\
A \{ \{ k_{j} \}; \Xi_{N-1} \} \\
\end{pmatrix}
\otimes
\cdots
\otimes
\begin{pmatrix}
 1  \\
A \{ \{ k_{j} \}; \Xi_1 \} \\
\end{pmatrix}.
\end{align}

The expression above is a $2^N$-vector, where the component $\sigma_0$ denotes the null part outside the total fiducial phase space. The acceptance (indicator) functions $A : \{k_j\} \rightarrow \{0,1\}$ encoding the Bernoulli trials are
\begin{align}
A_i(\{k_j\}; \Xi_i) =
\begin{cases}
    1, & \text{if $\exists k_j \in \Xi_i$ }\\
    0, & \text{otherwise}.
\end{cases}
\end{align}
The acceptance function returns $1$ if any of the final state particles was inside the phase-space domain $\Xi_i$ defined for $B_i$ and returns $0$, if none were inside the domain. This function can be constructed with simple phase space cuts suitable for analytical calculations. In practice, they are most easily calculated with soft QCD Monte Carlo event generators, where arbitrary fiducial phase space cuts are possible. We abstract out the detailed singularity and theory structure connections with the phase space cuts out at this point, they are theory specific, but certainly collinear and infrared singularities may used to design a specific structure for the acceptance functions.

Experimental efficiency and resolution functions work as a mixing map $\sigma_c \mapsto \sigma_{c'}$, which is a highly complicated stochastic non-linear operator modulating the pure acceptance indicator function. For the expectation values, it can be represented with a finite linear \textit{folding} operator acting on the final states partial cross section vector as $\mathcal{F} : \bm{\sigma} \rightarrow \tilde{\bm{\sigma}}$, where $\bm{\sigma}$ denotes the partial cross sections at the fiducial phase space particle level and $\tilde{\bm{\sigma}}$ denotes the visible partial cross sections at the detector level. Both can have almost arbitrary definitions, depending on the problem. The hierarchy goes as follow
\begin{equation}
d\mathcal{O} \xrightarrow[\forall k ]{{\sum_{n'} \int_{\Omega_{n'}}}} \bm{\sigma} \xrightarrow[- \sigma_E, + \sigma_L]{{\mathcal{F}}} \tilde{\bm{\sigma}},
\end{equation}
where the first map is the `theory' or `generator' and the second one is the `detector', to illustrate. The zero-th component $\sigma_0$ of $\bm{\sigma}$ denotes the invisible fiducial cross section in the fiducial phase space division (geometrically invisible), only accessible theoretically or via indirect measurements. The inverse operator $\mathcal{F}^{-1}$ is the so-called \textit{unfolding} process, which can be naively solved with a $2^N-1$ matrix inversion, obtained via detector simulation. The $\sigma_E$ denotes the integrated efficiency loss, which is not the same as pure acceptance loss $\sigma_0$. Integrated leakage from outside the fiducial to inside the visible is denoted with $\sigma_L$, which may happen due to e.g. material re-scattering -- this can be a non-negligible effect with very forward detectors. We emphasize here that the unfolding is for efficiency and smearing induced corrections, not to extrapolate the detector geometry.

As an example of the indicator functions: a simple division of the (pseudo)rapidity into $N$ intervals with suitable $p_t$ thresholds per interval. Experimentally, the intervals may overlap over pseudorapidity such that $[\eta_{i,\min}, \eta_{i,\max}] \cup [\eta_{j,\min},\eta_{j,\max}] \neq \{0\}$ for a detector pair $(ij)$, and thus making them trivially correlated. It is possible, in principle, via suitable non-overlapping fiducial definition to take this into account in the unfolding procedure which de-correlates the data. Another option is simply to take care that no overlap happens physically. Also, the detectors may have gaps over $\eta$ such that $[\eta_{i,\min}, \eta_{i,\max}] \cup [\eta_{j,\min},\eta_{j,\max}] = \{0\}$. 
However, it is worth pointing out that the formalism presented here makes no assumption of statistical independence of any kind, either implicit or explicit.

\subsection{Abramovski-Gribov-Kancheli cutting rules}

We collect here some illustrating results of the AGK cutting rule calculus \cite{AGK}. By no means this is a complete description, but being of more exposing nature. These rules are somewhat similar to more well known Cutkosky-Landau cutting rules based on the $S$-matrix unitarity, the rules which treat scattering amplitude discontinuities. The AGK rules were introduced in the Regge theory context, but they are actually flexible in terms of the underlying theory. What we need here is the reggeization, which happens in field theories under Regge limit but also in string theory. The AGK calculus results here are collected from a stringy black hole context \cite{addazi2017glimpses} and from QCD papers \cite{bartels1997space, bartels2005agk}. We assume the usual unitarity $\mathcal{S}$-matrix and $\mathcal{T}$ matrix related with
\begin{equation}
\mathcal{S} = 1 + i\mathcal{T}.
\end{equation}
Let us have the elastic scattering amplitude $\mathcal{T}_{AB}(s,t)$ with a Sommerfield-Watson integral transform representation in the complex angular momentum space as
\begin{equation}
\mathcal{T}_{AB}(s,t) = \int \frac{d\omega}{2i} \xi(\omega) s^{1+\omega} \mathcal{F}(\omega,t) \;\;\text{with} \;\; \xi(\omega) = \frac{\tau - e^{-i \pi \omega}}{\sin \pi\omega},
\end{equation}
where $\mathcal{F}(\omega,t)$ is the corresponding partial wave, $\omega = J - 1 \in \mathbb{C}$ being the conjugate variable to $s$ together with the signature factor $\xi$ for the positive or negative signature $\tau = \pm 1$. The positive signature case, such as Pomeron exchange, can be written more compactly as $\xi(\omega) = i + \tan(\frac{\pi}{2} \omega)$. Now assume exchange of $n$ Reggeons, that is, exchange of Regge trajectories.

The $n$-cut discontinuity over $\omega$ of $\mathcal{F}(\omega,t)$ can be written in terms of vertex functions $\mathcal{V}_n^{a}, \mathcal{V}_n^{b}$ and integrating over transverse momentum phase space as \cite{bartels2005agk}
\begin{align}
\textsc{DISC}_\omega^{(n)}[\mathcal{F}(\omega,t)] &= 2\pi i \int \frac{d\Omega_n}{n!} \delta (\omega - \sum_j \beta_j) \Pi(\{ \beta_j \}) \mathcal{V}^a_n( \{\mathbf{k}_j,\omega \}) \mathcal{V}^b_n(\{ \mathbf{k}_j,\omega \}) \\
d\Omega_n &\equiv (2\pi)^2 \delta^{(2)} \left( \mathbf{q} - \sum_{j=1}^n \mathbf{k}_j \right) \prod_{j=1}^n \frac{ d^2\mathbf{k}_j }{ (2\pi)^2 } \\
\Pi(\{\beta_j \}) &\equiv \text{Im}\left[-i\prod_j i\xi_j \right] = (-1)^{n-1} \frac{ \cos \left[ \frac{\pi}{2} \sum_j (\beta_j + \frac{1 + \tau_j}{2} \right]  }{ \prod_j \cos \left[ \frac{\pi}{2}(\beta_j + \frac{1 - \tau_j}{2}) \right] },
\end{align}
where $\beta(-\mathbf{k}^2) \equiv \alpha(-\mathbf{k}^2) - 1$ with $\alpha(t_j = -\mathbf{k}_j^2)$ being the Regge trajectory function, for the soft Pomeron often used parametrization is $\alpha(t) \simeq 1.08 + 0.25t$. The factor $n!$ comes from all possible exchange diagram planar and crossed non-planar orderings.

Now the amplitude for the $n$-cut exhange can be written using the Sommerfield-Watson transform and the cut discontinuity as \cite{bartels1997space}
\begin{align}
\mathcal{T}_{AB}^{n-\text{cut}}(s,t) 
&= \int \frac{d\omega}{2i} \xi(\omega) s^{1+\omega} \text{DISC}_\omega^{(n)} [ \mathcal{F}(\omega,t)] \\
&= \pi \int \frac{d\Omega_n}{n!} \xi( \sum_j \beta_j) s^{1+\sum_j \beta_j} \Pi(\{\beta_j\})  \mathcal{V}_n^a(\{ \mathbf{k}_j, \sum_i \beta_j \}) \mathcal{V}_n^b(\{ \mathbf{k}_j, \sum_j \beta_j \}).
\end{align}

Next we write down the Abramovski-Gribov-Kancheli sign alternating combinatorial factors \cite{bartels2005agk}
\begin{equation}
\boxed{
G^n_k = \begin{cases}
          \frac{2^n}{n!}(-1)^n + \frac{1}{n!}\Pi(\{ \beta_j \}), \text{ for } k = 0 \text{ cut Pomerons},\\
          (-1)^{n-k} \frac{ 2^n }{ (n-k)!k! }, \text{ for } 0 < k \leq n,
        \end{cases}}
\end{equation}
which we tabulate for illustration in Table \ref{table: AGK_table}.

\begin{table}[h]
\center
\begin{tabular}{c|ccccccc}
$n,k$ & 0 & 1 & 2 & 3 & 4 & 5 & 6 \\
\hline
1 & 0 & 1 & 0 & 0 & 0 & 0 & 0 \\ 
2 & 1 & -4 & 2 & 0 & 0 & 0 & 0 \\ 
3 & -3 & 12 & -12 & 4 & 0 & 0 & 0 \\ 
4 & 7 & -32 & 48 & -32 & 8 & 0 & 0 \\ 
5 & -15 & 80 & -160 & 160 & -80 & 16 & 0 \\ 
6 & 31 & -192 & 480 & -640 & 480 & -192 & 32 \\ 
\end{tabular}
\caption[Table caption text]{AGK factors $G_k^n$ for the even signature (Pomeron) case.}
\label{table: AGK_table}
\end{table}

Let us now point some integer sequence properties of the AGK series, which might be illuminating in our hyperspace context. The series $|G_{k=1}^n|$ is the number of edges in an $n$-hypercube. The series $|G_{k=2}^n|$ is the number of diagonals in an $n$-hypercube of length $\sqrt{2}$. The series $|G_{k=3}^n|$ is the number of 3-cycles in the halved $n$-cube graph. These suggest to us directly that in order to be (experimentally) sensitive to $n$ Pomeron exchanges, the number of rapidity slices $N$ should be at least $n+1$. Now the use of AGK cutting factors is that they allow us to represent the total discontinuity over $s$-space scattering amplitude as a weighted sum over all $k$-cut Pomeron contributions \citep{bartels2005agk}
\begin{align}
\nonumber
\mathcal{A}^n(s,t) &\equiv \text{DISC}_s[\mathcal{T}_{AB}^{n-\text{cut}}(s,t)] = \sum_{k = 0}^n a_k^n(s,t), \;\; \text{where} \\
a_k^n &= \frac{2 \pi i}{n!} \, G_k^n \int d\Omega_n s^{1+\sum_i \beta_i}\mathcal{V}_n^a(\{\mathbf{k}_j,\sum_i \beta_i \}) \mathcal{V}_n^b(\{\mathbf{k}_j,\sum_i \beta_i \}).
\end{align}

The last step in our short discussion on this topic is the impact parameter $b$ space picture and eikonalization. The eikonal opacity function in the impact parameter space is obtained using the (inverse) Fourier transform of single discontinuity integrated over $\omega$ \cite{bartels2005agk}
\begin{equation}
\Omega(s,b) \equiv \frac{1}{\pi i} \int \frac{ d^2 \mathbf{q} }{ (2\pi)^2 }
e^{-i \mathbf{b} \cdot \mathbf{q}} \int d\omega \, \text{DISC}_\omega [ \mathcal{F}(\omega,-\mathbf{q}^2) ] \, s^\omega
\end{equation}
with the amplitude recovered in the $s$-space by (forward) Fourier transform of the $n$-times exponentiated amplitude, by \textit{assuming factorization} of each exchange as $\mathcal{V}_n^a = [\mathcal{V}_1]^n$, giving
\begin{equation}
\mathcal{A}_k^n(s,t) = 4is \frac{ (-1)^{n-k} }{ k!(n-k)! } \int d^2\mathbf{b} e^{i\mathbf{b} \cdot \mathbf{q}} \, [\Omega(s,b)]^n
\end{equation}
with summation $n \geq k$ resulting in
\begin{equation}
\mathcal{A}_k(s,t) = 4is \int d^2 \mathbf{b} e^{i\mathbf{b} \cdot \mathbf{q}} P(s,b).
\end{equation}
Where the probability is a sum over the $\mathcal{S}$-matrix elements \cite{bartels2005agk}
\begin{equation}
P(k,s,b) \equiv \sum_{n \geq k} [\mathcal{S}(s,b)\mathcal{S}(s,b)^\dagger]_k^n = \frac{[\Omega (s,b)]^k }{ k! } e^{-\Omega(s,b)},
\end{equation}
gives a Poisson distribution with $\langle k(s,b) \rangle = \text{Var}[k(s,b)] = \Omega(s,b)$.
That is, the number of $k$-cut Pomerons at given energy squared $s$ and impact parameter $b$ is Poisson distributed in this picture. Now, in typical Monte Carlo implementations one needs also the parton densities to pick the longitudinal momentum fractions from vertex factors $\mathcal{V}_1^{a,b}$ for each exchanged Pomeron and at least triple-Pomeron interactions with diffractive cuts, in addition (hard) QCD matrix elements, parton shower and fragmentation.

\begin{figure}[H]
\vspace{1em}
\centering
$\begin{array}{ll}
\includegraphics[width=65mm]{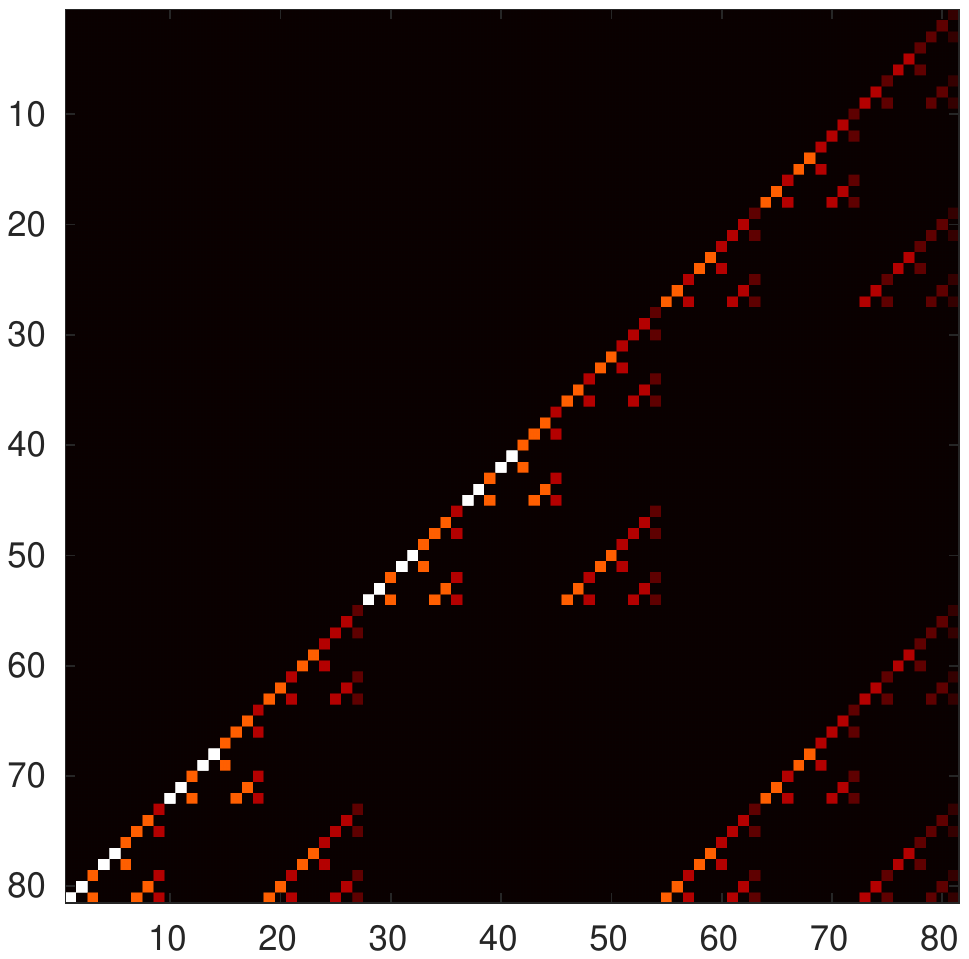} & \hspace{3.5em} \includegraphics[width=65mm]{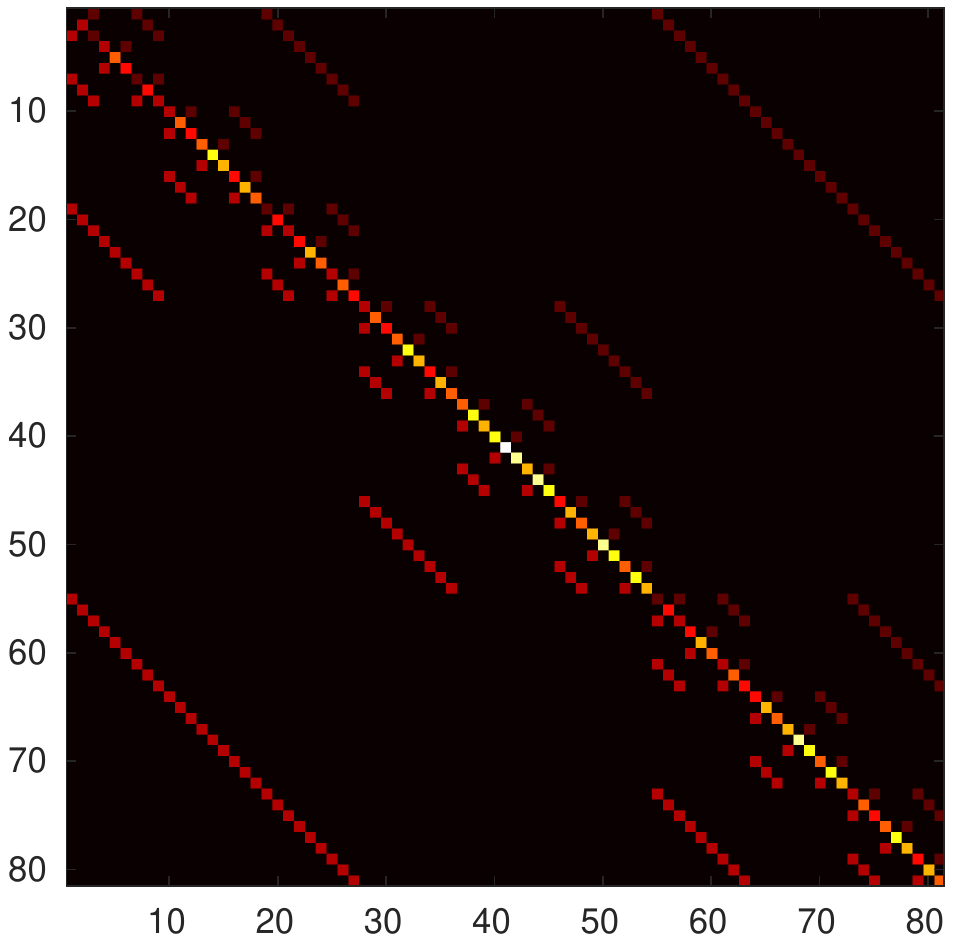} \\ \end{array}$
\caption{Tensor product chain $M^{\otimes \,4}$ (on left) and tensor sum chain $\frac{1}{4}M^{\oplus \,4}$ (on right). Black to red to white denote values in [0,1].}
\label{fig: AGKmatrix}
\end{figure}

As an example, we shall refer to the description of QGSJet-II event generator for details \cite{ostapchenko2011monte}, which implements advanced Reggeon calculus for soft and hard interactions. The scope of this paper is to provide new algebraic methods for measuring the observables. We use the Poissonian AGK distribution as our motivation for the later construction. We shall also note here that a left stochastic matrix devised by Ryskin and Bartels \cite{bartels1997space}, describes the `non-diagonal' cut transitions necessary for rapidity gap configurations
\begin{equation}
M =
\left(\begin{array}{ccc}
 0  &  0  & \frac{1}{2} \\
 0  &  1  & 0 \\
 1  &  0  & \frac{1}{2}
\end{array}\right),
\end{equation}
which has $\mathbf{x} = [1,-4,2]^T$ as its eigenvector. That is, for every rapidity interval, these ratios hold, and the transition matrix describes the stochastic probability of transition from different configurations per chain. In Figure \ref{fig: AGKmatrix} we illustrate the nature of this matrix under the  Kronecker product $A \otimes B$ and the Kronecker sum defined as $A\oplus B = A \otimes I_A + I_B \otimes B$, where $I$ is an identity matrix. The product matrix belongs to class of inclusion-exclusion matrices where as the sum matrix resembles a hypercube adjacency (connectivity) matrix. Later we learn how these concepts appear in the incidence algebra.

\subsection{Rapidity gaps as a spacing distribution}

Typical is that `non-diffractive events' have exponentially decreasing maximum rapidity gap distribution $\sim \exp(-\ell_c \Delta y)$, where the correlation length $\ell_c \sim 1$ depends on the average multiplicity density. This can be derived assuming $n$ independent final state emissions over rapidity. Diffractive events are those which have `anomalously' large gaps. The rapidity gap itself is a kinematic consequence given the excited (dissociative) proton system mass is small enough and the initial state boost large enough, however, the $M^2$ distribution itself is driven dynamically. The standard triple Pomeron limit distributions obtained via generalized multibody unitarity together with Regge asymptotics give $d\sigma / dM^2 \sim 1/(M^2)^{\alpha_P(0)}$, where the $\alpha_P(0)$ is the effective Pomeron intercept $\alpha_P(0) \equiv 1 + \Delta_P \sim 1.08$. Then by standard kinematic\footnote{Beam rapidity is $y_\pm = \pm \ln(\sqrt{s}/m_p)$ and the diffractive system spans $y_{\text{span}} = \ln(M^2/m_p^2)$.} change of a  variable $\phi^{-1}(\Delta y) \equiv s \exp(-\Delta y) = M^2$, we obtain the single diffractive rapidity gap distribution
\begin{align}
\label{eq: diffraction}
\nonumber
P_D(\Delta y) \sim \frac{d\sigma}{d\Delta y} &= \frac{d\sigma}{dM^2}(\phi^{-1}(\Delta y)) \left| \frac{d\phi^{-1}}{d \Delta y} \right| \\
&= \frac{1}{(s\exp(-\Delta y))^{\alpha_P(0)}} \left| \frac{-s}{e^{\Delta y}} \right| = e^{\Delta y(\alpha_P(0) - 1)}.
\end{align}
Above, we neglected Mandelstam $t$-dependence of the triple Regge expressions by taking $t \rightarrow 0$, not relevant for this discussion. Now Equation \ref{eq: diffraction} is a slowly rising exponential, instead of decreasing such as for the random non-diffractive processes. In the case of double diffractive process, the equivalent derivation goes via Jacobian determinant $\det J(\Delta y, y_0) = sm_p^2 e^{-\Delta y}$, from $\phi_i^{-1}(\Delta y, y_0) = (sm_p^2 e^{-\Delta y \pm 2y_0})^{1/2} = M_i^2$ with $i = 1,2$ for the two systems and $\Delta_y^{DD} = -\ln \left(M_1^2 M_2^2/(sm_p^2) \right)$ and $y_0 = \frac{1}{2} \ln(M_1^2/M_2^2)$. This gives us
\begin{align}
\label{eq: doublediffractive}
\nonumber
P_{DD}(\Delta y) \sim \frac{d\sigma}{d\Delta y dy_0} &= \frac{d^2\sigma}{dM_1^2dM_2^2}(\phi_{1,2}^{-1}(\Delta y, y_0)) \left| \det(J) \right| \\
&= \prod_i \left[ \frac{1}{(sm_p^2\exp(-\Delta y \pm 2y_0))^{\alpha_P(0)/2}} \right] \frac{sm_p^2}{e^{\Delta y}} \\
&\sim e^{\Delta y(\alpha_P(0) - 1)},
\end{align}
where the dependence on $y_0$ vanishes and the functional dependence is the same as in the single diffractive case. The boundary conditions are different, because the both edges of the gap are moving in the case of double diffractive. Also, we can count only distances between particles (type I), or with respect to the minimum or maximum of the rapidity interval, which may be taken for example with respect to the detector geometric boundary (type II). One must remember that any unitarization scheme may modify these results. That is, the triple Pomeron expressions are not unitarity under $s \rightarrow \infty$, which is a long standing problem, with numerous partially plausible solutions (eikonalization, triple Pomeron decoupling, ...). Currently, there is no detailed data enough to gain new insights on this problem neither the non-perturbative QCD techniques are advanced enough to solve it. However, new LHC data may change the situation.

At this point, we refer the reader to \href{https://mcplots.cern.ch}{mcplots.cern.ch} for the measured rapidity gap distributions at the LHC with different minimum $p_t$ thresholds and Monte Carlo event generator comparisons. The large (diffractive) gaps spanning several units of rapidity originate from charge neutral exchanges, such as exchange of color neutral gluon systems (pomerons) or photons. The interaction rate of gap events may be screened by additional elastic or inelastic interactions, this is known as gap survival discussion. An important property is that the final states down to arbitrary infrared cutoff, not experimentally accessible, modifies the observed rates of rapidity gap events. This corresponds e.g. to extra emission of soft gluons. Thus, the rapidity gap events are not clearly `infrared safe' without explicit $p_t$ threshold definitions. Later in this work, in Figure \ref{fig:ptsimulation}, we show with a toy Monte Carlo simulation how the combinatorial gap rates run as a function of $p_t$ thresholds for four different particle densities per rapidity slice. We see that it is possible, given different $p_t$ thresholds and particle densities, to run the relative event rates of different final state topologies to wildly different.

For the non-diffractive events, now let $Y_0,Y_1,\dots,Y_n$ be $n+1$ rapidity intervals spanned by an $n$ particle final state proceed e.g. via `string breaking process'. To calculate statistics of these, we may use the Darling's contour integral technique from mathematical statistics \cite{darling1953class} via Laplace transform
\begin{equation}
\mathbb{E}[f_0(Y_0)f_1(Y_1)\dots f_n(Y_n)] = \frac{n!}{2\pi i} \int_{c-i\infty}^{c+i\infty} e^z \prod_{j=0}^n \int_0^\infty dz\, dr_j \, e^{-r_j z} f_j(r_j).
\end{equation}
The path of integral is along the straight line $\text{Re}\{z\} = c$ and $f_j(x)$ are arbitrary real valued functions. Using this expectation integral, one can derive numerous results regarding the spacing statistics. For example:
\vspace{2em}
\\
\noindent
\textsc{MaxGap}: The probability that all rapidity gaps are smaller than $\beta$ is
\begin{equation}
\label{eq: maxgap}
P(Y_j < \beta \; \forall j) = \sum_{0 \leq j < 1/\beta}
\begin{pmatrix}
n + 1 \\
j
\end{pmatrix}
(-1)^j (1-\beta j)^n.
\end{equation}
\\
\textsc{MinGap}: The probability that all rapidity gaps are larger than $\alpha$ is
\begin{equation}
\label{eq: mingap}
P(Y_j > \alpha \; \forall j) = (1 - (n + 1)\alpha)^n \;\;\; \text{s.t.} \;\;\; \alpha  < 1/(n+1).
\end{equation}
Both of these are derived in the Darling's paper \cite{darling1953class} under the uniform distribution assumptions. We see that given uniformly distributed rapidities, the number of final states gives physical boundaries for the probabilities, visible in the inequalities in Equation \ref{eq: maxgap} and \ref{eq: mingap}. Now clearly, one can consider higher-order spacing statistics such as distributions of the largest and the second largest gap and so on. Clearly, our incidence algebra construction is an integrating embedding of these taking also into account the translation (boost) position on the rapidity axis.

Our construction includes, in the limit $N \rightarrow \infty$, exactly the `traditional' rapidity gap description. That is, one can recover the one dimensional rapidity gap distribution from the $N$-dimensional combinatorial distribution. Moreover, in practice $N$ needs not to be infinite due to random correlation length over rapidity which smears out useful information at very small gap values.

\subsection{Hypergraph and topology}

\begin{figure}[H]
\centering
\includegraphics[scale=0.6]{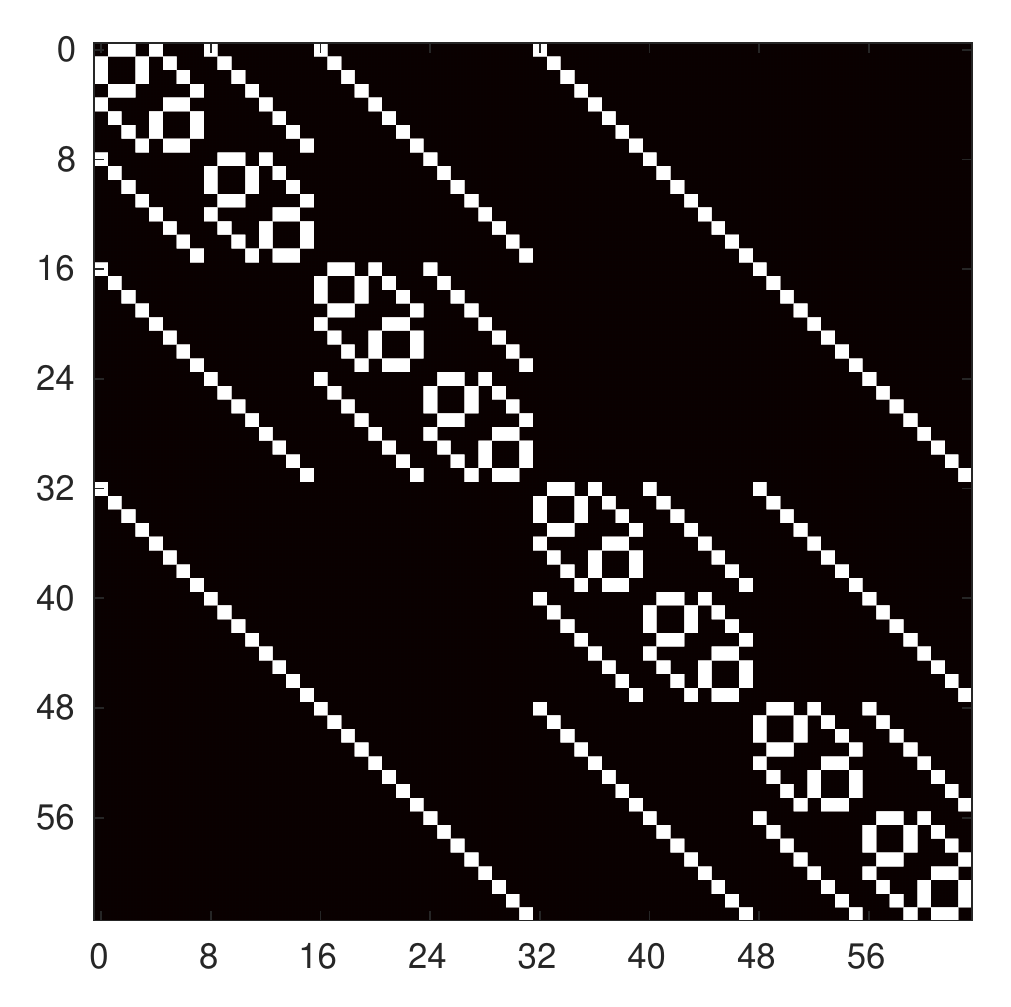}
\caption{Binary vector space illustrated as an $N = 6$ hypercube graph with its corresponding adjacency matrix (black is 0, white is 1).}
\label{fig: d6}
\end{figure}

In Figure \ref{fig: d6} we demonstrate the vector space as an equivalent hypercube $G(E,V)$. Each vertex in set $E$ corresponds to one particular final state configuration and partial cross section component. The number of edges in the graph is $N2^{N-1}$, also counting the number of non-zero phase space elements (bits). The total number of subspaces is $2825$, for $N = 6$. That is, it is radically more rich space than a single dimensional projection of rapidity gap distributions. In the following chapters a complete algebraic construction for any finite $N$ will be given.

We point out that our definition of `event topology' is reasonably well defined even if slightly abstract at first sight. Regarding the topological invariants of the hypercube, the Euler characteristic gives us $V - E + F = 2 - 2\gamma(G_N)$ where $F$ being the number of faces and the minimum genus of $N$-cube for $N \geq 2$ is \cite{beineke1965genus}
\begin{equation}
\gamma(G_N) = (N - 4)2^{N-3} + 1.
\end{equation}
That is, if $N \leq 3$, then $\gamma(G_N) = 0$ and we can embed the planar graph without edge crossings on a sphere, $N = 4$ requires a one hole torus and then exponentially more complicated topologies are generated. The connection between final state geometry (kinematics) and topology is now well defined -- more complicated `event topologies' require larger $N$. An interesting construction would be the limit $N\rightarrow \infty$, that is, from finite dimensional hypercubes to infinite dimensional construction. In practice, approximation schemes such as neural networks could be useful proxies for that.

\section{Posets, incidence algebra and M\"obius inversion}
\label{sec:incidencealgebra}

In order to have the necessary mathematical preliminaries to attack our problems further, we need to first define here a few basic concepts well known in combinatorics. For more information, we refer reader to the original papers on incidence algebras by G.C. Rota \cite{rota1964foundations} and a textbook on combinatorics by R. Stanley \cite{stanley1986enumerative}.

Let $P$ be a set. A binary relation $\leq$ between elements $x,y,z \in P$ satisfying
\begin{align}
&1. \text{ Reflexivity: } x \leq x \text{ for all } x \in P \\
&2. \text{ Antisymmetry: if } x \leq y \text{ and } y \leq x, \text{ then } x = y \\
&3. \text{ Transitivity: if } x \leq y \text{ and } y \leq z, \text{ then } x \leq z 
\end{align}
is called a \textit{partial order}. Thus, a set equipped with a partial order is called a \textit{poset}.

Now $(P,\leq)$ is a poset. An \textit{incidence algebra} is constructed by first defining an interval $[x,y] = \{ z \in P \, | \, x \leq z \leq y \}$. Then the incidence algebra on $P$ is
\begin{equation}
I(P) = \{ f : P \times P \rightarrow \mathbb{R} \,|\, f(x,y) = 0 \text{ unless } x \leq y \}.
\end{equation}
Thus, it gives us the precise definition of inclusion operation. The incidence algebra has two operations of addition and multiplication defined as
\begin{align}
&\text{1. Addition}\,:\,(f + g)(x,y) = f(x,y) + g(x,y) \\
&\text{2. Multiplication}\,:\,(f \ast g)(x,y) = \sum_{x \leq z \leq y} f(x,z)g(z,y)
\end{align}
so the addition is point-wise and the `multiplication' is a convolution sum. Thus, division can be understood as a deconvolution operation.

\begin{figure}[H]
\vspace{1em}
\centering
$\begin{array}{ll}
\includegraphics[width=65mm]{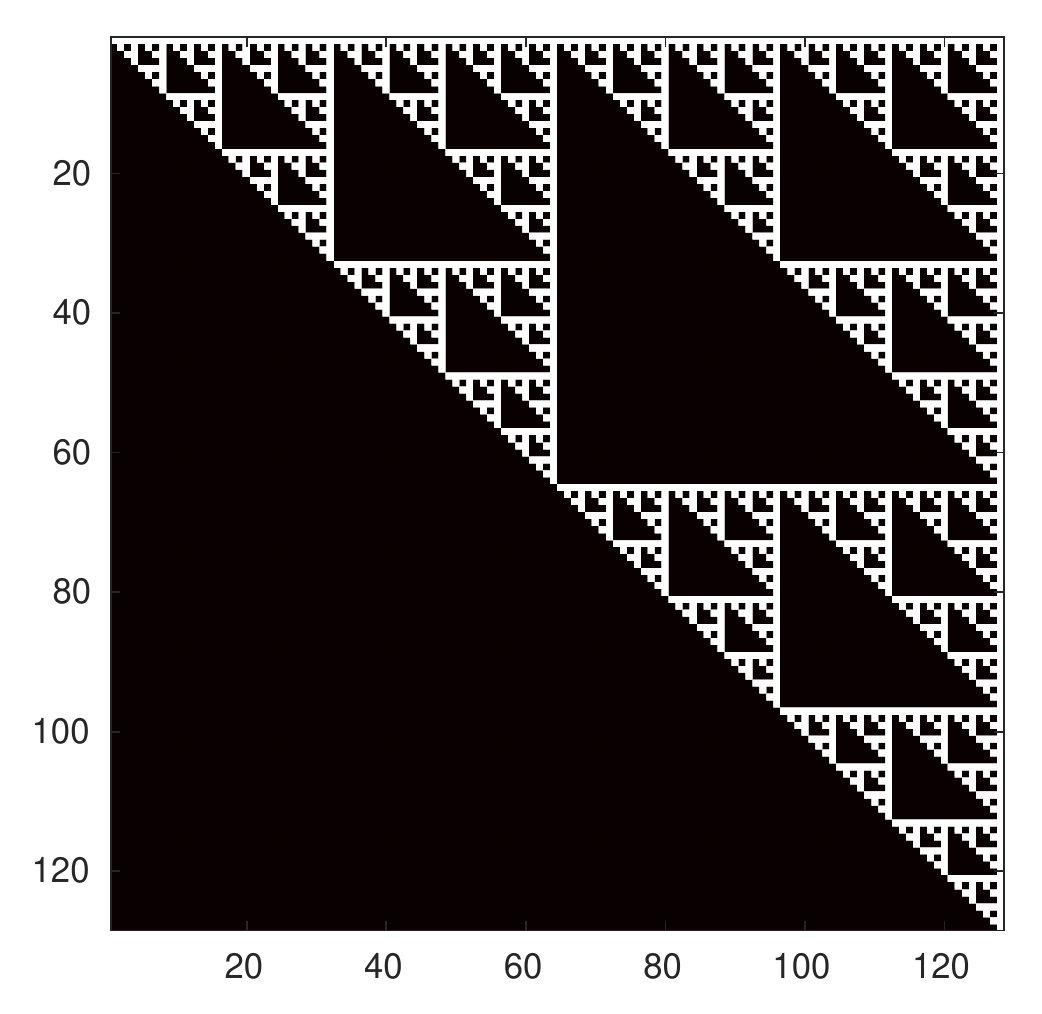} & \hspace{3.5em} \includegraphics[width=65mm]{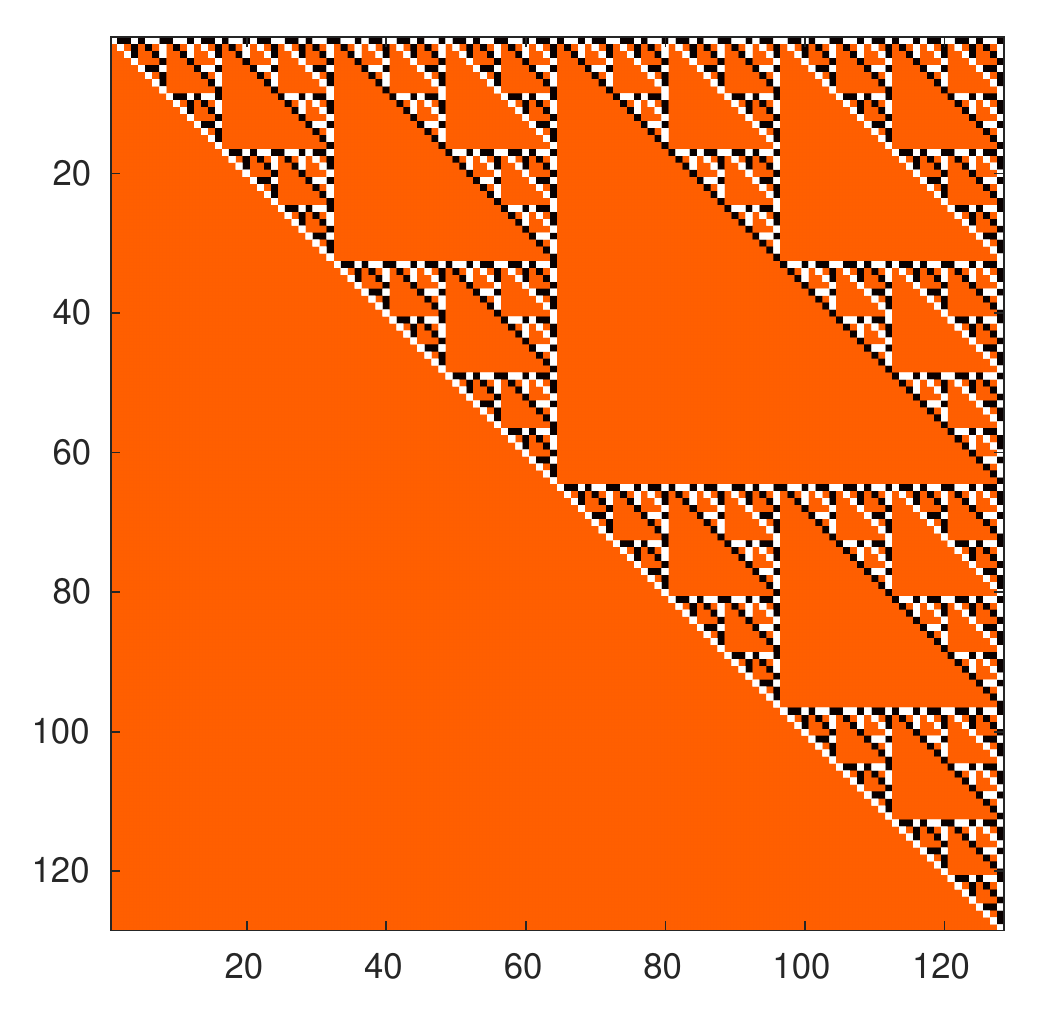} \\ \end{array}$
\caption{Matrix $\zeta_{128\times 128}$ and $\zeta_{128 \times 128}^{-1}$ for $N = 7$. On left black is 0 and white is 1. On right, orange is 0, black is -1 and white is 1.}
\label{fig: zetamatrix}
\end{figure}

In order to proceed, we need a $\zeta$-function\footnote{This has a relation with the Riemann $\zeta$-function, which explains the name.} defined as
\begin{equation}
\zeta(x,y) = 
\begin{cases}
    1, & \text{if $x \leq y$}.\\
    0, & \text{otherwise}.
\end{cases}
\end{equation}
and its inverse $\zeta^{-1}$ is the so-called M\"obius function $u$ (as in number theory) defined with a recursive relation
\begin{align}
&u(x,y) =
\begin{cases}
    1, &\text{if } x = y \\
    -\sum_{x \leq z \leq y} u(x,z) &\text{for} \; x < y \\
    0, &\text{otherwise}.
\end{cases}
\end{align}

The identity or delta-function works as expected in $I(P)$: $\delta(x,y) = 1$ if $x = y$, and $0$ otherwise. In what follows, the $\zeta$-function can be represented with an upper triangular matrix of size $2^N \times 2^N$ with Boolean elements, unit diagonal and determinant 1. Thus, the M\"obius matrix is found easily by inverting the $\zeta$-matrix which results in an upper triangular matrix with unit diagonal and alternating signed elements.

\subsection{The M\"obius inversion theorem}

The \textit{M\"obius inversion theorem} goes as follows \cite{stanley1986enumerative}. Let $(P,\leq)$ be a finite poset and $f$ and $g$ elements of the incidence algebra $I(P)$. The theorem states two equivalent formulas in direct $g \mapsto f$ and inverse $f \mapsto g$ directions
\begin{align}
f(x,y) &= \sum_{x \leq z \leq y} g(x,z) &&\Leftrightarrow\;\;\;\;  f = g \ast \zeta \\
g(x,y) &= \sum_{x \leq z \leq y} f(x,z) u(z,y) &&\Leftrightarrow\;\;\;\; g = f \ast u.
\end{align}
Intuitively, multiplying by $\zeta$ can be understood as integration and by $u$ as differentiation operation. The same convolution products are easily written as
\begin{equation}
\mathbf{f} = \zeta \mathbf{g} \; \Leftrightarrow \; \mathbf{g} = \zeta^{-1} \mathbf{f},
\end{equation}
using a representation with column vectors $\mathbf{f,g}$ and triangular square matrices $\zeta, \zeta^{-1}$.  When increasing the dimension $N$, the recursive Sierpinski triangle fractal structure in these matrices will manifest itself. This is shown in Figure \ref{fig: zetamatrix}.

\subsection{The principle of inclusion-exclusion}

The Principle of Inclusion Exclusion (PIE) is the M\"obius inversion for subsets \cite{stanley1986enumerative}. The basic `sieving' principle of PIE itself has been known for long. Now let different scattering cross section domains for each Bernoulli observable be represented with subsets $D_1,D_2,\dots,D_N \subseteq D$. By defining two functions $f$ and $g$ on the poset $(P,\subseteq)$ with a vector space $\mathbb{F}_2^{|P|}$ as
\begin{align}
f(\chi) &= \left|\bigcap_{i \in \chi} D_i \right| \\
g(\chi) &= \left|\{ x \in D \, | \, x \in D_i \text{ for all } i \in \chi \text{ holds } x \notin D_j \text{ for all } j \notin \chi \} \right|,
\end{align}
where $\chi \subseteq P$. This can be written in the forward direction with
\begin{equation}
f(\Upsilon) = \sum_{\chi \subseteq \Upsilon \subseteq P} g(\chi).
\end{equation}
Now the inverse, the PIE, follows directly as
\begin{equation}
g(\chi) = \sum_{\chi \subseteq \Upsilon \subseteq P} (-1)^{|\Upsilon - \chi|}f(\Upsilon)
\end{equation}
%Now denoting $g(\varnothing) = |D - \bigcup_{i = 1}^N D_i|$, the M\"obius inversion gives us the principle of inclusion-exclusion
%\begin{align}
%\left|D - \bigcup_{i = 1}^N D_i \right| = g(\varnothing) &= \sum_{\varnothing \subseteq Y} u(\varnothing, Y) f(Y) \\
%&= \sum_Y (-1)^{|Y|} \left|\bigcap_{i \in Y} D_i \right|
%\end{align}
with the earlier introduced M\"obius function
\begin{equation}
u(\chi,\Upsilon) = 
\begin{cases}
    (-1)^{|\Upsilon - \chi|}, & \text{if $\chi \subseteq \Upsilon$}\\
    0, & \text{otherwise}.
\end{cases}
\end{equation}
Next we apply these tools.

\subsection{Algebraic representations}

The vector valued Bernoulli distributions can be now written using
\begin{equation}
\label{eq:zetarepresentation}
\mathbf{p} = \begin{pmatrix}
 1  &   -1 \\
 0  &   1 \\
\end{pmatrix}^{\otimes \,N}
\mathbf{m} \, \Leftrightarrow \, \mathbf{m} = \begin{pmatrix}
 1  &   1 \\
 0  &   1 \\
\end{pmatrix}^{\otimes \,N} \mathbf{p},
\end{equation}
where we immediately recognize on left $\zeta^{-1}$ and on right the $\zeta$-matrix obtained via recursive use of the Kronecker tensor product. This is the form given in \cite{teugels1990some}, but Teugels did not point out the connection to the incidence algebras or M\"obius functions. The form is very similar to ones often used in quantum mechanics with finite degrees of freedom, for example the Hilbert space decompositions into subsystems.
\noindent
Using the Kronecker tensor products, the probability vector $\mathbf{p}$ is
\begin{align}
p_c &=
\left\langle
\left(\begin{array}{cc}
 1  &   -1 \\
 0  &   1 \\
\end{array}\right)^{\otimes \,N}
\left(\begin{array}{c}
 1  \\
 B_N  \\
\end{array}\right)
\otimes
\left(\begin{array}{c}
 1  \\
 B_{N-1}  \\
\end{array}\right)
\otimes
\cdots
\otimes
\left(\begin{array}{c}
 1  \\
 B_1  \\
\end{array}\right)
\right\rangle_c.
\end{align}
The polynomial vector of \textit{ordinary moments} $\mathbf{m}$ is
\begin{align}
m_c &=
\left \langle \prod_{i = 1}^N B_i^{b^c_i} \right \rangle
=
\left\langle
\left(\begin{array}{c}
 1   \\
 B_N \\
\end{array}\right)
\otimes
\left(\begin{array}{c}
 1  \\
 B_{N-1} \\
\end{array}\right)
\otimes
\cdots
\otimes
\left(\begin{array}{c}
 1   \\
 B_1 \\
\end{array}\right)
\right\rangle_c.
\end{align}
The polynomial vector of \textit{central moments} $\bm{\delta}$ is
\begin{align}
\nonumber
\delta_c &= 
\left\langle \prod_{i = 1}^N (B_i - \langle B_i \rangle)^{b^c_i} \right\rangle \\
&=
\left\langle
\left(\begin{array}{c}
 1   \\
 B_N-\langle B_N \rangle \\
\end{array}\right)
\otimes
\left(\begin{array}{c}
 1  \\
 B_{N-1}-\langle B_{N-1} \rangle \\
\end{array}\right)
\otimes
\cdots
\otimes
\left(\begin{array}{c}
 1   \\
 B_1-\langle B_1 \rangle \\
\end{array}\right)
\right\rangle_c,
\end{align}
where $c$ is given by the binary expansion, $0 \leq c \leq 2^N - 1$ and $b^c_i \in \{0,1\}$ of $\mathbf{b}^c$. The central moments describe the multipoint correlations $(\#\, 2^N-N-1)$ between any 2 or more subspaces in phase-space (such as rapidity intervals) and $B_i$ are the corresponding Bernoulli random variables. The correlation functions $\langle B_i \dots B_j \rangle$ factorize to a product of $\langle B_i\rangle \dots \langle B_j\rangle$ if no correlations are present, as usual. This property can be used to test factorization of physics, such as factorization properties of (soft) QCD amplitudes.

\subsection{Subspace decompositions}

Now we drop the zero vector case $c = 0$, which corresponds to the case $B_1 = B_2 = \dots B_N = 0$. This gives no contribution being the additive identity of the algebra, physically being outside the fiducial phase space of the definition. Thus the vector $\mathbf{p}$ will be then $2^N-1$ dimensional, and $\sum p_c = 1$ still by definition. By slight abuse of notation, we will continue with this convention for the rest of the paper unless otherwise mentioned. Now analogously to Equation \ref{eq:zetarepresentation}, we write
\begin{equation}
\Lambda \mathbf{p} = \mathbf{r},
\end{equation}
where $\mathbf{r}$ is an auxiliary vector which collects the additive subspace probabilities of different mutually exclusive final states and $\Lambda$ is a full rank (invertible) square Boolean matrix of size $2^N-1$. The number of different subspaces of varying dimensionality is given by $q$-binomials of Equation \ref{eq:qbinomial} in Appendix \ref{sec:appendix}.

Example $N = 2$: Denote the Bernoulli probability of the observable $B_1$ with $P_{B_1}$ and of $B_2$ with $P_{B_2}$. The probability of events obeying fiducial binary combinations $\mathbf{b}_1 = (1,0) \equiv {[B_1]\wedge [\neg{B_2}]}$, $\mathbf{b}_2 = (0,1) \equiv {[\neg{B_1}] \wedge [B_2]}$ and $\mathbf{b}_3 = (1,1) \equiv {[B_1] \wedge [B_2]}$ can be obtained by inverting a system
\begin{align}
\label{eq: N2}
\begin{pmatrix}
1 & 0 & 1 \\
0 & 1 & 1 \\
1 & 1 & 1
\end{pmatrix}
\begin{pmatrix}
P_{1,0} \\
P_{0,1} \\
P_{1,1}
\end{pmatrix}
&=
\begin{pmatrix}
P_{B_1} \\
P_{B_2} \\
P_{{B_1}\vee {B_2}} = P_{B_1} + P_{B_2} - P_{{B_1}\wedge {B_2}} \equiv 1
\end{pmatrix}, \;
\Lambda_{3\times3}^{-1} &=
\begin{pmatrix}
 0  &  -1  &   1 \\
-1  &   0  &   1 \\
 1  &   1  &  -1 
\end{pmatrix}.
\end{align}

Example $N = 3$: Observables are now ${B_1}$, $B_2$, ${B_3}$. We get in a completely analogous way as in the $N = 2$ case
\begin{align}
\label{eq: N3}
\begin{pmatrix}
     1  &   0  &   1  &   0  &   1  &   0  &   1 \\
     0  &   1  &   1  &   0  &   0  &   1  &   1 \\
     0  &   0  &   0  &   1  &   1  &   1  &   1 \\
     1  &   1  &   1  &   0  &   1  &   1  &   1 \\
     1  &   0  &   1  &   1  &   1  &   1  &   1 \\
     0  &   1  &   1  &   1  &   1  &   1  &   1 \\
     1  &   1  &   1  &   1  &   1  &   1  &   1 \\
\end{pmatrix}
\begin{pmatrix}
P_{1,0,0} \\
P_{0,1,0} \\
P_{1,1,0} \\
P_{0,0,1} \\
P_{1,0,1} \\
P_{0,1,1} \\
P_{1,1,1}
\end{pmatrix}
&=
\begin{pmatrix}
P_{B_1} \\
P_{B_2} \\
P_{B_3} \\
P_{{B_1}\vee {B_2}} = P_{B_1} + P_{B_2} - P_{{B_1}\wedge {B_2}}\\
P_{{B_1}\vee {B_3}} = P_{B_1} + P_{B_3} - P_{{B_1}\wedge {B_3}}\\
P_{{B_3}\vee {B_2}} = P_{B_2} + P_{B_3} - P_{{B_2}\wedge {B_3}}\\
P_{{B_1}\vee {B_2}\vee {B_3}} = P_{B_1} + P_{B_2} + P_{B_3} - Z \equiv 1
\end{pmatrix},
\end{align}
\begin{equation}
\Lambda_{7\times7}^{-1} =
\begin{pmatrix}
     0  &   0  &   0  &   0 &    0  &  -1  &   1 \\
     0  &   0  &   0  &   0 &   -1  &   0  &   1 \\
     0  &   0  &  -1  &   0 &    1  &   1  &  -1 \\ 
     0  &   0  &   0  &  -1 &    0  &   0  &   1 \\
     0  &  -1  &   0  &   1 &    0  &   1  &  -1 \\
    -1  &   0  &   0  &   1 &    1  &   0  &  -1 \\
     1  &   1  &   1  &  -1 &   -1  &  -1  &   1 \\
\end{pmatrix},
\end{equation}
where $Z = P_{{B_1}\wedge {B_2}} + P_{{B_1}\wedge {B_3}} + P_{{B_2}\wedge {B_3}} - P_{{B_1}\wedge {B_2} \wedge {B_3}}$.

\begin{figure}[H]
\vspace{1em}
\centering
$\begin{array}{ll}
\includegraphics[width=65mm]{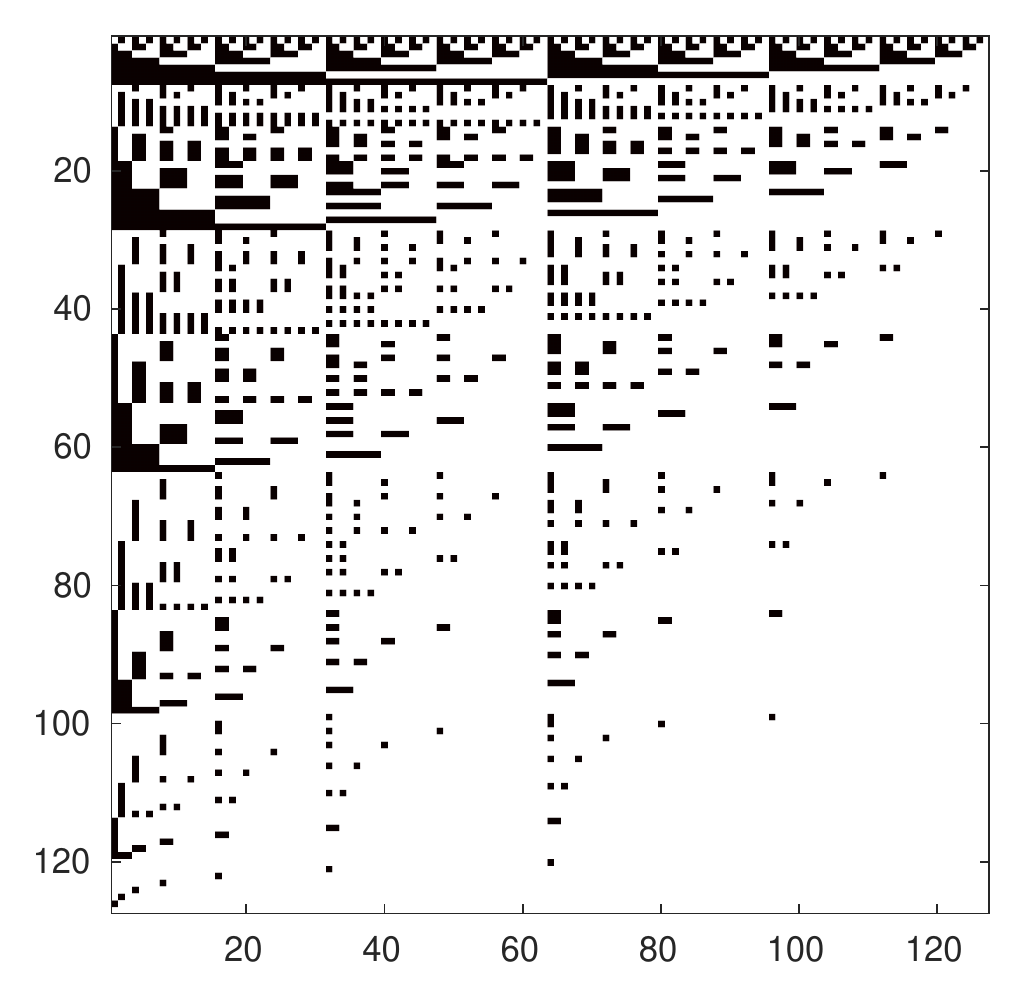} & \hspace{3.5em} \includegraphics[width=65mm]{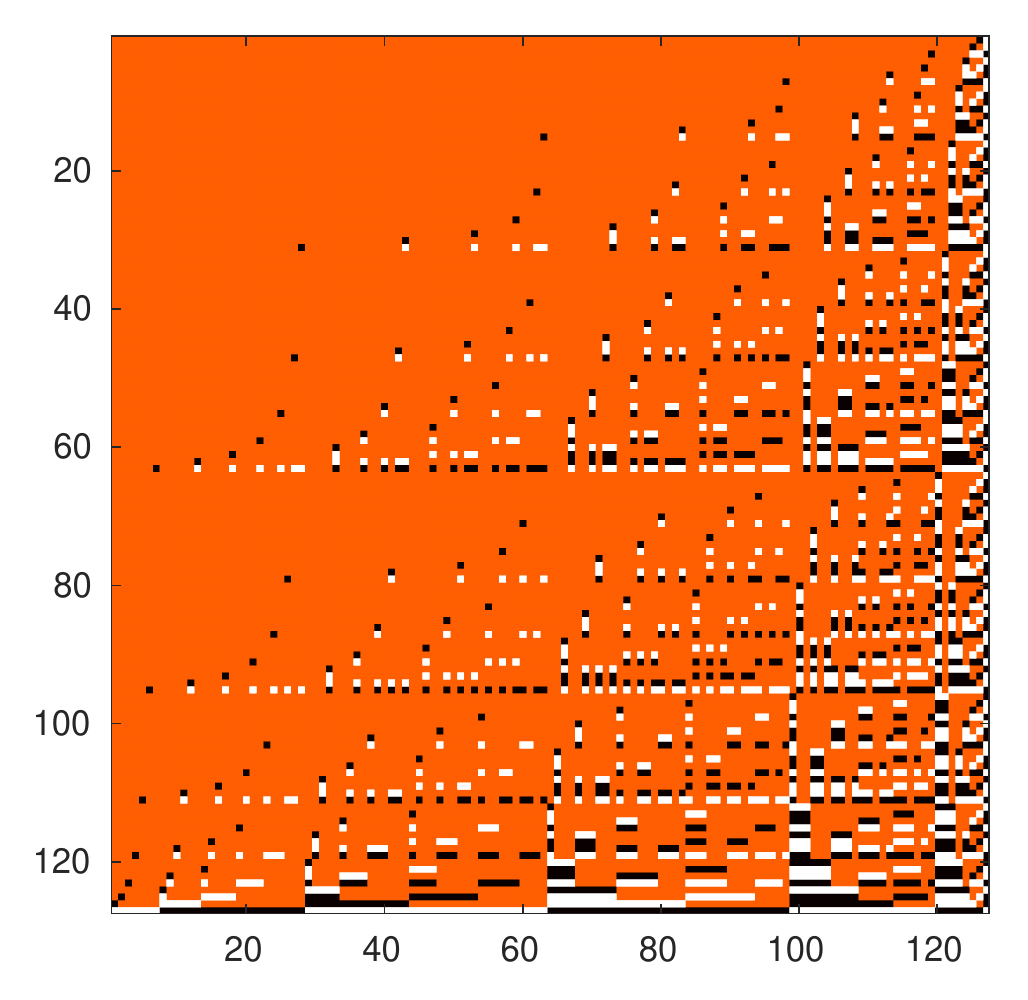} \\ \end{array}$
\caption{Matrix $\Lambda_{127\times 127}$ and $\Lambda_{127 \times 127}^{-1}$ for $N = 7$. On left black is 0 and white is 1. On right, orange is 0, black is -1 and white is 1.}
\label{fig: binarymatr}
\end{figure}

We can proceed in a same way up to an arbitrary number $N$ of Bernoulli observables, which span a number of $2^N-1$ non-zero combinations. To summarize: the idea is to generate a system of all linear combinations of event rates described by a full rank (invertible) matrix $\Lambda$. This matrix and its inverse will be used directly in the solution to the compound inverse problem. The matrix $\Lambda$, or more precisely the inverse of it, can be generated recursively utilizing the principle of inclusion-exclusion 
\begin{equation}
P(\bigcup_{i = 1}^N D_i) = \sum_{i = 1}^N \left( (-1)^{i-1} \sum_{I \subset \{1,\dots,N\}, |I| = i} P(D_I) \right).
\end{equation}
The inner sum runs over all subsets $I$ constructed by the indices $1,\dots,N$, with cardinality of $I$ identically $i$ elements. Also above we have defined
\begin{equation}
D_I := \bigcap_{i \in I} D_i.
\end{equation}
Figure \ref{fig: binarymatr} shows an algorithmically generated example where the fractal structure is again evident. Finally the relation between the inverses of $\zeta$-matrix and $\Lambda$ matrix can be written as
\begin{equation}
[\Lambda^{-1} (\Lambda^{-1})^T]_{i,j} =\begin{cases}
               [(\zeta^{-1})^T \zeta^{-1}]_{i+1,j+1}, \;\; \text{ when } 1 \leq i,j \leq 2^N-1 \\
[(\zeta^{-1})^T \zeta^{-1}]_{i+1,j+1}  - 1, \;\; \text{ if } i=j=2^N-1. \\
            \end{cases},
\end{equation}
which is evident by careful inspection of the definitions.

Because our space $\mathbb{F}_2^N$ is linear, all subspaces $\mathbb{F}_2^M$ with $1\leq M < N$ are naturally included in it as $\mathbb{F}_2^M \subseteq \mathbb{F}_2^N$, and we used that property. Additional funny remark is that the probability to have $n \times n$ matrix over $\mathbb{F}_2$ with determinant $\neq 0$ is 1/4. This is easily proven.

%\begin{figure}[h]
%\begin{center}
%\includegraphics[width=65mm]{fishmatrix.eps}
%\end{center}
%\caption{Matrix $\Lambda\Lambda^T$.}
%\label{fig: fishmatrix}
%\end{figure}

\newpage

\section{Measurements}
\label{sec:measurements}

Regarding proper observables and measurements we could start to write down suitable theory formulations from which one could perhaps derive semi-analytical distributions for the Bernoulli observables, at least with toy models. However, here our abstraction level is different. This construction is essentially a new model independent definition of high energy diffraction: different correlations over space-time observables, in terms of rapidity and transverse momentum (and multiplicity), are embedded in the construction and predictions can be obtained directly with any Monte Carlo event generator. We rely only on the observable fiducial final state information. To point out, our construction has some similarity with the coherence definition by Glauber in quantum optics \cite{glauber1963quantum}.

\subsection{Partial cross sections}

To be concrete, a practical measurement procedure of the combinatorial partial cross sections, which may be called also vector fiducial cross sections, goes as follows
\vspace{1em}
\\
\noindent
\textbf{1.} \textsc{Input:} Use minimum bias or zero-bias (bunch cross-over) triggered data. \\
\textbf{2.} \textsc{Beam Background Filter}: Filter event by event the beam-gas and satellite interactions via detector time-domain cuts, if possible. Tune the beam background cuts using a combination of Monte Carlo (pure beam-beam) and data in a way that beam-beam interaction efficiency is preserved. \\
\textbf{3.} \textsc{Fiducial Cut Definition}: Apply fiducial cuts for pseudorapidity $\eta$ and transverse momentum $p_t$. A single fixed set of cuts gives $2^N-1$ non-zero observable binary combinations. A sliding threshold cuts give a `trajectory of observables'. \\
\textbf{4.} \textsc{Residual Beam Background Correction}: Correct residual beam-gas background at statistical level using beam-empty, empty-beam, empty-empty trigger masks: $
N \leftarrow N - \sum_{j} \alpha_j N_j$, where weight factors $\alpha_j$ are obtained from the $j$-th trigger mask statistics with pre-determined and/or random trigger downscaling and $N$ is the number of events. \\
\textbf{5.} \textsc{Pile-Up}: Correct pile-up via statistical M\"obius inversion if the run was with high instantaneous luminosity.  \\
\textbf{6.} \textsc{Luminosity}: Map event counts to visible cross section units via integrated luminosity, calibrated via van der Meer scans, see Appendix \ref{luminosity}. \\
\textbf{7.} \textsc{Multidimensional Unfolding}: Unfold visible partial cross sections to fiducial partial cross sections. One may use `standard' unfolding algorithms. \\
\textbf{8.} \textsc{Output:} A vector $\bm{\sigma}$ (set) of fiducial cross sections.
\\
\vspace{1em}

\noindent In addition, one may now use the unfolded partial cross sections to obtain highly constraining multidimensional fit extractions of diffractive cross sections and the Pomeron parameters or the maximum diffractive system mass limit, sometimes called the `coherence limit' \cite{goulianos1983diffractive}. Also re-projections of other observables are possible. For highly efficient algorithms, see Appendix \ref{sec:Fstar_projection} and \ref{sec:fitalgorithms}.

\begin{figure}[H]
\centering
\includegraphics[scale=0.65]{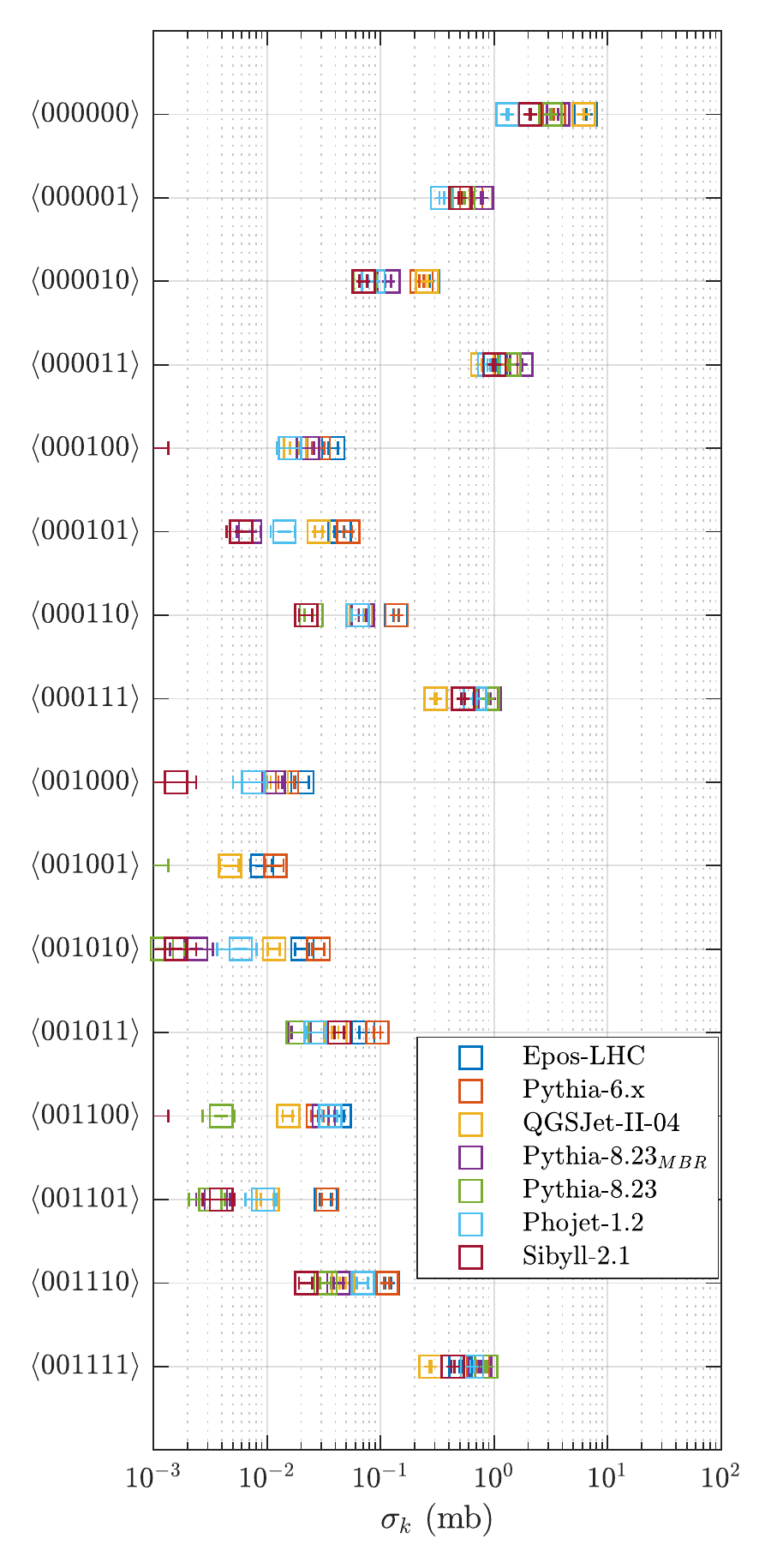}
\includegraphics[scale=0.65]{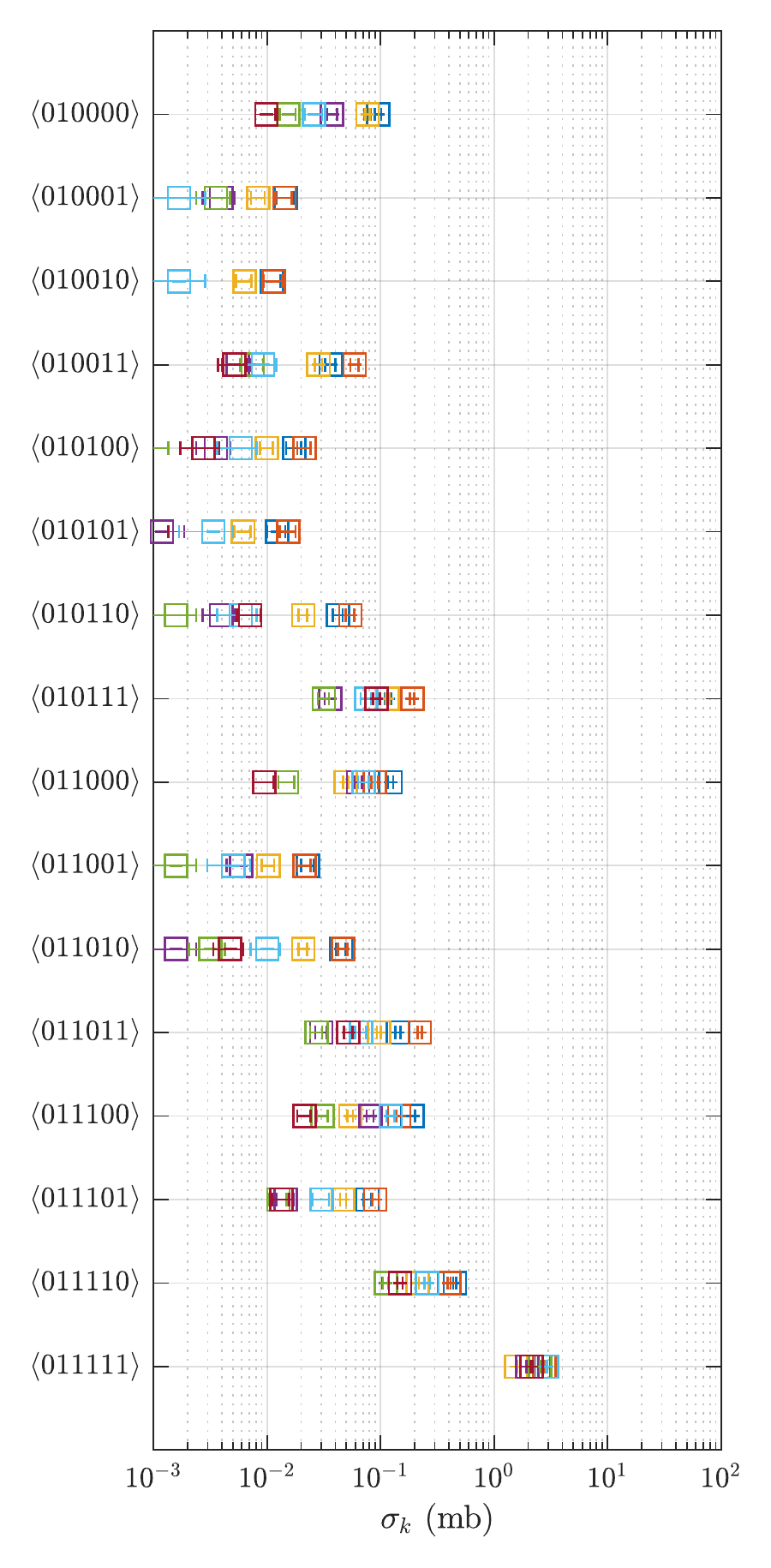}
\includegraphics[scale=0.65]{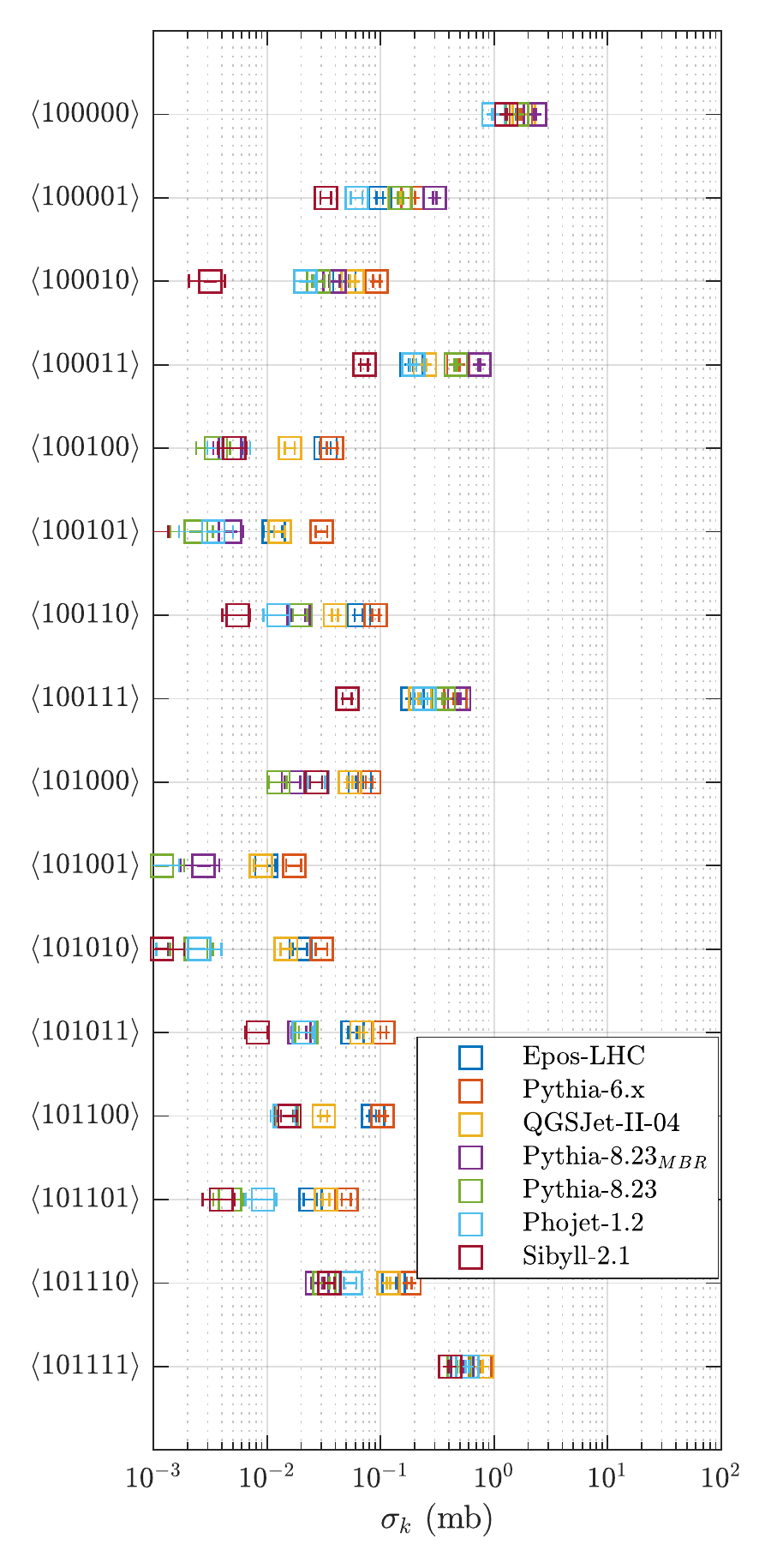}
\includegraphics[scale=0.65]{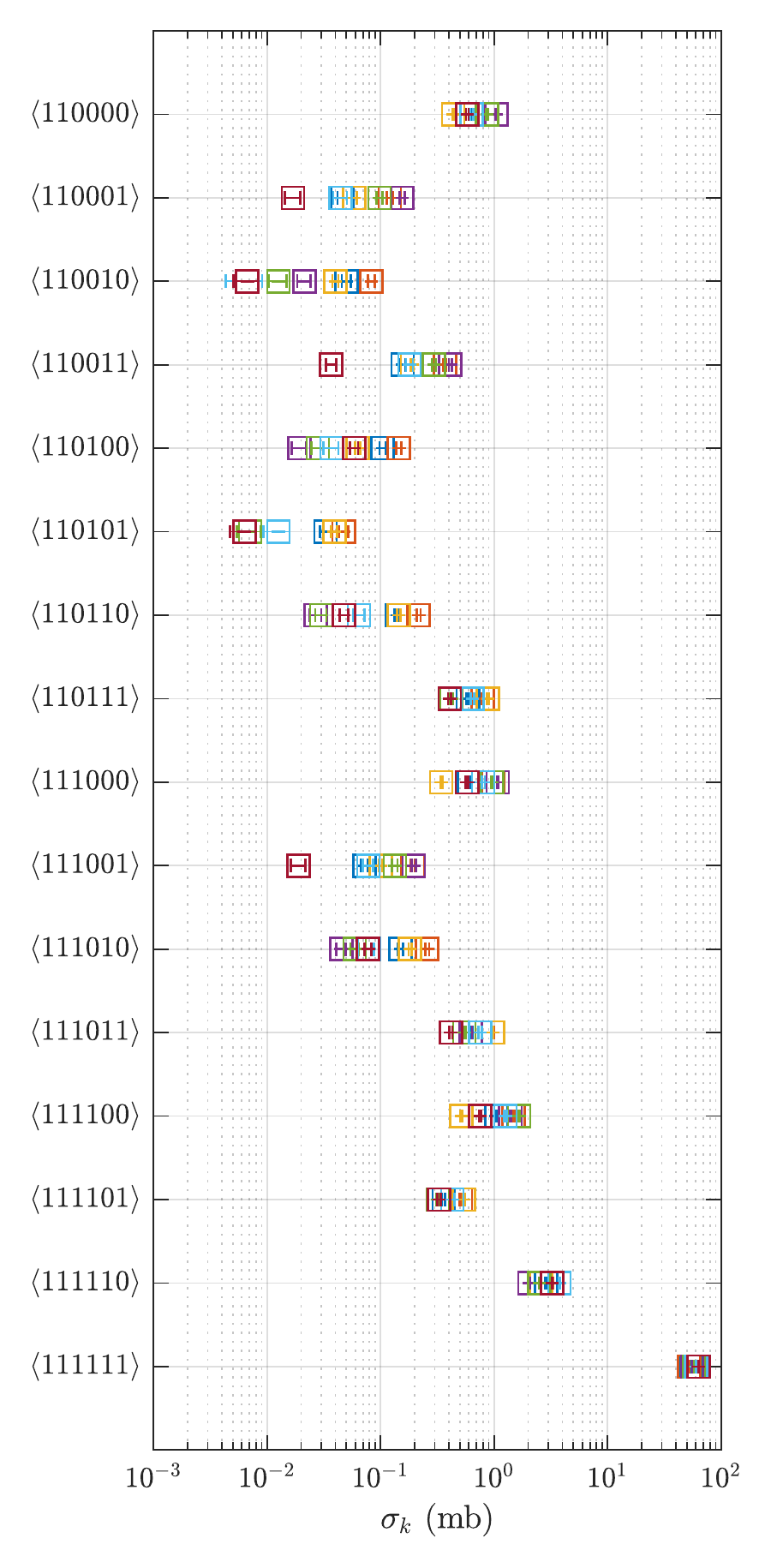}
\caption{Partial cross sections evaluated with proton-proton minimum bias Monte Carlo event generators at $\sqrt{s} = 13$ TeV. Statistical uncertainties are indicated with error bars. Total inelastic cross section set for all models to 80 mb.}
\label{fig: MCfiducial}
\end{figure}

In Figure \ref{fig: MCfiducial} we have several Monte Carlo model \cite{ahn2009cosmic, ciesielski2012mbr, sjostrand2015introduction, sjostrand2006pythia, bopp1998rapidity, pierog2015epos, ostapchenko2011monte} fiducial proton-proton cross sections at $\sqrt{s} = 13$ TeV with cuts suitable for the ALICE experiment. The fiducial acceptance domains for charged particles with $N=6$ are
\begin{equation}
\eta \in [-7.0, -4.9], [-3.7, -1.7], [-2.0, 0],[0, 2.0],[2.8, 5.1],[4.8, 6.3] \; \wedge \; p_t > 0.05 \; \text{GeV},
\end{equation}
motivated by the nominal acceptances of AD, V0 and SPD detectors of the ALICE experiment. A large variation between models is seen, representing different tunes and design choices. In certain sense, the vector fiducial cross sections are closer to classic diffraction measurements (aperture + screen), than the typical rapidity gap counting. Regarding the multidimensional unfolding, we recommend using several reference MC models together with the detector simulation. One should also characterize the unfolding regularization parameter strength in a data-driven way, for example, using the Shannon information entropy
\begin{equation}
S = -\sum_i P_i \ln P_i,
\end{equation}
where $P_i \equiv \sigma_i / \sum_j \sigma_j$, by calculating the entropy of the unfolded spectrum and the relative entropy between the unfolded and raw combination rates
\begin{equation}
D = \sum_i P_i \ln \left( \frac{ P_i }{ \tilde{P}_i   }\right).
\end{equation}

Naturally, pure simulation studies should be done in addition. The non-trivial systematics behind efficiency corrections are made here very explicit by the multidimensional unfolding, which accounts for both efficiency losses and the flow of event topologies from one combination to another. This flow is not just because of efficiency losses but also due to major material re-scattering and other detector effects, often hardly understood with single dimensional rapidity gap counting measurements.

To this end we point out that in principle one could extend the fiducial measurement vector space from the GF(2) finite field $\mathbb{F}_2^N$ to $\mathbb{F}_q^N$, where $q$ is an integer power of a prime number. The prime power is a finite field requirement. We see that this could be useful regarding different rapidity slices containing different extreme (low or high) multiplicity densities in the same event, not necessarily hard scale or jets driven, for example.

\newpage

\subsection{Fractal marginal distributions}

We proceed as before, but now measure all single dimensional distributions available per binary combination, such as the multiplicity $P(N_{ch})$ and transverse momentum $P(p_t)$ distributions, once the slicing is done over (pseudo)rapidity. For example, assume $N = 3$ slices over pseudorapidity. For combination $\langle 0,0,1 \rangle$ we have one non-zero component $P(N_{ch})_{\langle 0,0,[1] \rangle}$, for $\langle 0,1,1 \rangle$ we get $P(N_{ch})_{\langle 0,1,[1] \rangle}$, $P(N_{ch})_{\langle 0,[1],1 \rangle}$ and similarly for transverse momentum. As an illustrative example, we tabulate those in Tables \ref{table: 1Dsubmarginals} and \ref{table: 2Dsubmarginals} with $N=3$. The number $a(n)$ of non-zero distributions per vector combination is enumerated with $a(0) = 0, a(2n) = a(n), a(2n+1) = a(n) + 1$, producing a fractal sequence $0,1,1,2,1,2,2,3,\dots$. That is simply the Hamming weight. The total number of distributions is $N2^{N-1}$, the number of edges in an $N$-hypercube.
\\
\begin{table}[h]
\center
\begin{tabular}{c||c|c|c|}
1 & \cellcolor{gray!25} $\emptyset$ & \cellcolor{gray!25} $ \emptyset$ & \cellcolor{green!25} $f(\mathcal{O}_x)_{\langle 0,0,[1] \rangle}$ \\ 
\hline 
2 & \cellcolor{gray!25} $\emptyset$ & \cellcolor{green!25} $f(\mathcal{O}_x)_{\langle 0,[1],0 \rangle}$ & \cellcolor{gray!25} $\emptyset$ \\ 
\hline 
3 & \cellcolor{gray!25} $\emptyset$ & \cellcolor{green!25} $f(\mathcal{O}_x)_{\langle 0,[1],1 \rangle}$ & \cellcolor{green!25} $f(\mathcal{O}_x)_{\langle 0,1,[1] \rangle}$ \\ 
\hline 
\vdots &  & \vdots &  \\
\hline
8 & \cellcolor{green!25} $f(\mathcal{O}_x)_{\langle [1],1,1 \rangle}$ & \cellcolor{green!25} $f(\mathcal{O}_x)_{\langle 1,[1],1 \rangle}$ & \cellcolor{green!25} $f(\mathcal{O}_x)_{\langle 1,1,[1] \rangle}$ \\
\hline 
\hline
$\Sigma \downarrow $ & $\underbrace{f(\mathcal{O}_x)_{\langle [1],-,- \rangle}}_{\int_{\Delta \eta_1} d\eta}$ & $\underbrace{f(\mathcal{O}_x)_{\langle -,[1],- \rangle}}_{\int_{\Delta \eta_2} d\eta}$ & $\underbrace{f(\mathcal{O}_x)_{\langle -,-,[1] \rangle}}_{\int_{\Delta \eta_3} d\eta}$
\end{tabular}
\caption[Table caption text]{Direct combinatorially embedded marginal 1D-distributions of the observable $\mathcal{O}_x$ with $N=3$ slices over pseudorapidity.}
\label{table: 1Dsubmarginals}
\end{table}

\begin{table}[h]
\center
\begin{tabular}{c||c|c|c|c}
1 & \cellcolor{gray!25} $\emptyset$ & \cellcolor{gray!25} $\emptyset$ & \cellcolor{green!25} $f(\mathcal{O}_x,\mathcal{O}_y)_{\langle 0,0,[1] \rangle}$ \\ 
\hline 
2 & \cellcolor{gray!25} $\emptyset$ & \cellcolor{green!25} $f(\mathcal{O}_x,\mathcal{O}_y)_{\langle 0,[1],0 \rangle}$ & \cellcolor{gray!25} $\emptyset$ \\ 
\hline 
3 & \cellcolor{gray!25} $\emptyset$ & \cellcolor{green!25} $f(\mathcal{O}_x,\mathcal{O}_y)_{\langle 0,[1],1 \rangle}$ & \cellcolor{green!25} $f(\mathcal{O}_x,\mathcal{O}_y)_{\langle 0,1,[1] \rangle}$ \\ 
\hline
\vdots &  & \vdots &  \\
\hline 
8 & \cellcolor{green!25} $f(\mathcal{O}_x,\mathcal{O}_y)_{\langle [1],1,1 \rangle}$ & \cellcolor{green!25} $f(\mathcal{O}_x,\mathcal{O}_y)_{\langle 1,[1],1 \rangle}$ & \cellcolor{green!25} $f(\mathcal{O}_x,\mathcal{O}_y)_{\langle 1,1,[1] \rangle}$ \\ 
\hline
\hline
$\Sigma \downarrow$ & $\underbrace{f(\mathcal{O}_x,\mathcal{O}_y)_{\langle [1],-,- \rangle}}_{\int_{\Delta \eta_1} d\eta}$ & $\underbrace{f(\mathcal{O}_x,\mathcal{O}_y)_{\langle -,[1],- \rangle}}_{\int_{\Delta \eta_2} d\eta}$ & $\underbrace{f(\mathcal{O}_x,\mathcal{O}_y)_{\langle -,-,[1] \rangle}}_{\int_{\Delta \eta_3} d\eta}$ \\
\end{tabular}
\caption[Table caption text]{Direct combinatorially embedded marginal 2D-distributions of observables $(\mathcal{O}_x,\mathcal{O}_y)$ with $N=3$ slices over pseudorapidity.}
\label{table: 2Dsubmarginals}
\end{table}

We emphasize here that this is just cut-and-count with bookkeeping, a well defined fiducial measurements. The problem of unfolding from detector level to particle level, however, can be non-trivial. We believe that these fractal distributions together with combinatorial partial cross sections can be key handles to test the AGK rules and (topological) fluctuations, perhaps driven by anomalous QCD color field configurations. A related theory discussion may be in \cite{lappi2006some, gelis2008high}.

\subsection{Threshold gapflow trajectories}

The idea here is a straightforward one. One measures $2^N-1$ fiducial vector cross sections over rapidity as a function of a threshold condition. The threshold condition can be global over rapidity, or local. For example, minimum transverse momentum threshold, minimum multiplicity threshold, minimum flavor content threshold (e.g. strangeness).

\begin{figure}[H]
\centering
$\begin{array}{ccc}
\includegraphics[width=75mm]{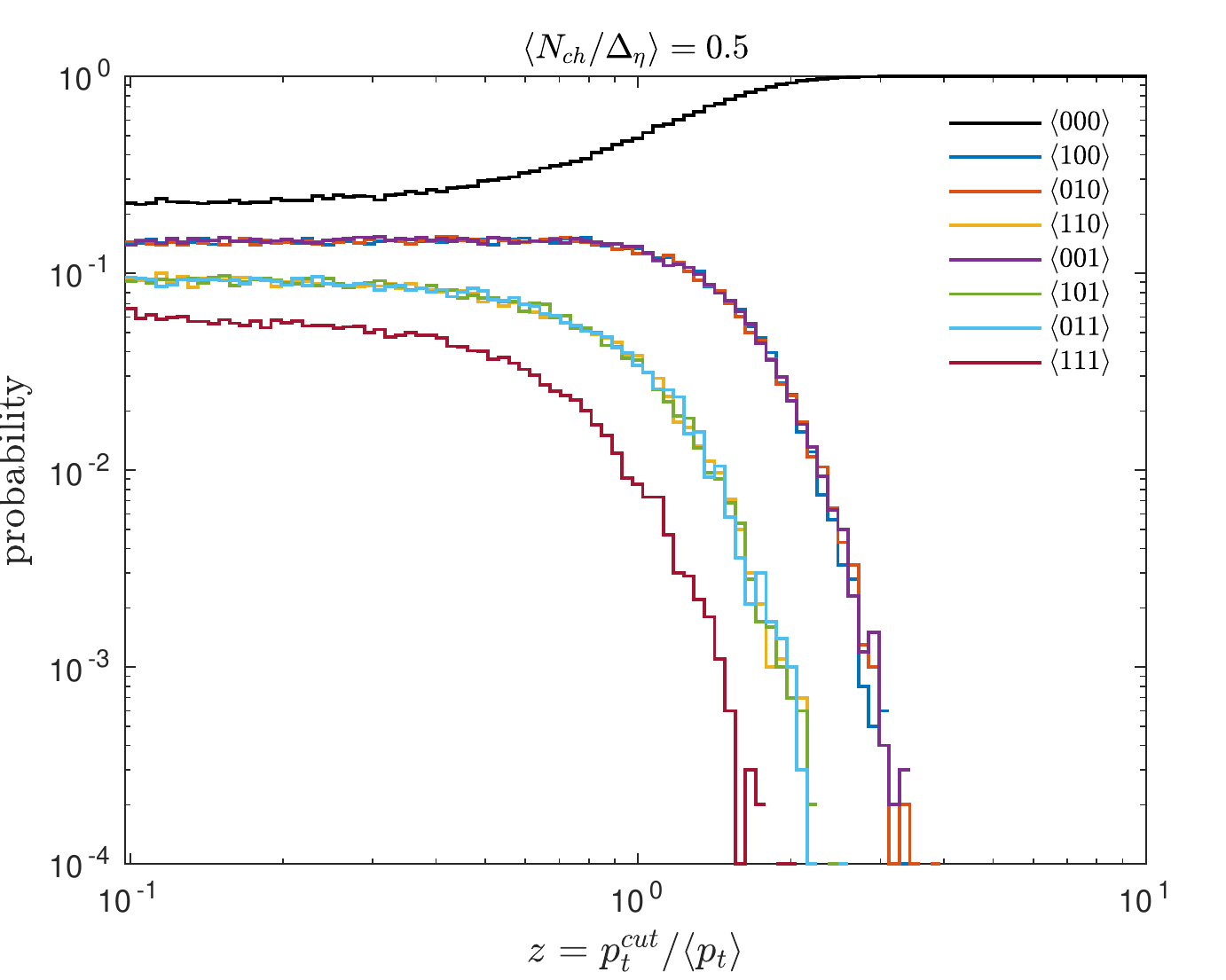} &
\includegraphics[width=75mm]{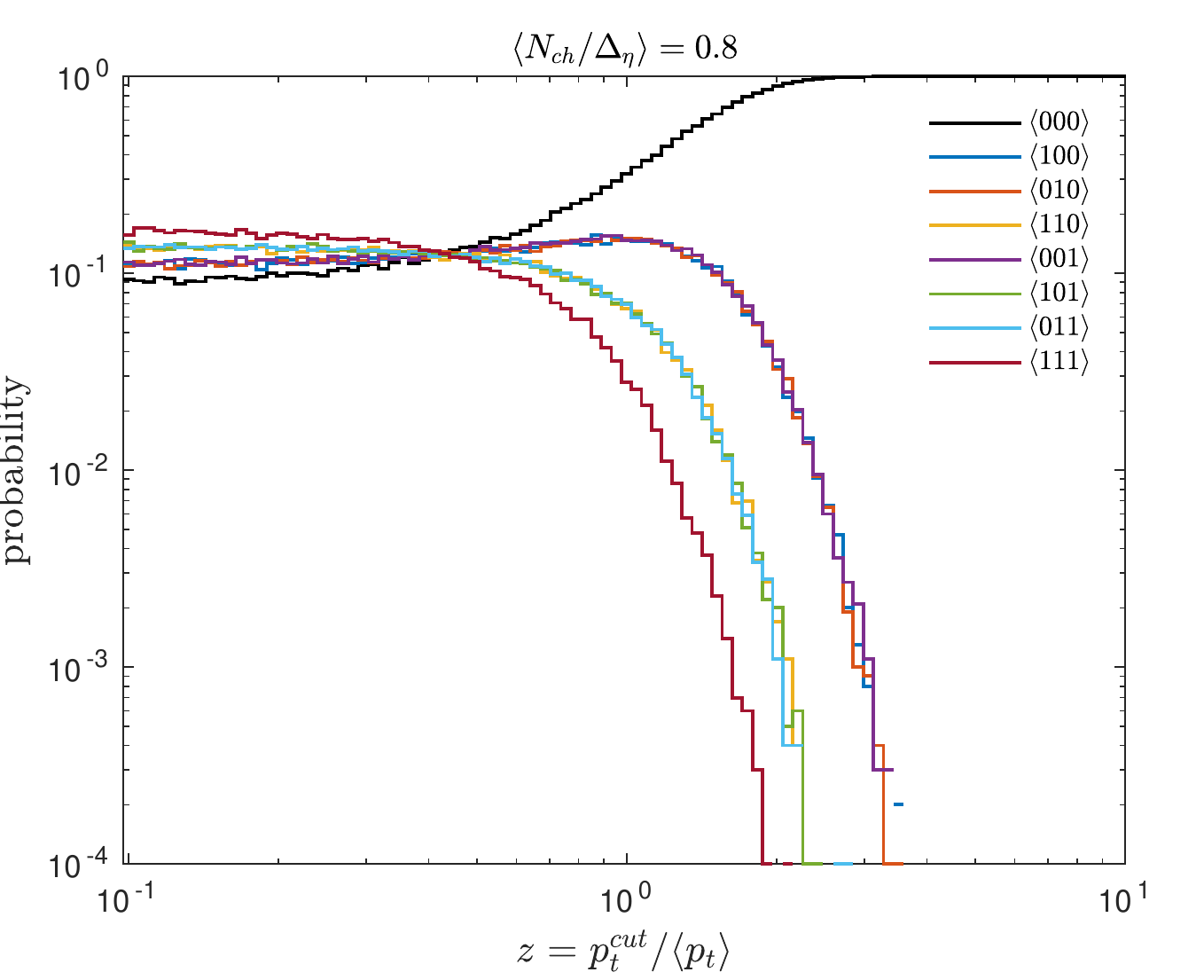} \\
\includegraphics[width=75mm]{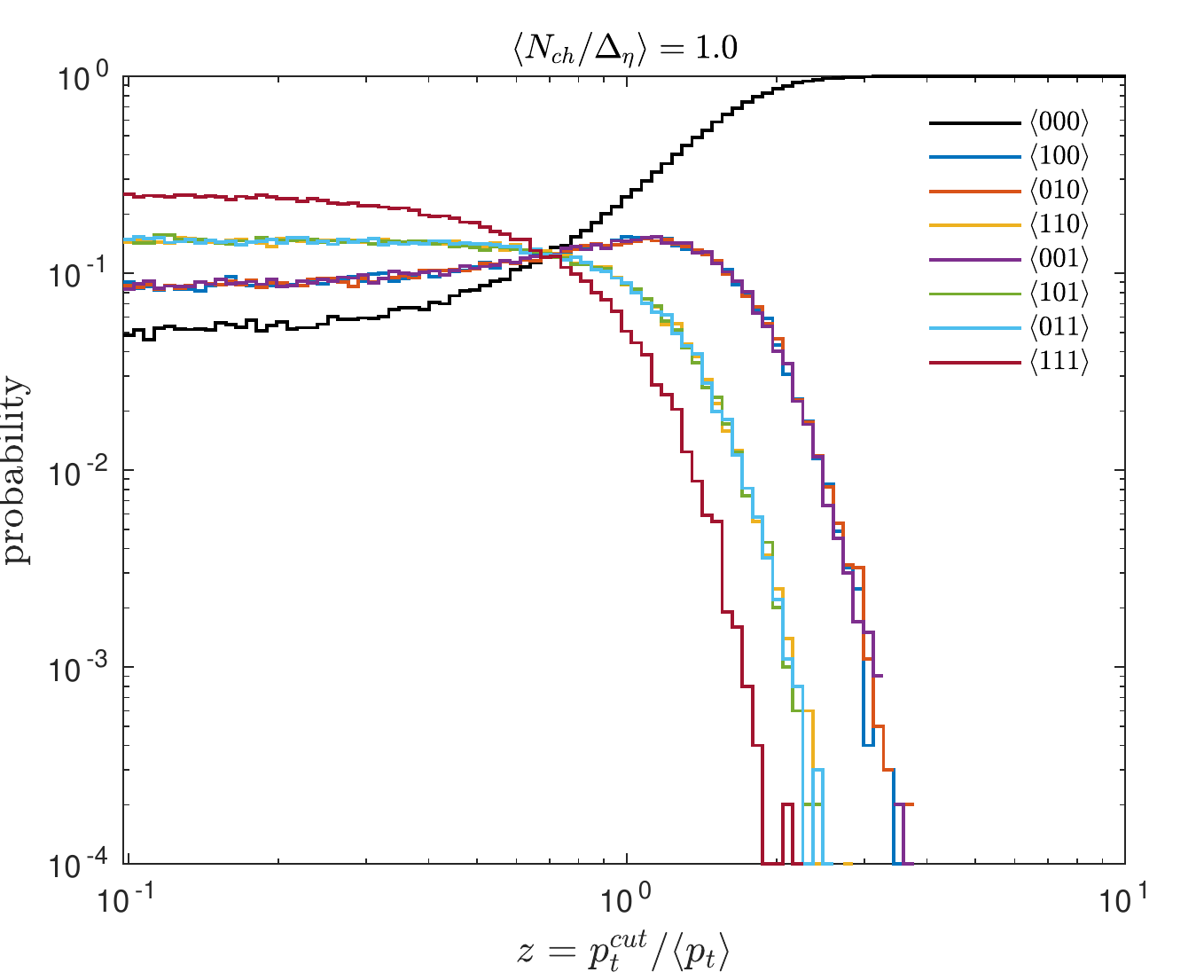} & \includegraphics[width=75mm]{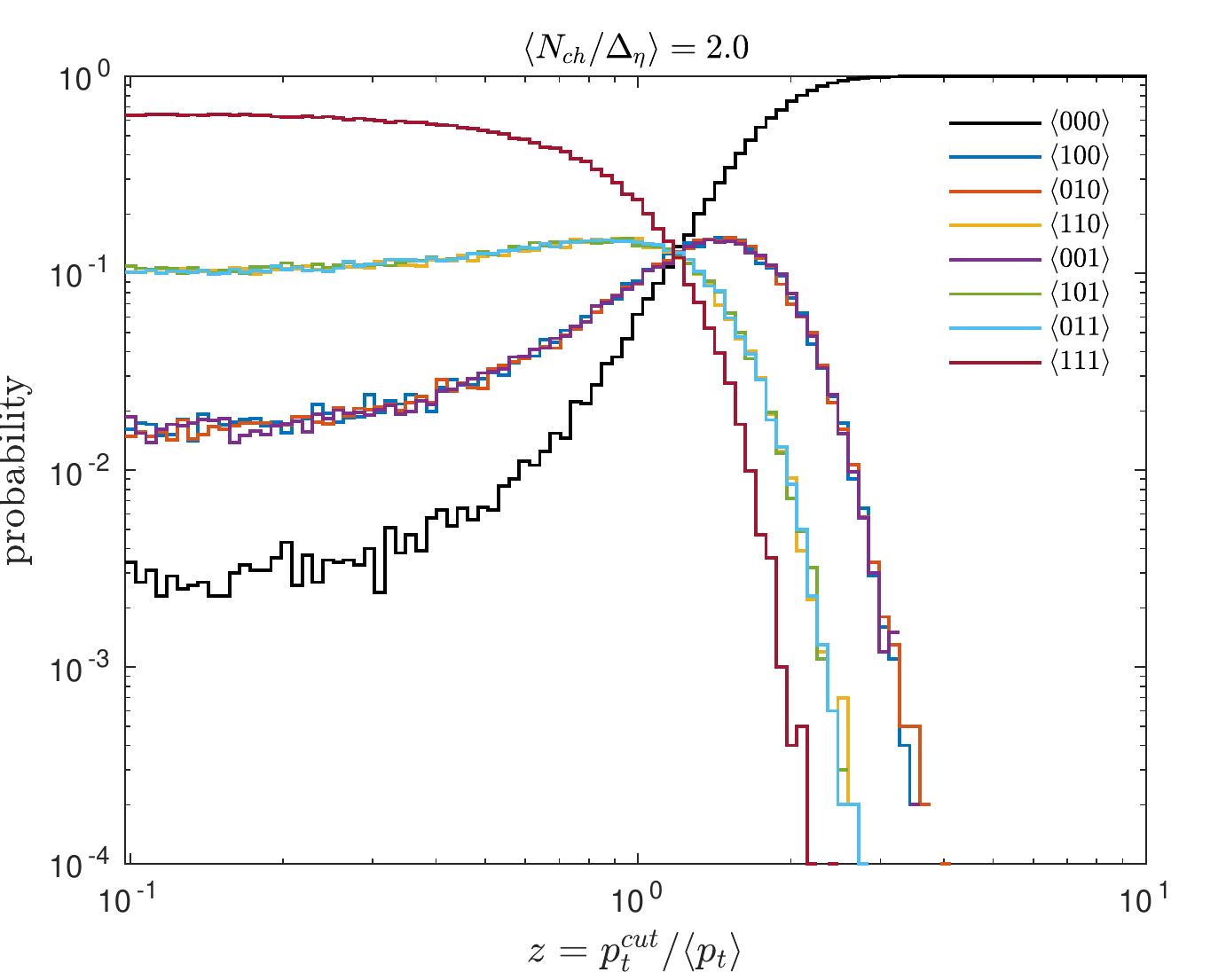}
\end{array}$
\caption{A toy simulation of `gapflow trajectories' as a function of normalized transverse momentum cutoff $z = p_t^{cut}/\langle p_t \rangle$ with four different particle density.}
\label{fig:ptsimulation}
\end{figure}

The measurement is illustrated in Figure \ref{fig:ptsimulation}, where the transverse momentum threshold is changed for four different average particle densities, particles drawn from a simple Poisson distribution with the given mean value. Here we used simple soft exponential $p_t^2$ distribution with particles generated uniformly over rapidity with $N=3$ slices. The exponential $p_t^2$ results in a Rayleigh type $p_t$-distribution with Gaussian distributed $p_x$ and $p_y$ components. Now it is very easy to see that rapidity gaps are not invariant under transverse momentum cutoffs, a well known fact but sometimes forgotten truth. Depending on the particle density and the cutoff, different event topology rates may have a cross over. The transition to pQCD with power law tails will be interesting.

\newpage

\section{Combinatorial supercompound Poisson process}
\label{sec:PoissonProblem}

We start by assuming a Poisson fluctuating random variable $K \sim \frac{\mu^k}{k!} e^{-\mu}$ with mean $\mu$ and argument $k \in \mathbb{N}$ denoting the number of simultaneous interactions, such as multiparton interactions or multiple independent proton-proton interactions. Both cases are treated analogously, modulo the energy-momentum conservation differences. The probability generating function for Poisson is $G(z) = \mathbb{E}(z^X) = \exp(\mu(z-1))$ which is useful for deriving the compound moments. In general, the expectation values and variances of a random variable $C_{|K>0} = \sum_{j=1}^K X_j$ after compounding stochastics are obtained from the compound relations
\begin{align}
\langle C \rangle &= \mathbb{E}_{K>0}[\mathbb{E}_{C|K>0}(C)] = \langle K > 0 \rangle \langle X \rangle \\
\nonumber \text{Var}[C] &= \mathbb{E}_{K>0}[\text{Var}_{C|K>0}(C)] + \text{Var}_{K>0}[\mathbb{E}_{C|K>0}(C)] \\
&= \langle K > 0\rangle \text{Var}[X] + \text{Var}[K > 0](\langle X \rangle)^2.
\end{align}
The zero-truncated Poisson mean and variance of $K$ are given in Appendix \ref{sec:Poisson}. We need the zero truncated distribution, because the 0 case is not a physical observable. However, all the $N$-point moments necessary to describe the multidimensional case, i.e. the case $N > 1$, are \textit{not} handled with these simple formulas but only via the full machinery which follows. Processes of this type are known as compound processes.

For preliminaries, the ordinary binomial coefficient is
\begin{equation}
C^n_k =
\begin{pmatrix}
n \\
k \\
\end{pmatrix}
=
\frac{n!}{k!(n-k)!}, \;\;\; 0 \leq k \leq n
\end{equation}
which describes how many ways $k$ numbers can be chosen from a set of $n$ numbers. Counting Bernoulli random variables in a binomial setup, one obtains the Binomial distribution. By counting random variables in a multivariate Bernoulli setup, we obtain analogously the multivariate Binomial distribution. However, for the sake of practicality we use now the multinomial distribution with random variables $X_1,X_2,\dots,X_{2^N-1}$, which is as many as we have non-zero fiducial vectors in the binary space. The multinomial probability distribution is
\begin{align}
P_m(X_1 = x_1, \dots, X_{2^N-1} = x_{2^N-1} | \mathbf{p} ) = \frac{k!}{\prod_{c=1} x_c!}\prod_{c = 1}^{2^N-1} p_c^{x_c}\;\; \\
\text{with}\;\; \sum_{c = 1}^{2^N-1} x_c = k, \;\;\; x_c \in \mathbb{N} \;\; \text{and} \;\; \sum_c p_c = 1,
\end{align}
which represents sampling or occupying a different number $x_i$ of different fiducial vectors per mixed event, the number of vectors in superposition per event is the number $k$. The first term is the multinomial coefficient easily derived using the binomial coefficient, also known in Maxwell-Boltzmann statistics. The probability generating function is here given by $G(\mathbf{z}) = \left( \sum_{c = 1}^{2^N-1} p_c z_c \right)^k$, $\mathbf{z} \in \mathbb{C}^n$ which is useful, when expanded, in interpreting multinomial distribution as polynomial coefficients $p_c$ with unit sum.

Here \textit{distinguishability} means that we are able to separate between different final state vectors but not between vector of the same type. This is a sampling with replacement scheme. Sampling without replacement gives, for example, a multivariate hypergeometric distribution instead of the multinomial. Similar non-multinomial sampling scenarios are encountered with Fermi-Dirac and Bose-Einstein statistics, for example.

\subsection{Direct model}

We define a generative direct model $f: \Theta \rightarrow Y$, where $\Theta$ denotes the space of parameters and $Y$ the space of observables. The unknown parameters are in $\mathbf{p}$ = $[p_1,p_2,\dots,p_n]^T$ $\in \Theta$ and the compounding process diluted or enhanced combination rate measurements are in $\mathbf{y}$ = $[y_1,y_2,\dots,y_n]^T$ $\in Y$. The model estimates for the observed rates $y_c$ are generated, as a sum over product of combinatorially selected multinomial distribution terms $W_{ck}$ and Poisson probabilities $P_p(k = j)$ order by order as
\begin{equation}
\label{eq: matrixmodel}
\begin{pmatrix}
y_1 \\
y_2 \\
\vdots \\
y_n
\end{pmatrix}
=
\frac{1}{1-e^{-\mu}}
\begin{pmatrix}
W_{11} & W_{12} & W_{13} \dots W_{1\infty} \\
W_{21} & W_{22} & W_{23} \dots W_{2\infty}\\
   & \vdots           &       \ddots   \\
W_{n1} & W_{n2} & W_{n3} \dots W_{n\infty}\\
\end{pmatrix}
\begin{pmatrix}
P_p(k = 1 ; \mu)  \\
P_p(k = 2 ; \mu)  \\
P_p(k = 3 ; \mu) \\
\vdots \\
P_p(k = \infty ; \mu) \\
\end{pmatrix},
\end{equation}
where we have an infinite series. The (re)-normalization factor $P_p(0;\mu) = 1 - e^{-\mu}$ is needed, because we exclude (truncate) the null case $k = 0$ of the Poisson density. Solving the inverse of Eq. \ref{eq: matrixmodel} is a non-linear problem, because the matrix elements of $W$ depend on the unknown parameters. Now more explicitly written, the \textit{compound Poisson} model equation for each probability component 
\begin{align}
\nonumber y_c &= \frac{1}{1-e^{-\mu}}\sum_{k = 1}^\infty \frac{\mu^k}{k!} e^{-\mu} W_{ck} \\
\label{eq: directmodel}
&= \frac{e^{-\mu}}{1-e^{-\mu}}\sum_{k = 1}^\infty \frac{\mu^k}{k!} \left\lbrace \sum_{ \{x_j \} \subset \Omega_{ck}} \frac{k!}{\prod_{j=1}^n x_j!}\prod_{j = 1}^n p_j^{x_j} \right\rbrace,
\end{align}
with $\sum_c y_c = 1$. The multinomial term and its values of $x_j$ are evaluated over all valid combinations or partitions for $y_c$ from the set of $n$-tuples $\Omega_{ck}$, that is, those which are allowed by pileup poset combinatorics. Formally, this set of compositions can be enumerated with
\begin{equation}
\Omega_{ck} = \left\{ ({x_1,\dots, x_j, \dots, x_n} ) \, | \, \bigvee_j x_j \mathbf{b}_j = \mathbf{b}_i \text{ and } \sum_j x_j = k \right \},
\end{equation}
where the boolean $\bigvee$ operator takes care of `summing' the binary vectors $\mathbf{b}_j$ of multiplicity $x_j$ and thus evaluating the pileup compositions. The $p_j$ are the probabilities we want to find out by inverting the compound Poisson process, that is, the probabilities of each $n = 2^N-1$ fiducial final state. 

Example $N = 2$: a (heavily) truncated Poisson series at $\mathcal{O}(\mu^2)$ is
\begin{equation}
W_{3 \times 2} = 
\begin{pmatrix}
p_1 & P_m(2,0,0) \\
p_2 & P_m(0,2,0) \\
p_3 & P_m(1,1,0) + P_m(1,0,1) + P_m(0,1,1) + P_m(0,0,2)
\end{pmatrix}.
\end{equation}
Now a brute force inverse $f^{-1} : Y \rightarrow \Theta$ would be to use Eq. \ref{eq: directmodel} and then estimate parameters by minimizing a non-linear $\chi^2$-function between the generated $\hat{y}_c$ and measured $y_c$ values $\hat{\mathbf{p}} = \argmin_{\mathbf{p}} \chi^2$, where $\chi^2 = (\hat{\mathbf{y}} - \mathbf{y})^T \Sigma_{\mathbf{y}}^{-1} (\hat{\mathbf{y}} - \mathbf{y})$ and usually $\Sigma_{\mathbf{y}}^{-1} = \text{diag}(1/\delta y_c^2)$. This is a typical fit approach, which is the statistical Maximum Likelihood (ML) solution when the measurement noise (counting fluctuations) is approximated with Gaussian distribution. A ML solution taking into account the multinomial counting fluctuations can be formulated also. However, if the direct model is constructed just order by order in $k$, this approach is computationally extremely expensive at high values of $k$ (and $N$) due to factorial growth. This can be seen in Table \ref{table: multiplicities} in Appendix \ref{sec:appendix}. Instead of using this, we will formulate an `all-order' solution based on the M\"obius inversion.

\subsection{Solution for $N \leq 2$}
The first case $N = 1$ is trivial, because we have just one Bernoulli observable, no combinations involved and because of our normalization condition $\sum_c p_c = 1$. Thus, the most interesting quantity is directly the Poisson distribution $\mu$-value and probabilities for $k = 0,1,2,\dots$.

The case $N = 2$ is straightforward. An explicit formula for $y_1$ is obtained by noticing that the set $\Omega_{1k}$ allows only `autocompound' for all $k$. Thus $B_1 \equiv k$ and $B_2 \equiv B_3 \equiv 0$ for all values of $k$, which gives
\begin{align}
\nonumber y_1 &= \frac{e^{-\mu}}{1-e^{-\mu}}\sum_{k = 1}^\infty \frac{\mu^k}{k!} \left( \frac{k!}{k!0!0!} p_1^kp_2^0p_3^0 \right) \\
\nonumber &= \frac{e^{-\mu}}{1-e^{-\mu}}\sum_{k = 1}^\infty \frac{\mu^k}{k!} p_1^k \\
&= \frac{e^{\mu p_1}-1}{e^{\mu}-1}.
\end{align}
Above we used the definition of Taylor series for $e^x$. By calculating similarly for $y_2$ by replacing above $p_1$ with $p_2$, then we get $y_3$ using the conservation of probability
\begin{equation}
y_3 = 1 - \sum_{c=1}^2 \frac{e^{\mu p_c} - 1}{e^{\mu} - 1}.
\end{equation}
Now the inverse is given by
\begin{align}
\hat{p}_{c} = \frac{\ln ((e^\mu -1)y_c + 1)}{\mu}, \;\; \text{for } c = 1,2
\end{align}
and again by conservation of probability
\begin{align}
\hat{p}_3 = 1 - \sum_{c=1}^2 \frac{\ln ((e^\mu -1)y_c + 1)}{\mu},
\end{align}
which completes the solution.

\subsection{Solution for $N \geq 1$}

The arbitrary $N$ case follows recursively from the $N = 2$ case and can be done in practice by constructing the matrix $\Lambda$ and its inverse according to the incidence algebra described earlier. The main result follows. The combinatorial statistics is

\begin{equation}
\label{eq: thedirectformula}
\boxed{ \mathbf{y}(\mathbf{p}; \mu) = \Lambda^{-1} \frac{\exp[-\mu \Lambda \mathbf{p}]-1}{e^{-\mu}-1} }
\end{equation}
\\
\noindent
which has also an inverse
\begin{equation}
\label{eq: theformula}
\boxed{ \hat{\mathbf{p}}(\mathbf{y}; \mu) =  \Lambda^{-1} \frac{\ln\left[(e^{-\mu}-1)\Lambda\mathbf{y} + 1 \right]}{-\mu}. }
\end{equation}

The exponential and logarithm are taken vector entry wise, such that the Taylor series representation is using the Hadamard power: $\exp[\mathbf{v}] = \sum_{k=0}^\infty \frac{1}{k!} \mathbf{v}^{\odot k}$. Equations \ref{eq: thedirectformula}, \ref{eq: theformula} are relying on the matrix $\Lambda$ constructing linear combinations of the probabilities of elements in $\mathbb{F}_2^N$ and the property that multinomial probability distribution is factorized as a product of different terms. The multinomial distribution obeys the so-called grouping property: re-partitioning conserves the multinomial distribution. Similarly $\Lambda^{-1}$ deconstructs the linear combinations, and in between there is a non-linear transformation according to the compound Poisson statistics. One can understand this also in terms of convolution theorem which turns the convolution into a usual product in the transform (Fourier) domain or vice verse. Here the M\"obius inversion has the same role as Fourier decomposition.

%\begin{figure}[ht]
%\vspace{1em}
%\centering
%$\begin{array}{ll}
%\includegraphics[width=77mm]{FLATp2y.pdf} & \includegraphics[width=77mm]{FLATy2p.pdf}
%\end{array}$
%\caption{On left, the map $\mathbf{p} \mapsto \mathbf{y}$ with a "comb input" $\mathbf{p}$. On right, the map $\mathbf{y} \mapsto \mathbf{p}$ with a "comb input" $\mathbf{y}$.}
%\label{fig: levelsplit}
%\end{figure}

Example with $N = 3$ gives us a vector
\begin{align*}
\hat{\mathbf{p}} = \frac{1}{\mu}
\begin{pmatrix}
\ln (e_{-}^\mu y_1 + 1) \\
\ln (e_{-}^\mu y_2 + 1) \\
- \underset{c = 1,2}{\sum} \ln(e_{-}^\mu y_c + 1) + \ln(1 + \hspace{-0.6em} \underset{c = 1,2,3}{\sum} e_{-}^\mu y_c) \\
\ln (e_{-}^\mu y_4 + 1) \\
- \underset{c = 1,4}{\sum} \ln(e_{-}^\mu y_c + 1) + \ln(1 + \hspace{-0.6em} \underset{c = 1,4,5}{\sum} e_{-}^\mu y_c) \\
- \underset{c = 2,4}{\sum} \ln(e_{-}^\mu y_c + 1) + \ln(1 + \hspace{-0.6em} \underset{c = 2,4,6}{\sum} e_{-}^\mu y_c) \\
\mu + \hspace{-0.6em} \underset{c = 1,2,4}{\sum} \hspace{-0.6em} \ln(e_{-}^\mu y_c + 1) - \ln(1 + \hspace{-0.6em}\underset{c = 1,2,3}{\sum} e_{-}^\mu y_c)
- \ln(1 + \hspace{-0.6em} \underset{c = 1,4,5}{\sum} e_{-}^\mu y_c )
- \ln(1 + \hspace{-0.6em} \underset{c = 2,4,6}{\sum} e_{-}^\mu y_c )
\end{pmatrix},
\end{align*}
where by conservation of probability we chose to fix $y_7 = 1 - \sum_{c = 1}^6 y_c$ and for compacting the notation we set $e_{-}^\mu \equiv e^\mu - 1$. The fractal structure of the problem is visible again. Components $c = 1,2,4$ only undergo autocompound, thus no mixing between other vectors. These are the unit basis vectors $[1,0,\dots,0]^T, [0,1,\dots,0]^T$ and $[0,0,\dots,1]^T$ and cannot be thus constructed as a linear combination of other vectors. A practical remark when working with non-linear equations with exponential and logarithmic functions: numerical evaluations are prone to floating point accuracy problems at high $\mu$ values.

\subsection{Maximum Input Entropy}

The constraints used in the construction required that both $\mathbf{p}$ and $\mathbf{y}$ are probability distributions and that the problem satisfies the combinatorial incidence algebra, together with the multinomial statistics embedded in the Poisson process. Next we treat certain special solutions.

\begin{figure}[ht]
%\vspace{1em}
\centering
\includegraphics[width=75mm]{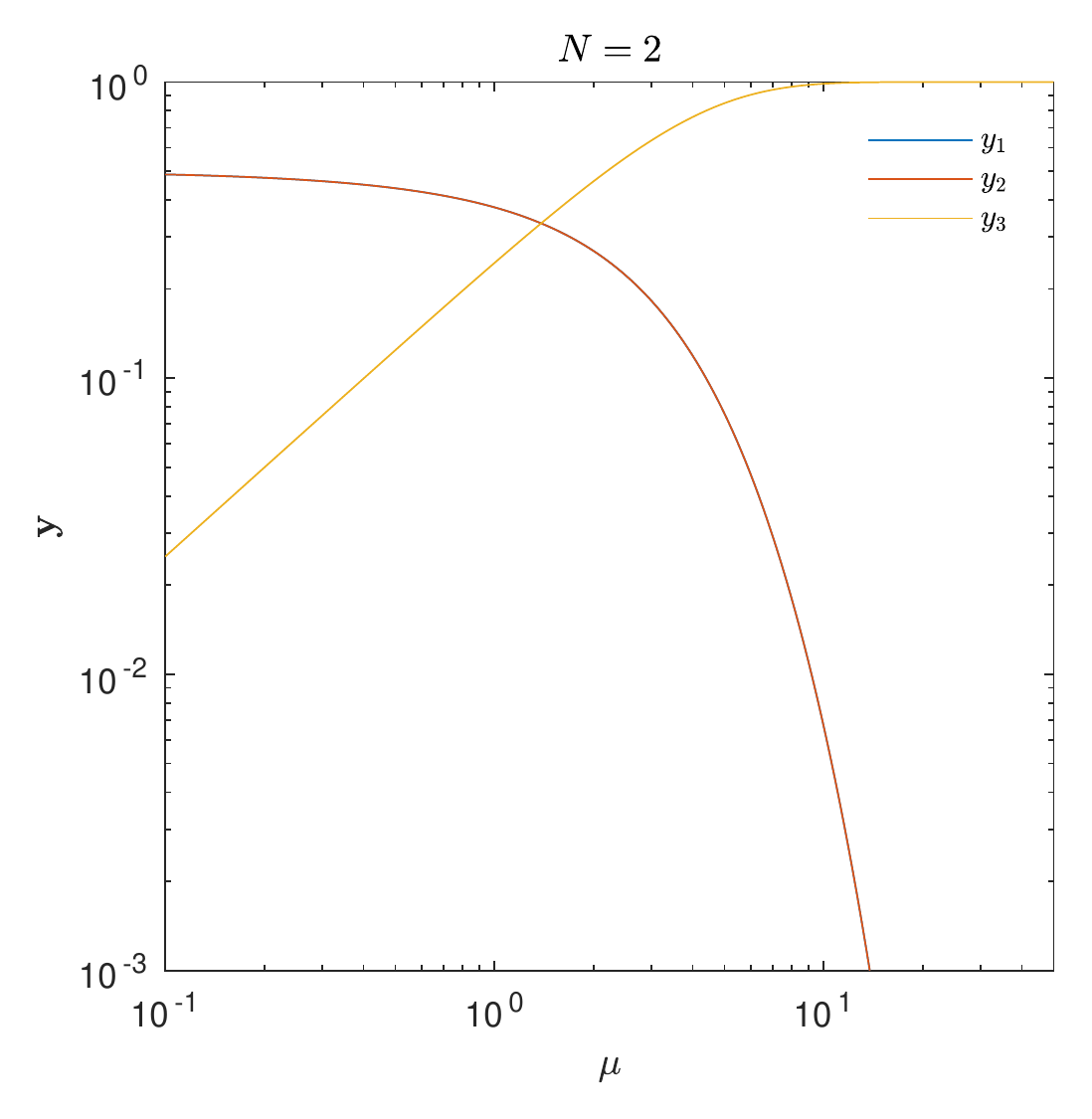}
\includegraphics[width=75mm]{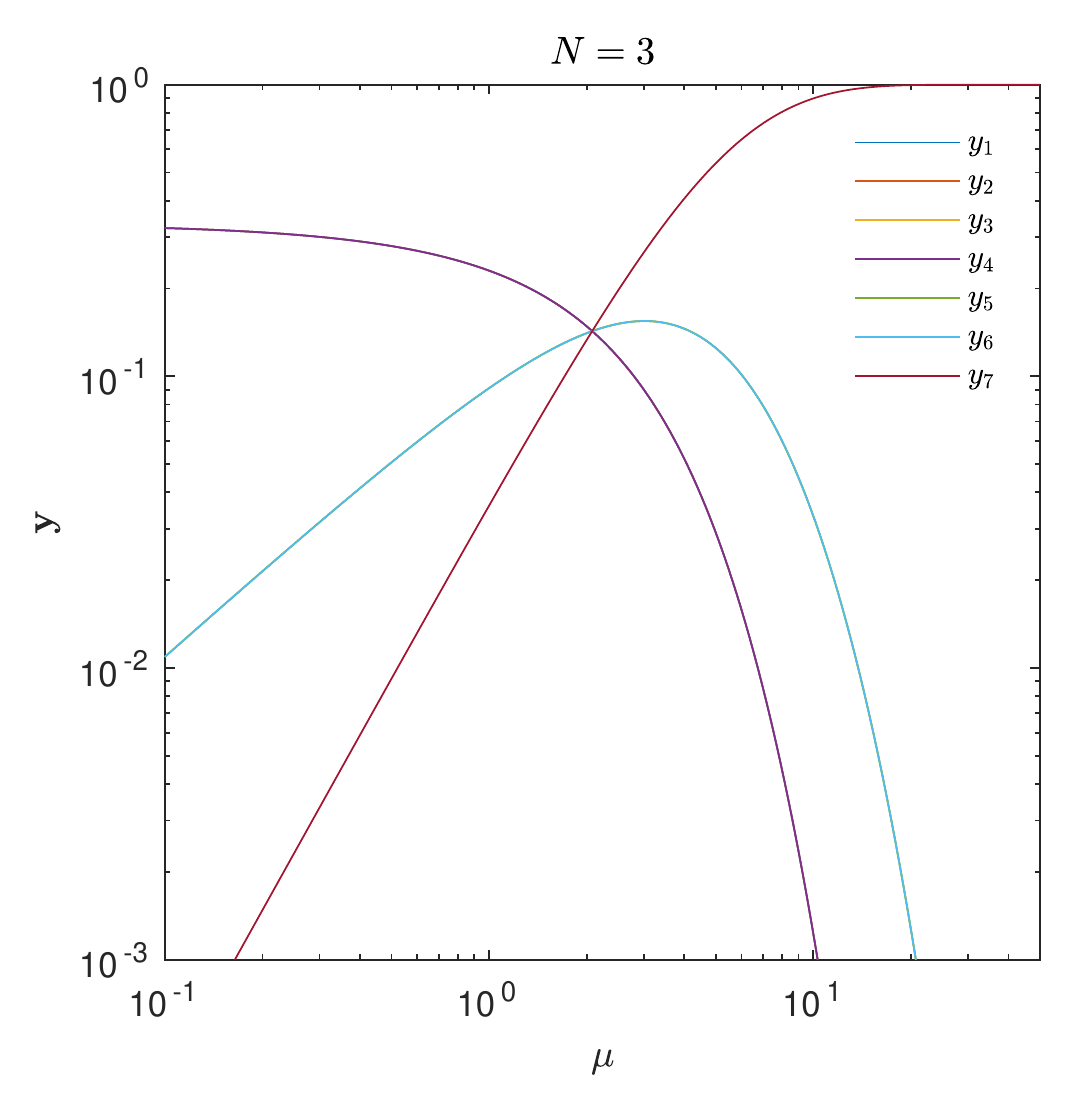} \\
\includegraphics[width=75mm]{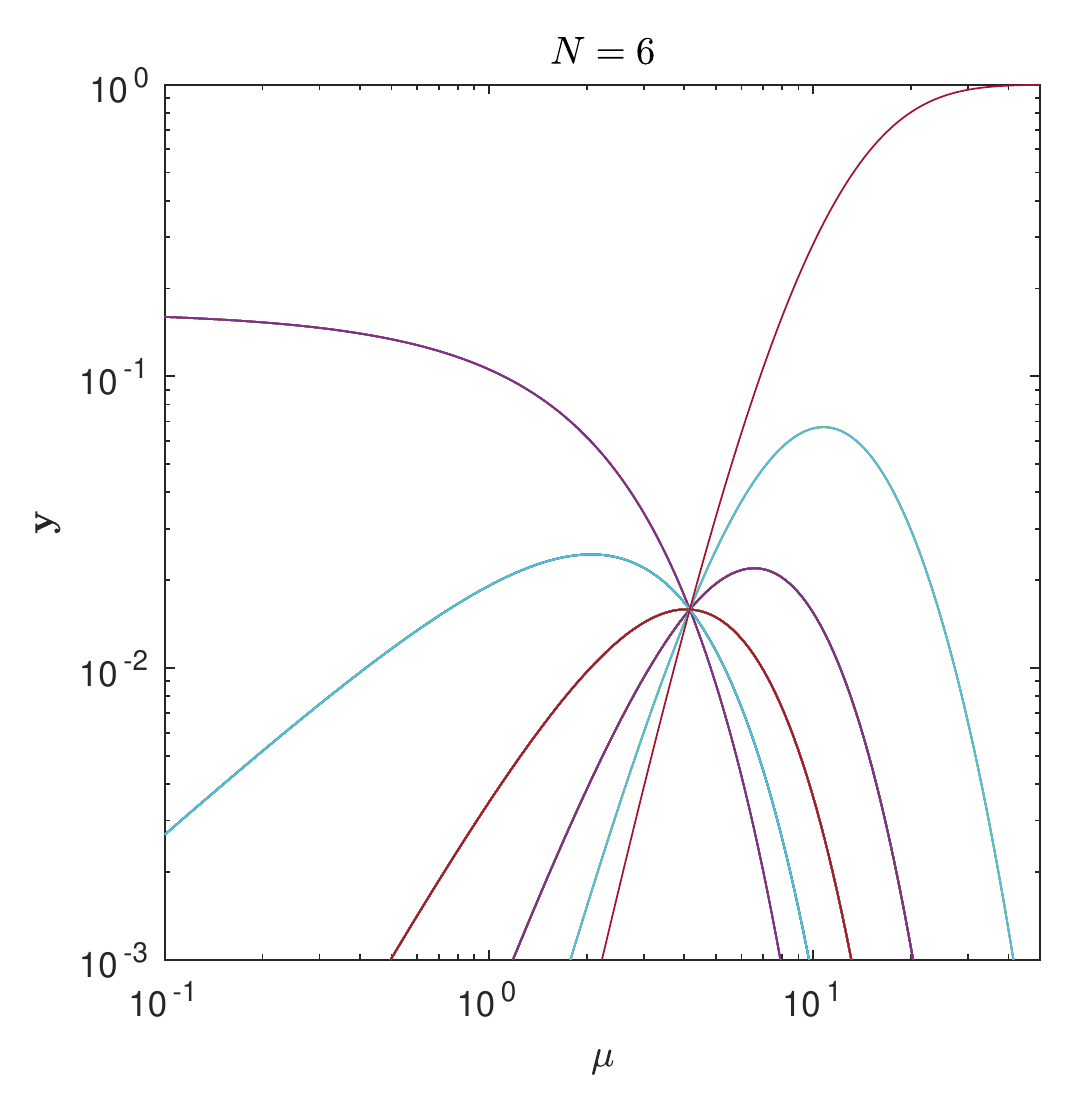}
\includegraphics[width=75mm]{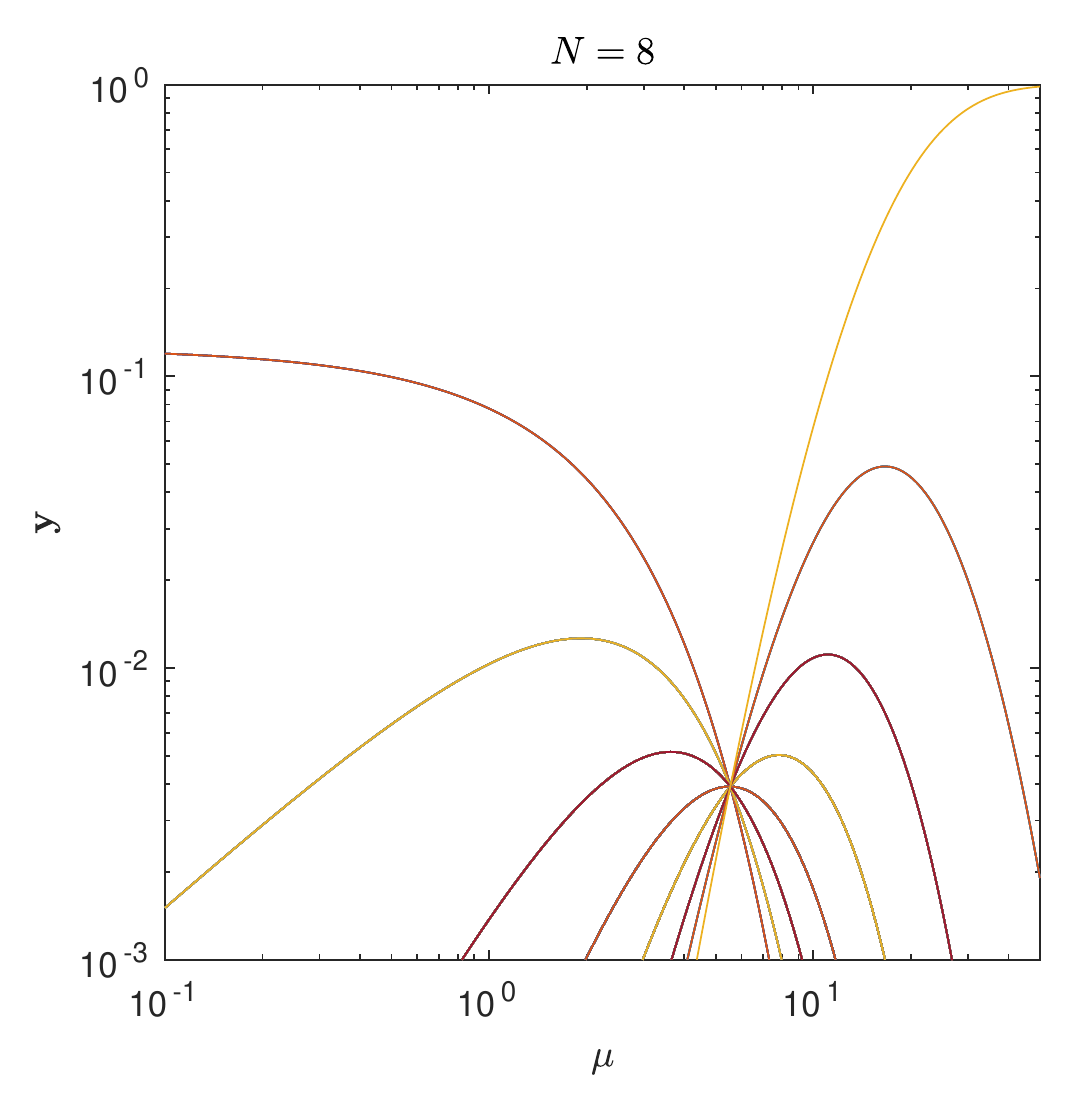}
\caption{Trajectories of $\mathbf{y}$ with the maximum entropy input initial condition $\mathbf{p} \equiv \mathbf{1}/n$.}
\label{fig:maxentropyP}
\end{figure}

One interesting case to study is the maximum input entropy case $\mathbf{p} \equiv \mathbf{1}/n$, that is, all different combinations are equally probably. We go through the case $N=3$ here explicitly. First we solve
\begin{align}
\nonumber
\ln(e^\mu_{-}y_1 + 1)/\mu &= \frac{1}{n} = \frac{1}{7} \\
y_{1} &= \frac{e^{\mu/7}-1}{e^\mu_{-}} \equiv y_2 \equiv y_4.
\end{align}
Then get we the third component by first substituting $y_1$
\begin{align}
\nonumber
[-2\frac{\mu}{7} + \ln(1+2(e^{\mu/7}-1)+e_{-}^\mu y_3) ]/\mu &= \frac{1}{7}  \\
y_{3} &= \frac{-2e^{\mu/7} + e^{3\mu/7} + 1}{e^\mu_{-}} \equiv y_5 \equiv y_6.
\end{align}
The last component is obtained by conservation of probability
\begin{align}
y_7 = 1 - y_{1:6} = \frac{3e^{\mu / 7}(1-e^{2\mu / 7})}{e_{-}^\mu} + 1,
\end{align}
where $y_{1:6}$ denotes a sum over all the given indices. It is interesting also to take a look at
\begin{align}
f(\mu) &= \sum_{c=3,5,6}y_c - \sum_{c=1,2,4} y_c = \frac{3 (2 -3e^{\mu / 7}+e^{3\mu / 7})}{e_{-}^\mu}.
\end{align}
Maximum input entropy solutions are represented in Figure \ref{fig:maxentropyP} as a function of $\mu$. We see that there is a fixed maximum output entropy point when $\mathbf{y} = \mathbf{1}/n$ with the critical $\mu$-value depending on $N$.

\subsection{Maximum output entropy}

\begin{figure}[H]
%\vspace{1em}
\centering
\includegraphics[width=75mm]{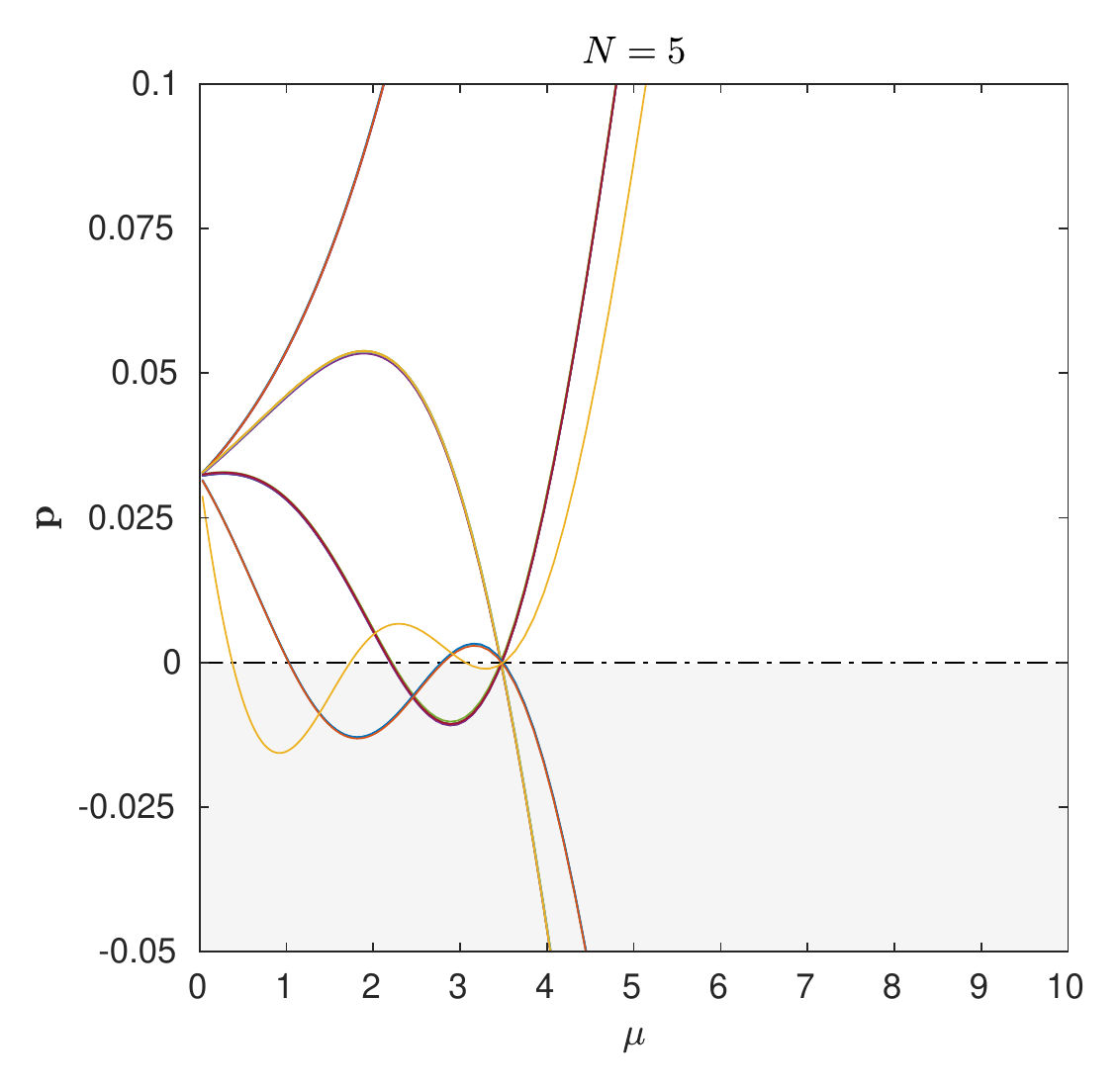}
\includegraphics[width=75mm]{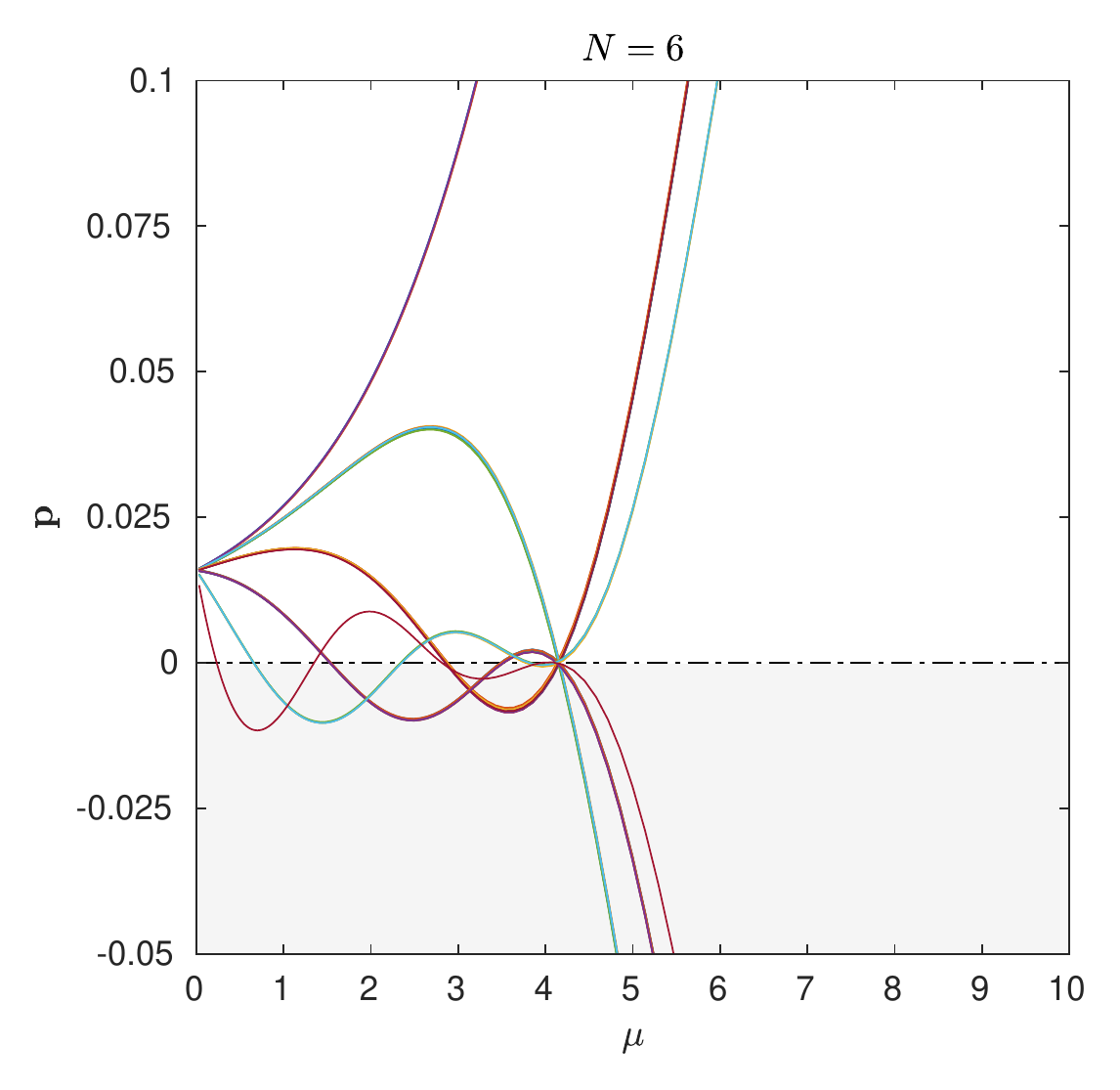}
\includegraphics[width=75mm]{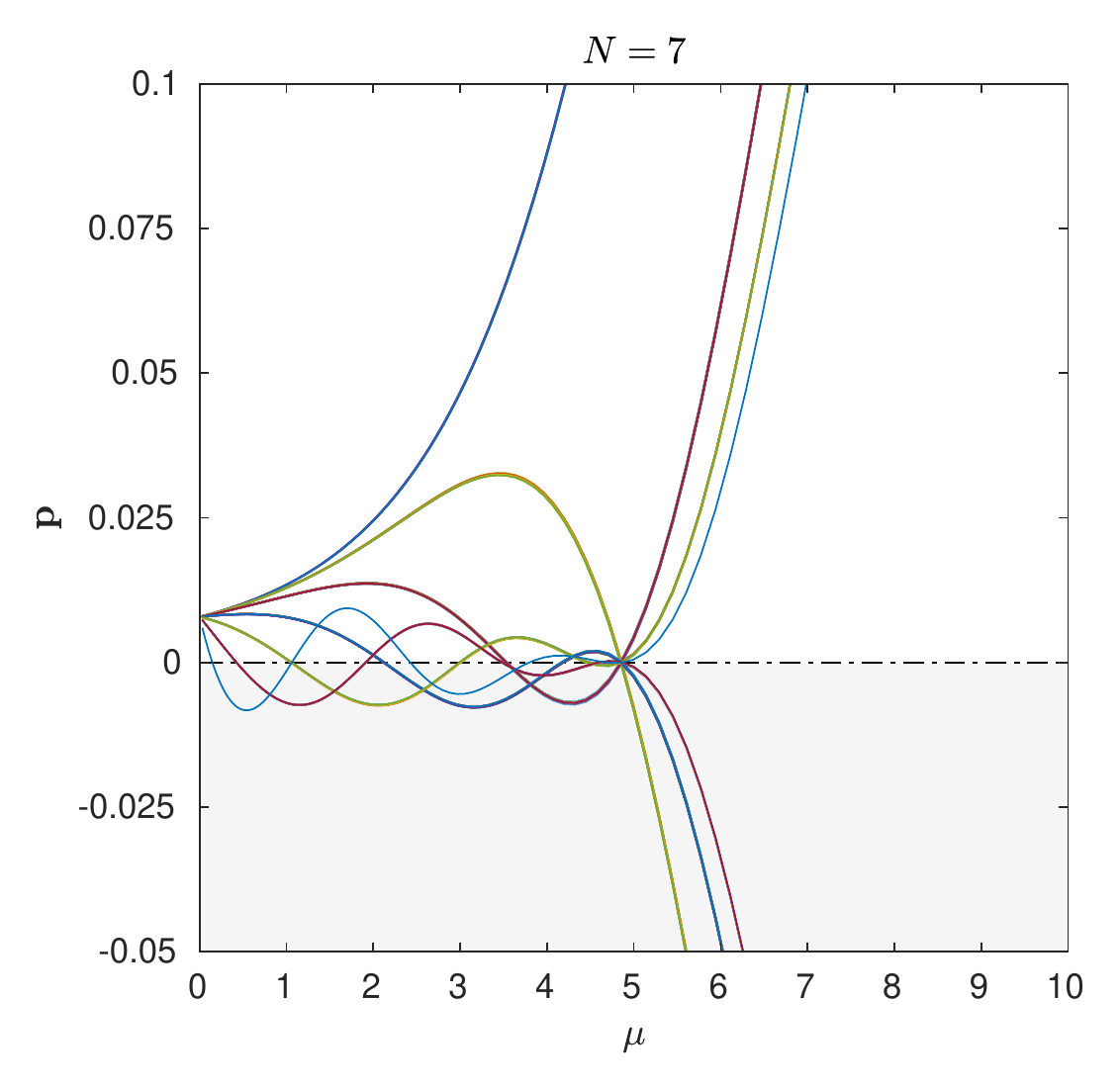}
\includegraphics[width=75mm]{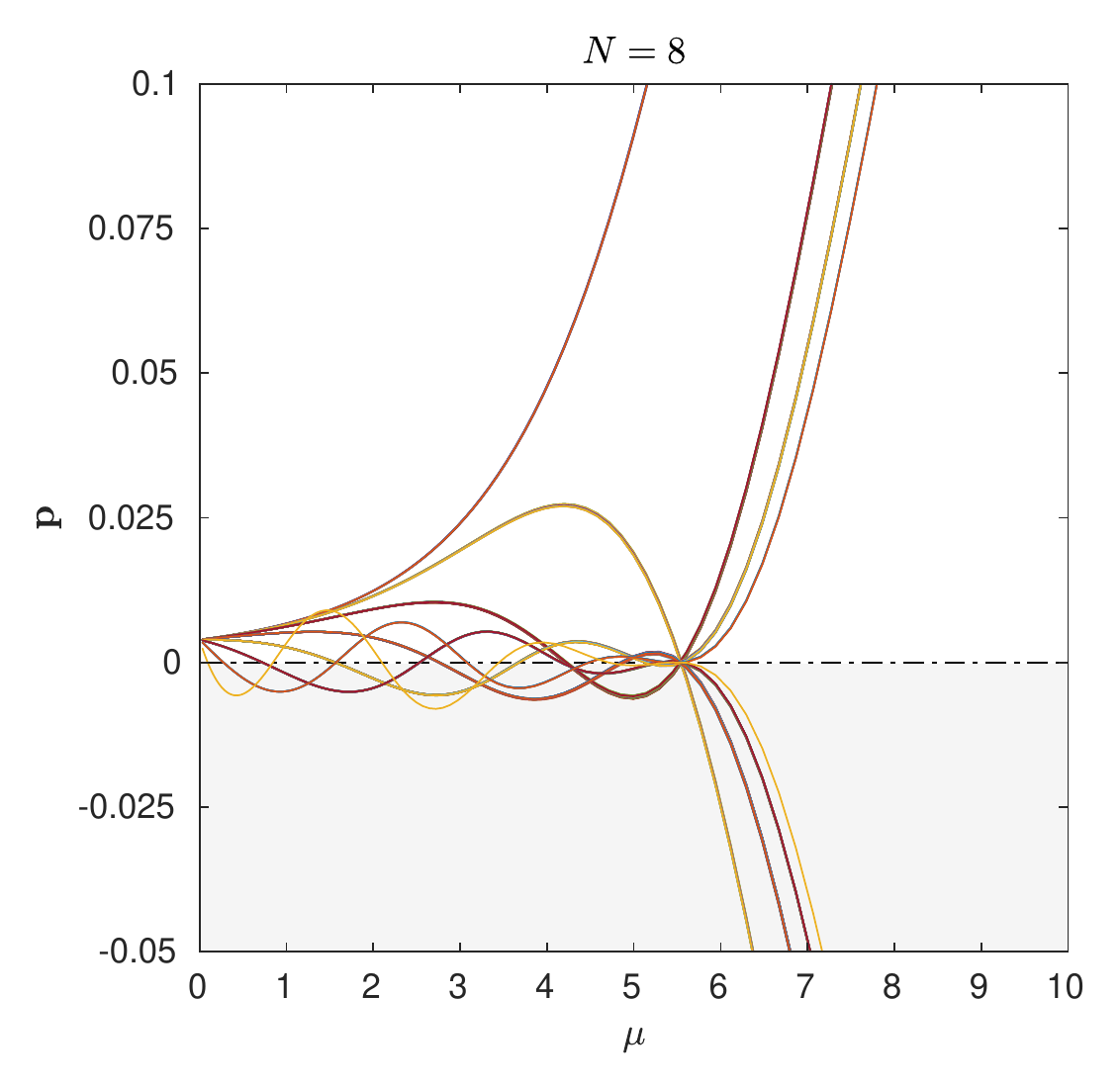}
\caption{Inverted trajectories of $\mathbf{p}$ with the maximum entropy output $\mathbf{y} \equiv \mathbf{1}/n$ boundary condition. Gray area shows non-positive definite domain.}
\label{fig:maxentropyY}
\end{figure}

Here we construct solutions with maximum output entropy. Examples of functional behavior are given in Figure \ref{fig:maxentropyY}. We see that it is not possible to obey the constraints of all probabilities being non-negative, thus we obtain $\mathbf{p}$ with negative elements when $\mu$ is high enough. An interesting oscillating phenomena appears with nodes at higher $\mu$ due to polynomial structure of our construction.

Negative values are seemingly unphysical, unless one extends the domain of the problem to include also `negative probabilities' -- sometimes called \textit{quasi-probability distributions}, utilized first by Wigner \cite{wigner1932quantum}. In Wigner's construction these are fundamentally non-classical distributions of quantum states. The construction here has certain analogous behavior, essentially because it is the vector $\mathbf{y}$ which are observables and the vector $\mathbf{p}$ contains the parameters of the states, which may have non-classical distributions.

Non-classical distributions of $\mathbf{p}$ here may thus be useful for quantum interference, absorption, shadowing and diffraction. Natural extension is to complex vector distributions. Negative probabilities as a useful intermediate step of calculations were also discussed in a well-known article by Feynman \cite{feynman1987negative} in the context of two non-commuting observables.

\begin{figure}[H]
%\vspace{1em}
\centering
\includegraphics[width=75mm]{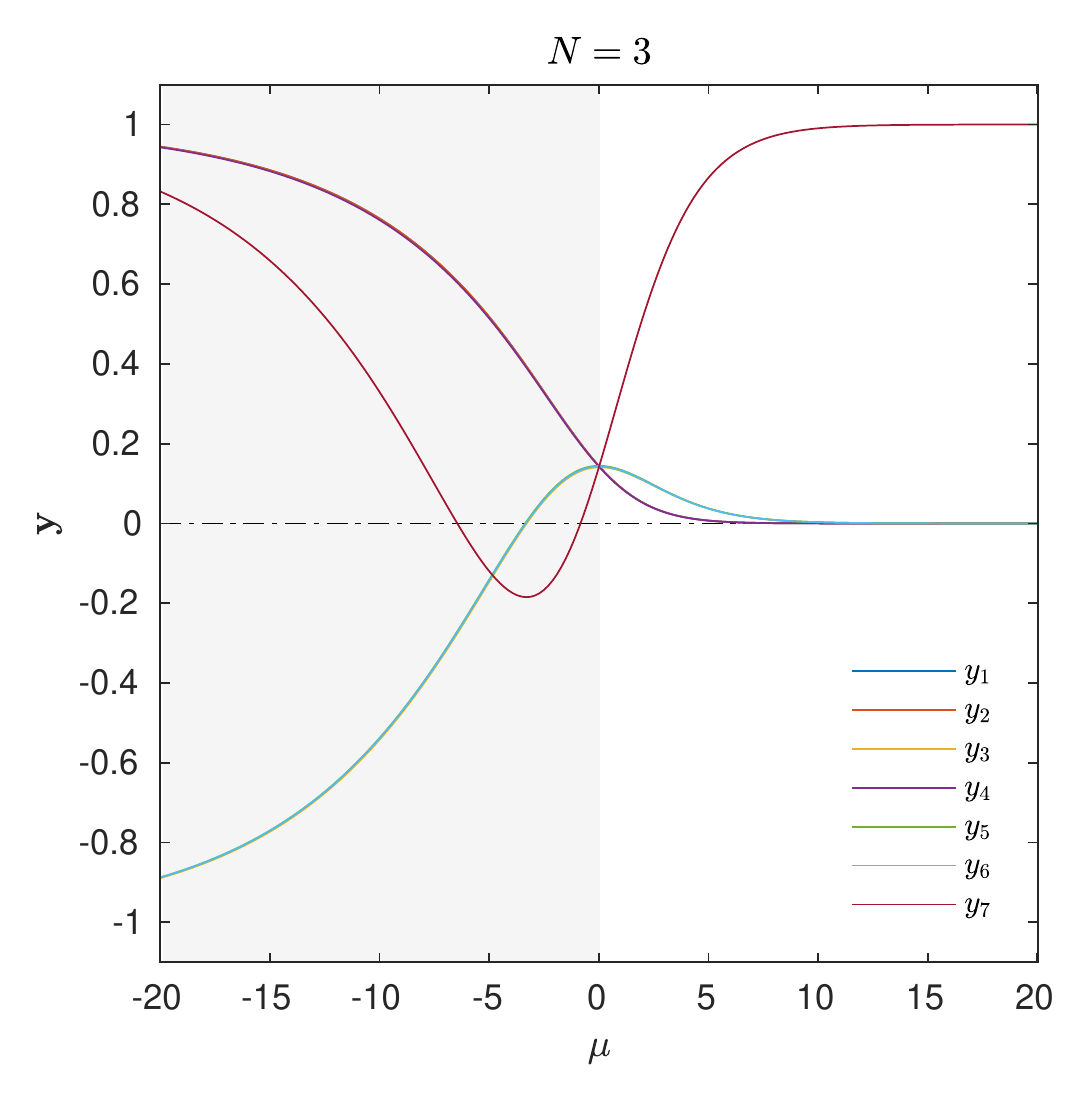}
\includegraphics[width=75mm]{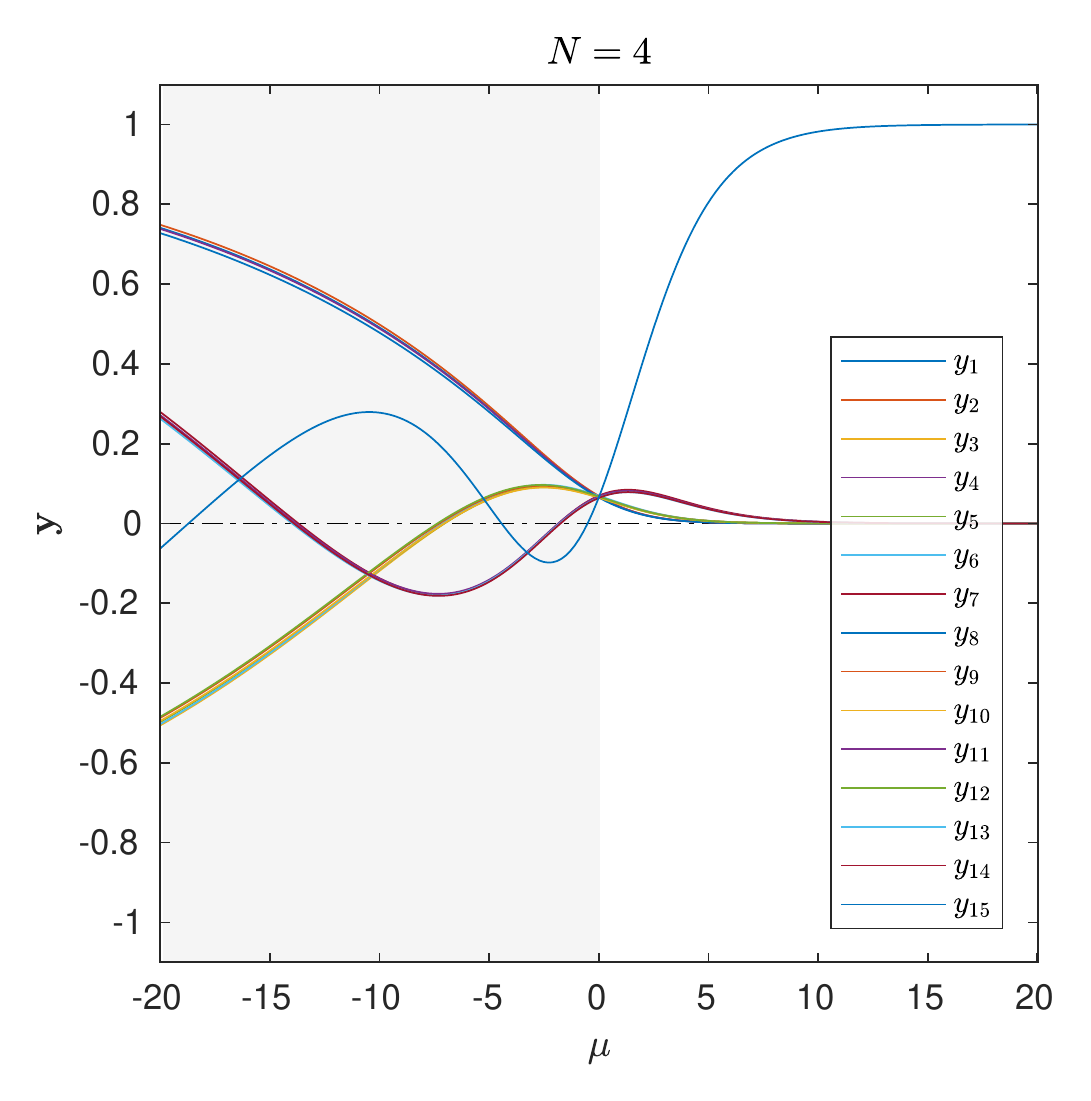} \\
\includegraphics[width=75mm]{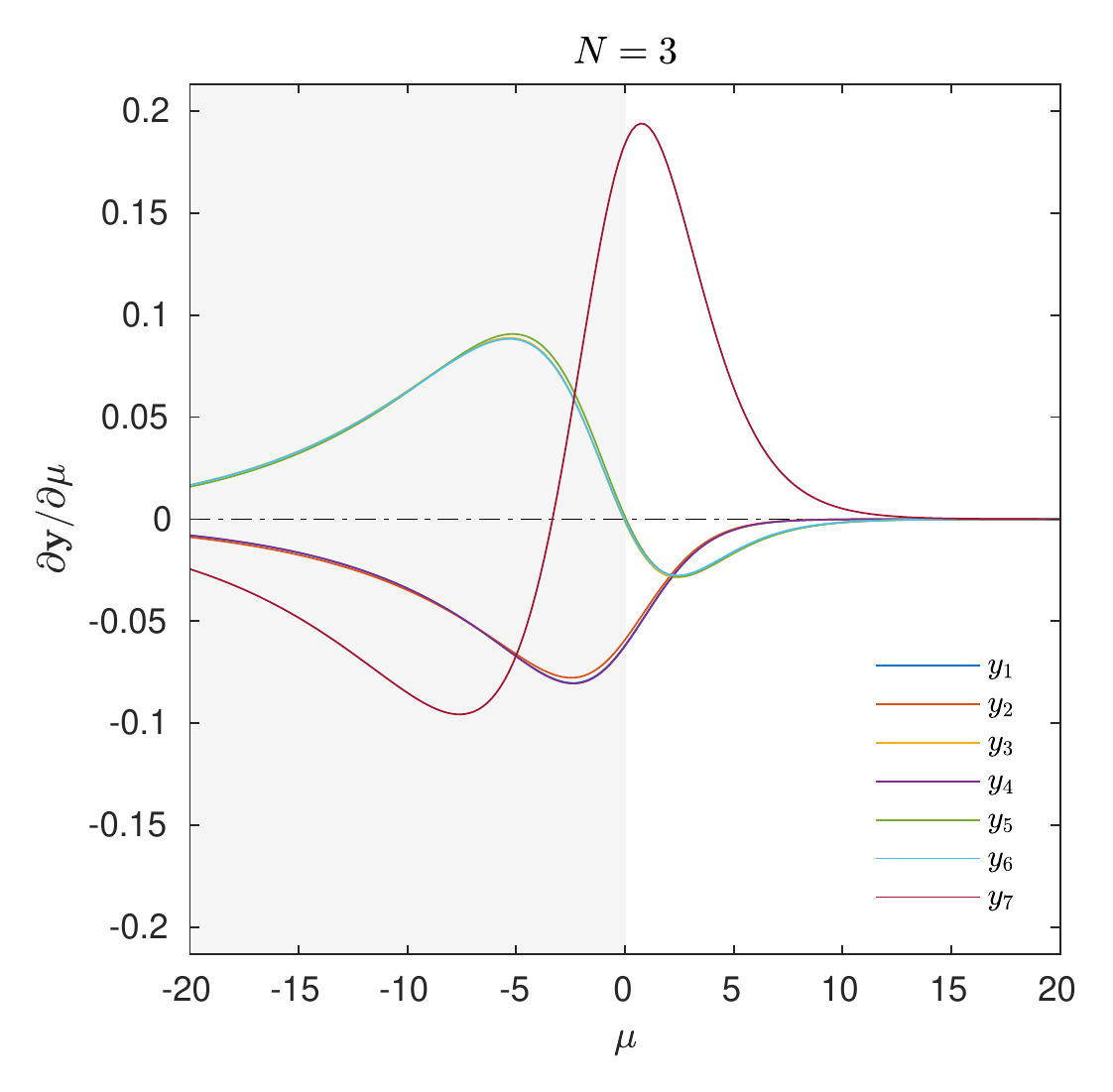}
\includegraphics[width=75mm]{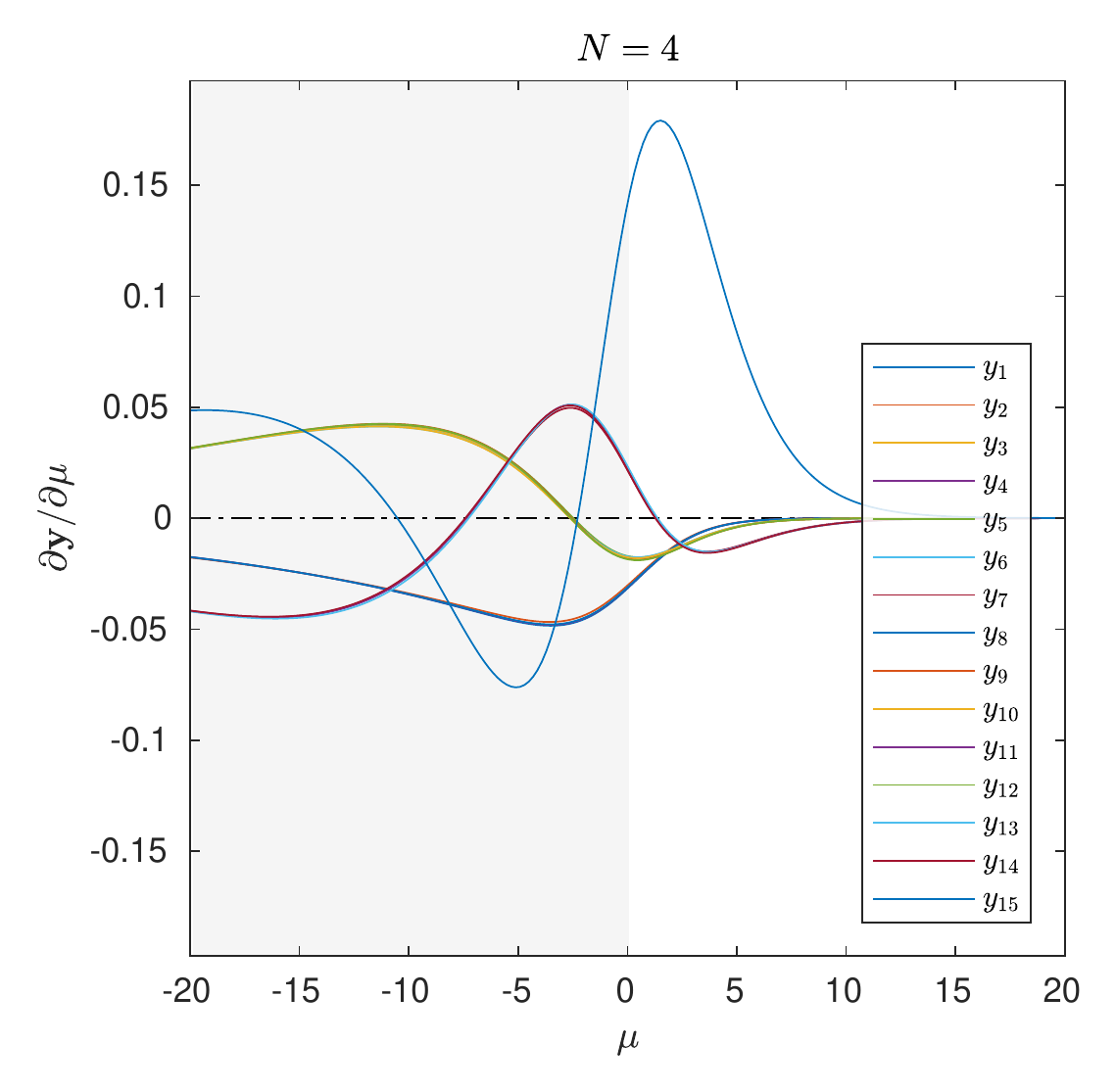} \\
\includegraphics[width=75mm]{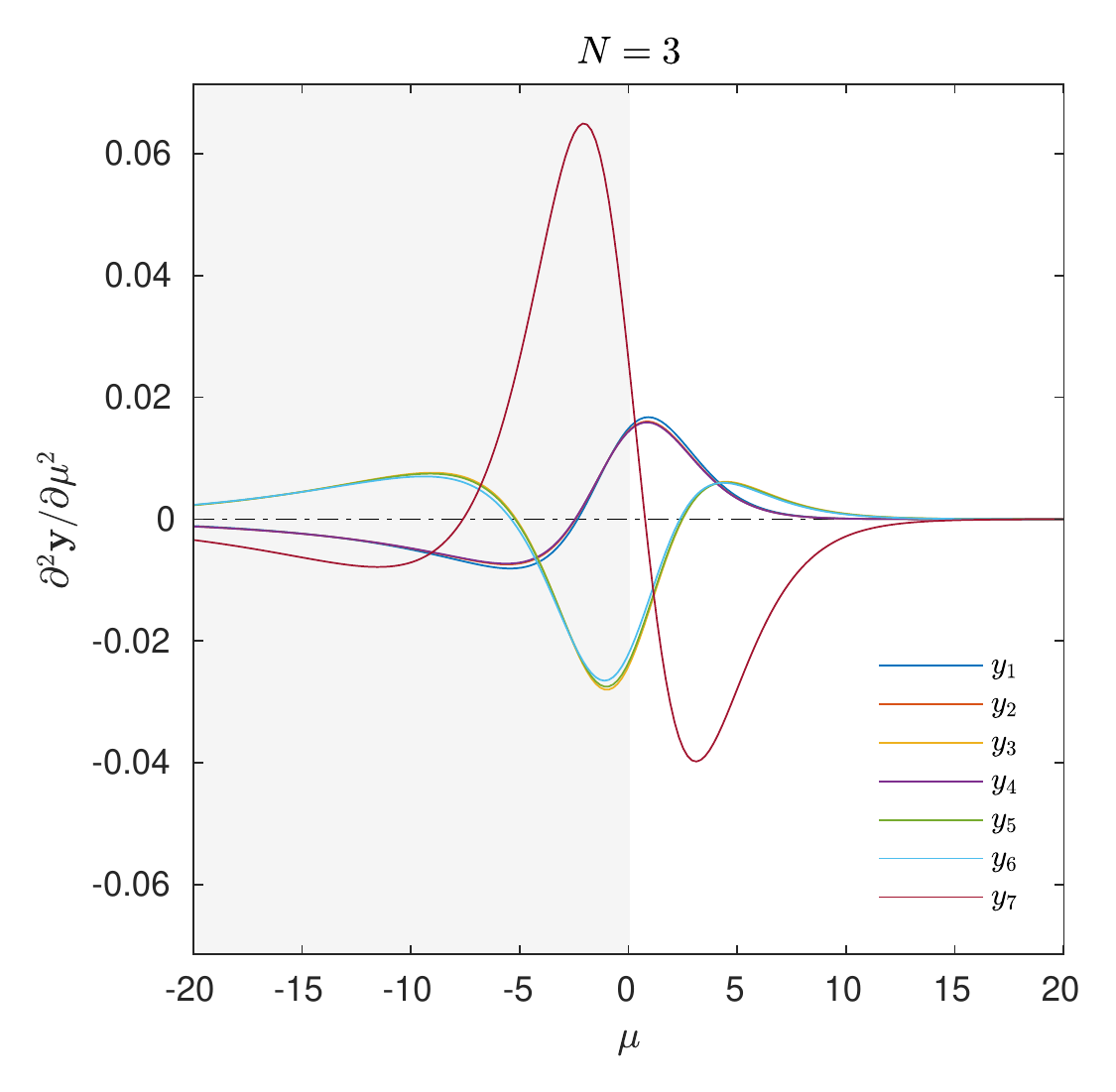} 
\includegraphics[width=75mm]{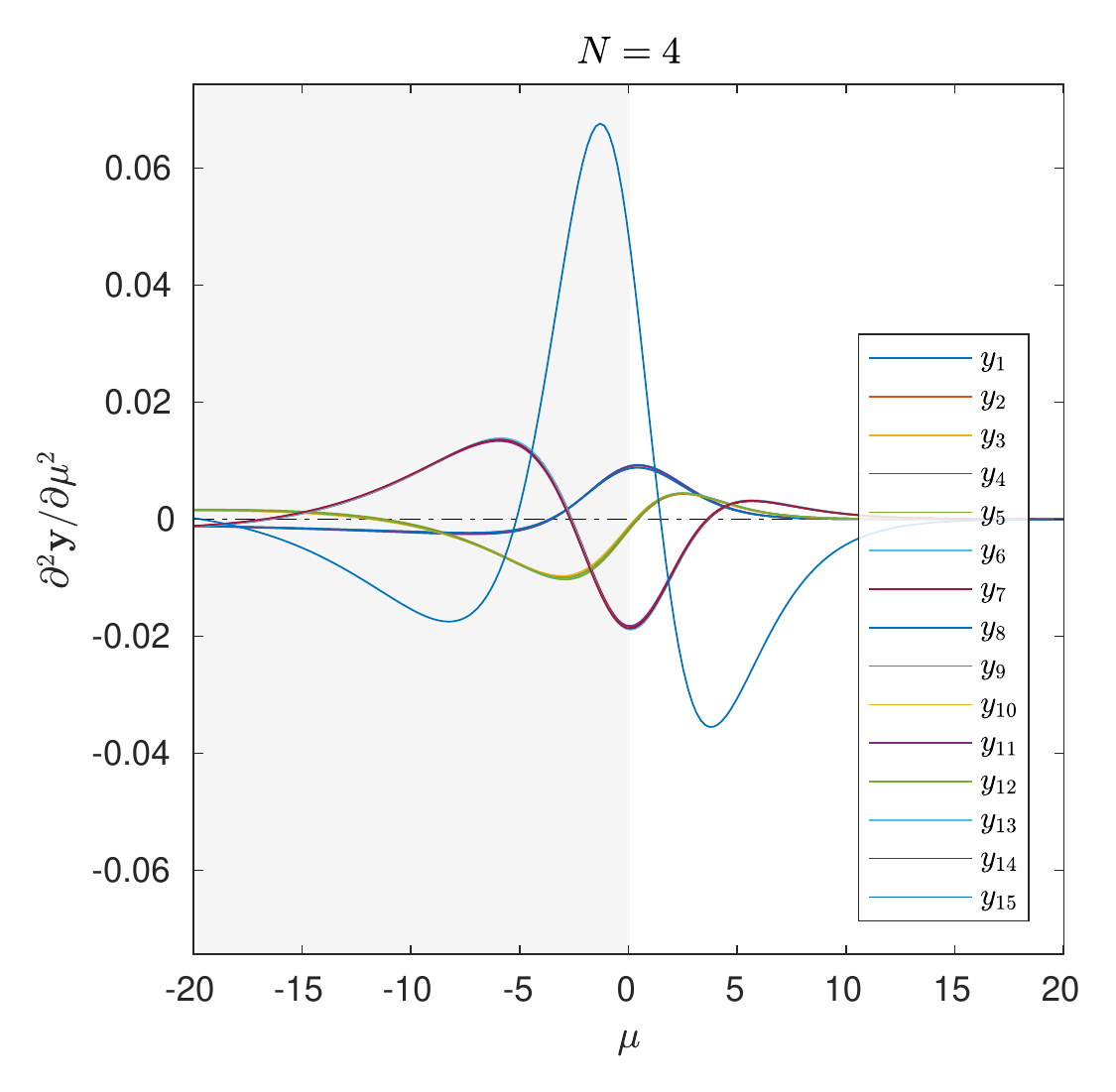}
\caption{Trajectories (1st row) and first and second order derivatives (2nd and 3rd rows) with the $\mathbf{p} \equiv \mathbf{1}/n$ input initial condition, for $N=3$ and $N=4$. We have added a tiny perturbation to the input, to distinguish visually between different parity degenerate components of trajectories.}
\label{fig:partialderivatives}
\end{figure}

\subsection{Master Equation}

The first order derivatives of $\mathbf{y}$ with respect to $\mu$ results in
\begin{equation}
\frac{\partial \mathbf{y}}{\partial \mu} = \Lambda^{-1} \left( \frac{\exp[-\mu \Lambda \mathbf{p}] - 1}{e^{\mu}(e^{-\mu}-1)^2} - \frac{ \Lambda \mathbf{p} \odot \exp[-\mu \Lambda \mathbf{p}]}{e^{-\mu}-1} \right)
\end{equation}
and the higher order derivatives follow similarly, with lengthy expressions which we evaluated with symbolic computer algebra. It is this expression which can be called the \textit{master equation} of describing the evolution of probabilities in terms of the $\mu$-value. Visualized, we obtain wavelet like behavior show in Figure \ref{fig:partialderivatives}. Now obvious future direction would be to probe a set of stochastic equations which would use these basic building blocks. In Figure \ref{fig:partialderivatives} we show also the `analytically continued' case with negative $\mu$-values for completeness and leave the possible applications out of discussion here. The values of $\mathbf{y}$ are positive with positive $\mu$ values, but obtain oscillating positive and negative values at negative $\mu$. The behavior is actually opposite, if we invert the case and study values of $\mathbf{p}$ at negative $\mu$ when $\mathbf{y} \equiv \mathbf{1}/n$.

\subsection{Semi-Bose-Einstein construction}

The construction above was the most relaxed one, in a sense that we allow for example $[0,1] + [0,1] \rightarrow [0,1]$ type vector combinations. However, one can construct a modified statistics by forbidding same the vector combination appearing more than once, but not limiting occupation number per vector dimension. This will limit a large number of possible combinations meaning all `autocompound' is forbidden. In even more strict construction, on the other hand, we would allow only one occupation per vector dimension. We recall that different canonical particle statistics are most easily derived using the combinatorial method presented in any advanced book on statistical mechanics.

\hspace{1em}
\\
\\
\noindent The case $N = 2$ is
\begin{align}
\nonumber y_1 &= P_{K=1}P_m(1,0,0) \\
\nonumber y_2 &= P_{K=1}P_m(0,1,0) \\
\nonumber y_3 &= P_{K=1}P_m(0,0,1) \\
\nonumber &+ P_{K=2}[P_m(1,1,0) + P_m(1,0,1) + P_m(0,1,1)] \\
&+ P_{K=3}P_m(1,1,1),
\end{align}
where the notation is the same as before.

\noindent The case $N = 3$ is
\begin{align}
\nonumber y_1 &= P_{K=1}P_m(1,0,0,0,0,0,0) \\
\nonumber y_2 &= P_{K=1}P_m(0,1,0,0,0,0,0) \\
\nonumber y_3 &= P_{K=1}P_m(0,0,1,0,0,0,0) \\
\nonumber&+ P_{K=2}[P_m(1,1,0,0,0,0,0) + P_m(1,0,1,0,0,0,0) + P_m(0,1,1,0,0,0,0)] \\ 
\nonumber &+ P_{K=3}P_m(1,1,1,0,0,0,0) \\
\nonumber y_4 &= P_{K=1}P_m(0,0,0,1,0,0,0) \\
\nonumber y_5 &= P_{K=1}P_m(0,0,0,0,1,0,0) \\
\nonumber &+ P_{K=2}\left[P_m(1,0,0,1,0,0,0) + P_m(1,0,0,0,1,0,0) +  P_m(0,0,0,1,1,0,0)\right] \\
\nonumber &+ P_{K=3}P_m(1,0,0,1,1,0,0)\\
\nonumber y_6 &= P_{K=1}P_m(0,0,0,0,0,1,0) \\
\nonumber &+ P_{K=2}\left[P_m(0,1,0,1,0,0,0) + P_m(0,0,0,1,1,0,0) + P_m(0,1,0,0,1,0,0)\right] \\
\nonumber &+ P_{K=3}P_m(0,1,0,1,0,1,0)\\
y_7 &= 1 - y_{1:6},
\end{align}
where the last component is obtained simply by conservation of probability. In this `Semi-Bosonic' case, we did not find such a simple closed-form formula as the matrix equation \ref{eq: theformula}, but solutions for any $N$ can be found simply algorithmically using symbolic computer algebra.

\subsection{Semi-Fermi-Dirac construction}

In the Semi-Fermi-Dirac construction, which is the most strict one we consider here, we allow only one occupation per vector element.

\vspace{1em}
\noindent The case $N = 2$ is
\begin{align}
\nonumber y_1 &= P_{K=1}P_m(1,0,0) \\
\nonumber y_2 &= P_{K=1}P_m(0,1,0) \\
y_3 &= P_{K=1}P_m(0,0,1) + P_{K=2}P_m(1,1,0).
\end{align}

\noindent The case $N = 3$ is
\begin{align}
\nonumber y_1 &= P_{K=1}P_m(1,0,0,0,0,0,0) \\
\nonumber y_2 &= P_{K=1}P_m(0,1,0,0,0,0,0) \\
\nonumber y_3 &= P_{K=1}P_m(0,0,1,0,0,0,0) + P_{K=2}P_m(1,1,0,0,0,0,0) \\
\nonumber y_4 &= P_{K=1}P_m(0,0,0,1,0,0,0) \\
\nonumber y_5 &= P_{K=1}P_m(0,0,0,0,1,0,0) + P_{K=2}P_m(1,0,0,1,0,0,0) \\
\nonumber y_6 &= P_{K=1}P_m(0,0,0,0,0,1,0) + P_{K=2}P_m(0,1,0,1,0,0,0) \\
y_7 &= 1 - y_{1:6}.
\end{align}
Again, the any $N$ case can be constructed algorithmically.

\begin{figure}[H]
%\vspace{1em}
\centering
\includegraphics[width=75mm]{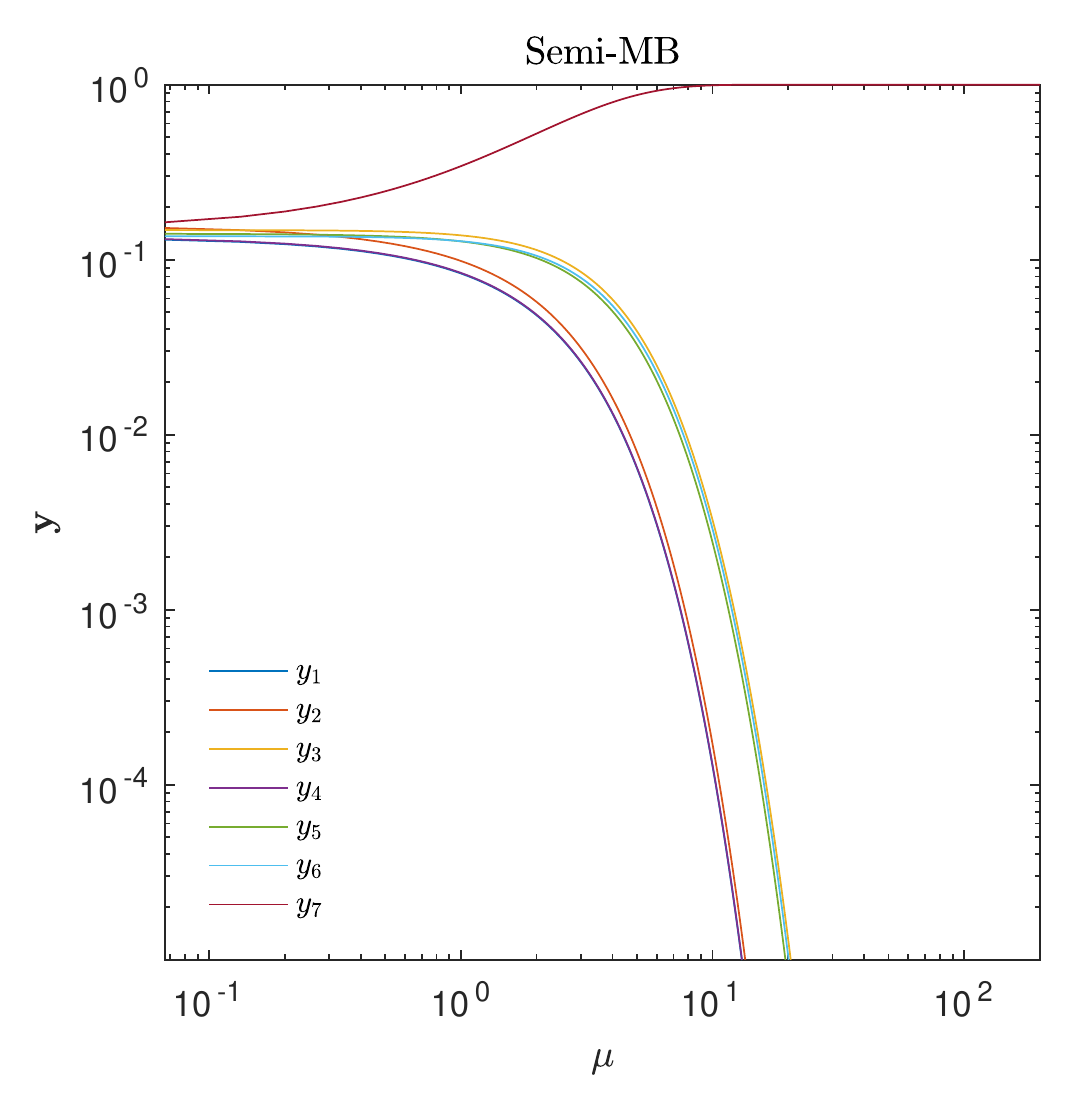}
\includegraphics[width=75mm]{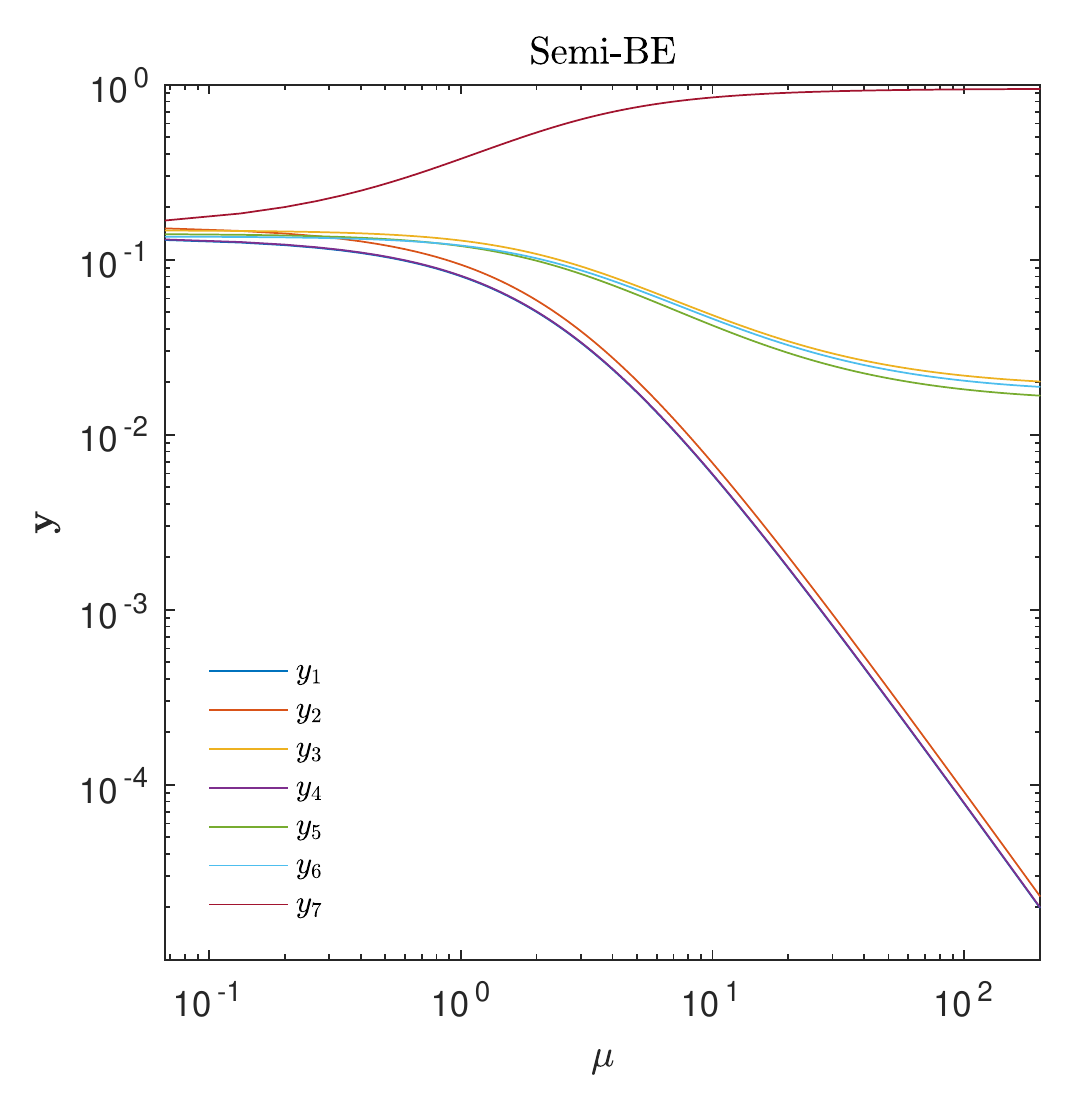}
\includegraphics[width=75mm]{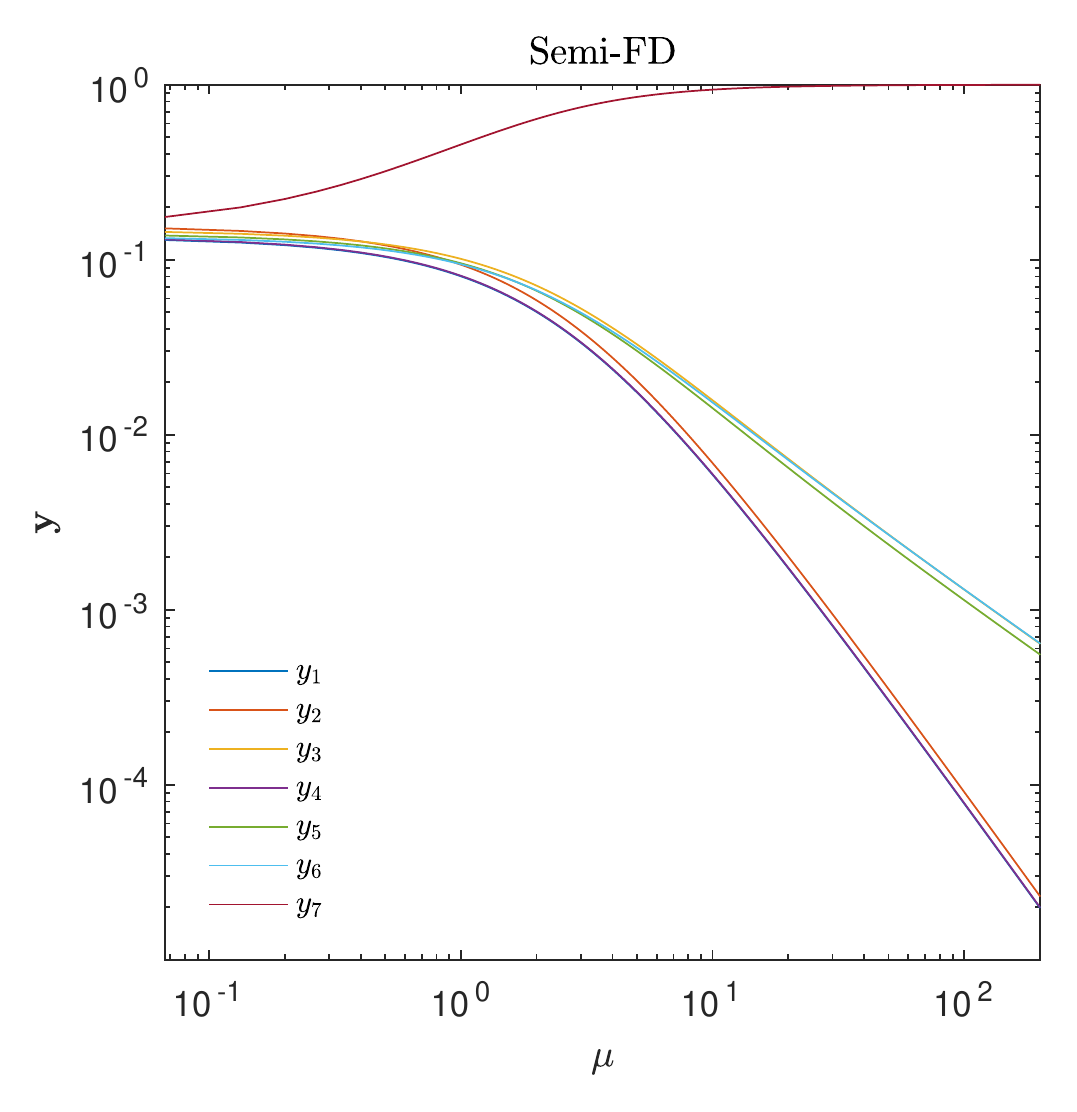}
\caption{Semi-Maxwell-Boltzmann, Semi-Bose-Einstein and Semi-Fermi-Dirac boundary conditions with maximum entropy input $\mathbf{p} = \mathbf{1}/n$ initial condition.}
\label{fig:MB_BE_FD_statistics_tails}
\end{figure}

The results are shown Figure \ref{fig:MB_BE_FD_statistics_tails}, where different constraints on the combinations result in different asymptotic behavior in the tails.

\newpage

\section{Invertability simulations}
\label{sec:simulations}

\begin{figure}[H]
\centering
$\begin{array}{ccc}
\includegraphics[width=70mm]{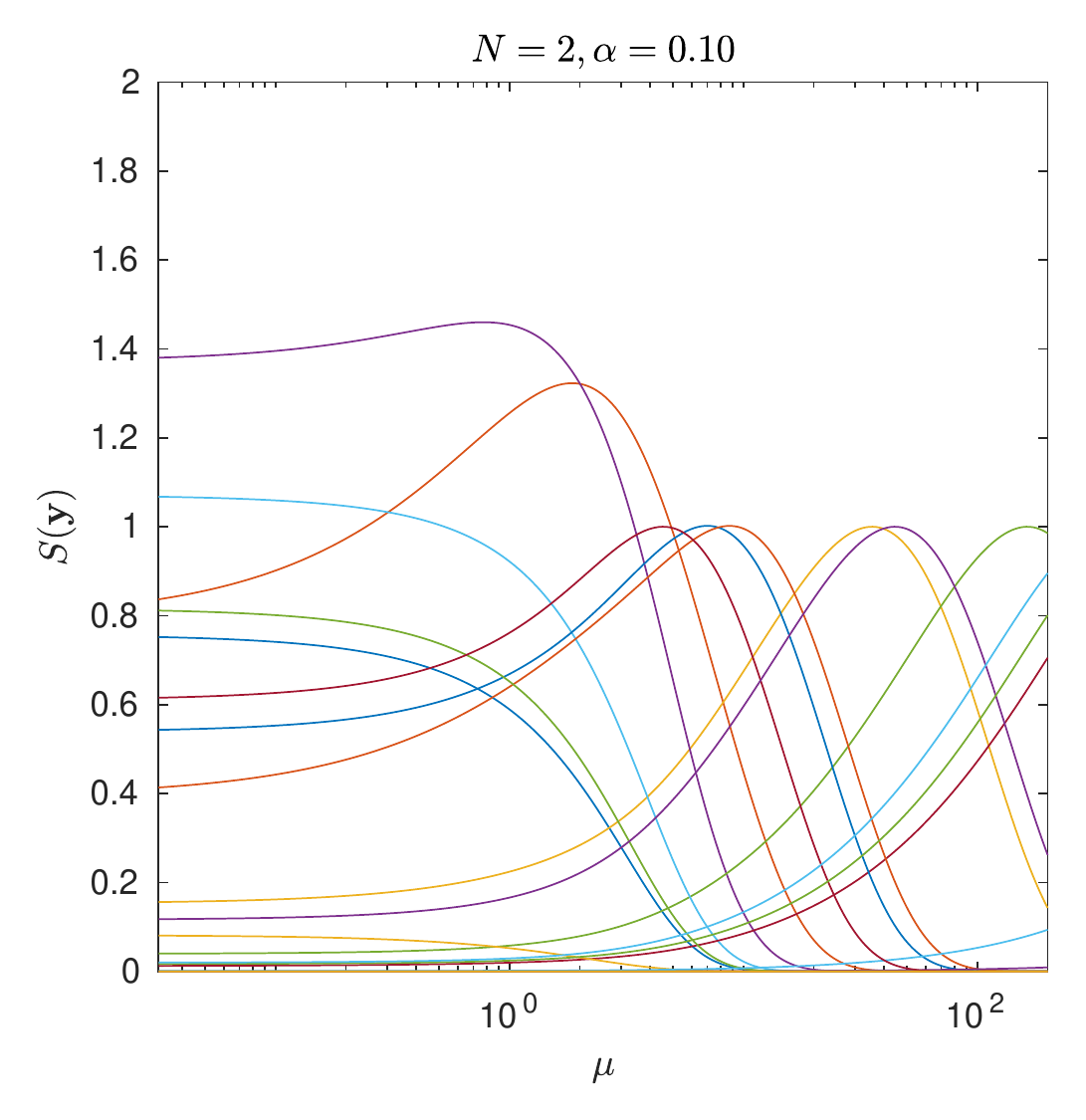} &
\includegraphics[width=70mm]{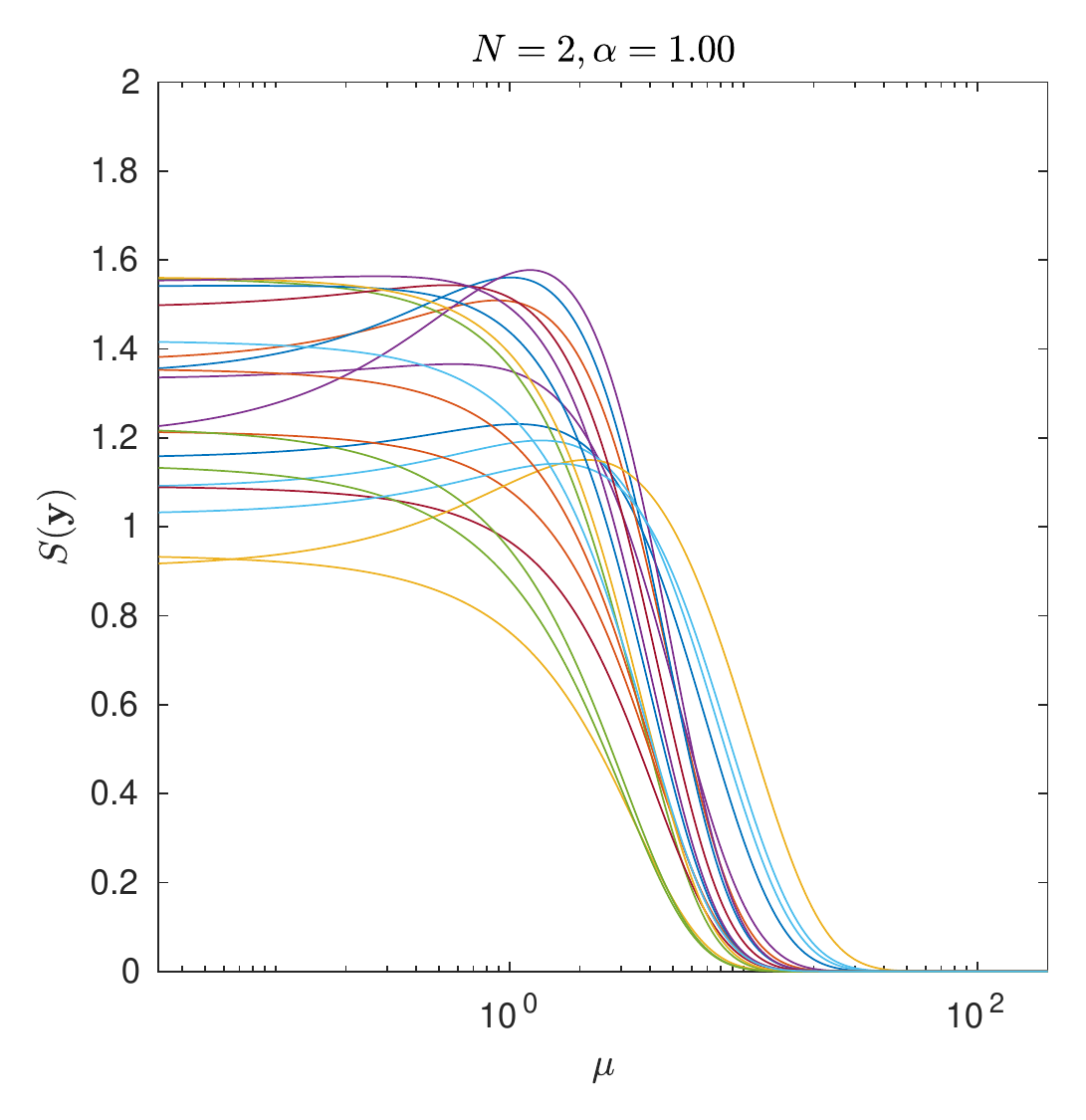} \\
\includegraphics[width=70mm]{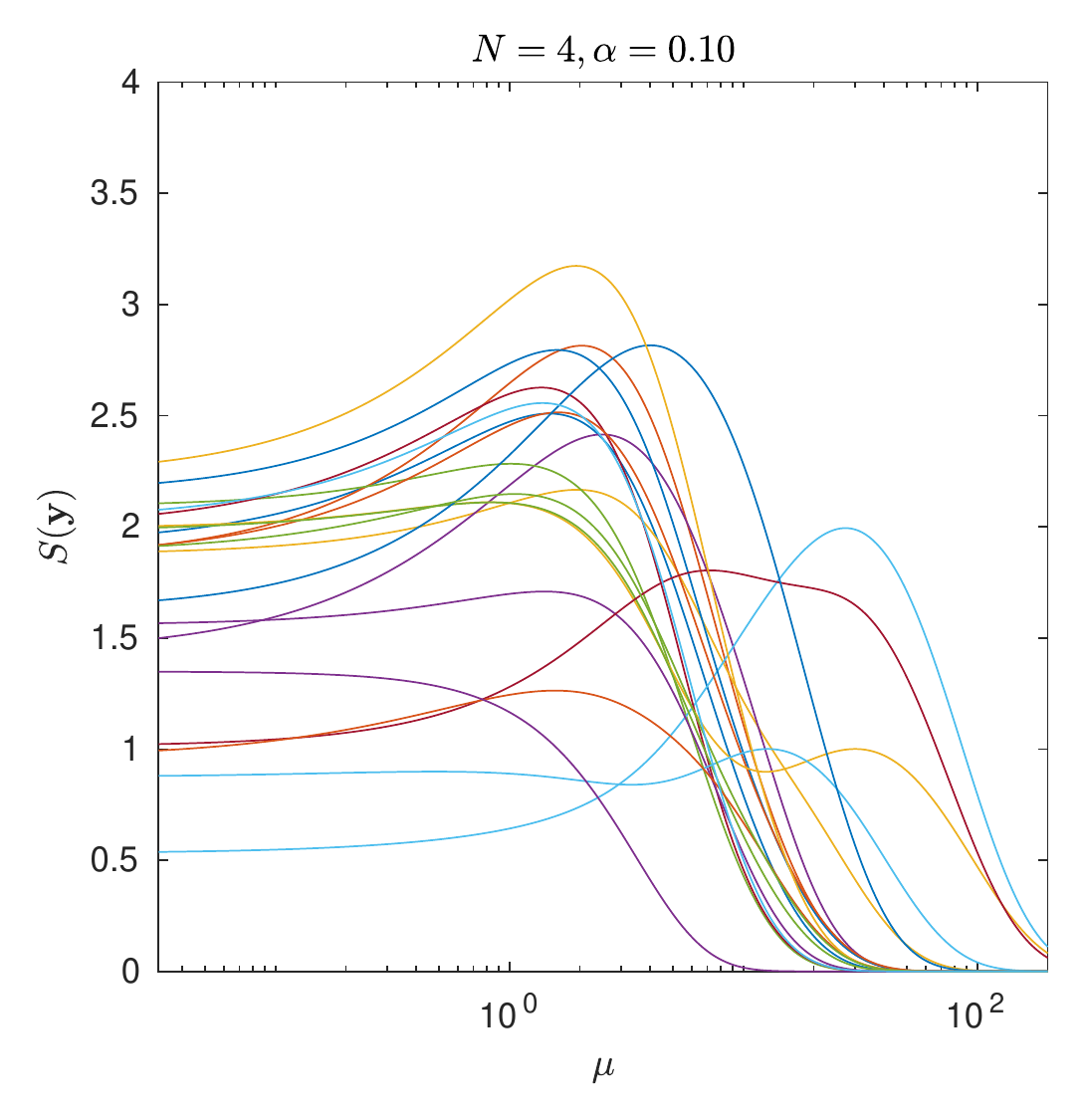} & \includegraphics[width=70mm]{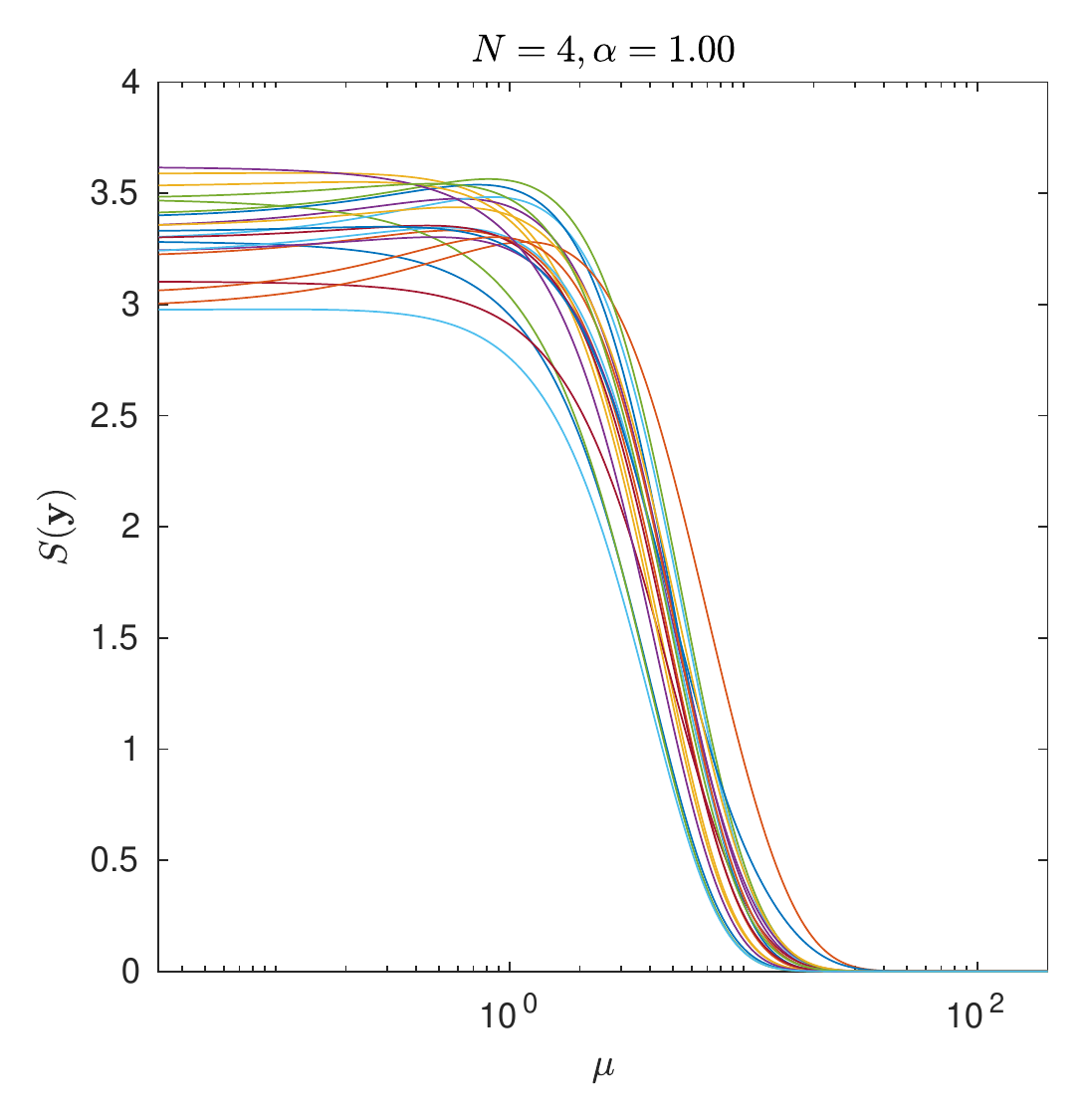} \\
\includegraphics[width=70mm]{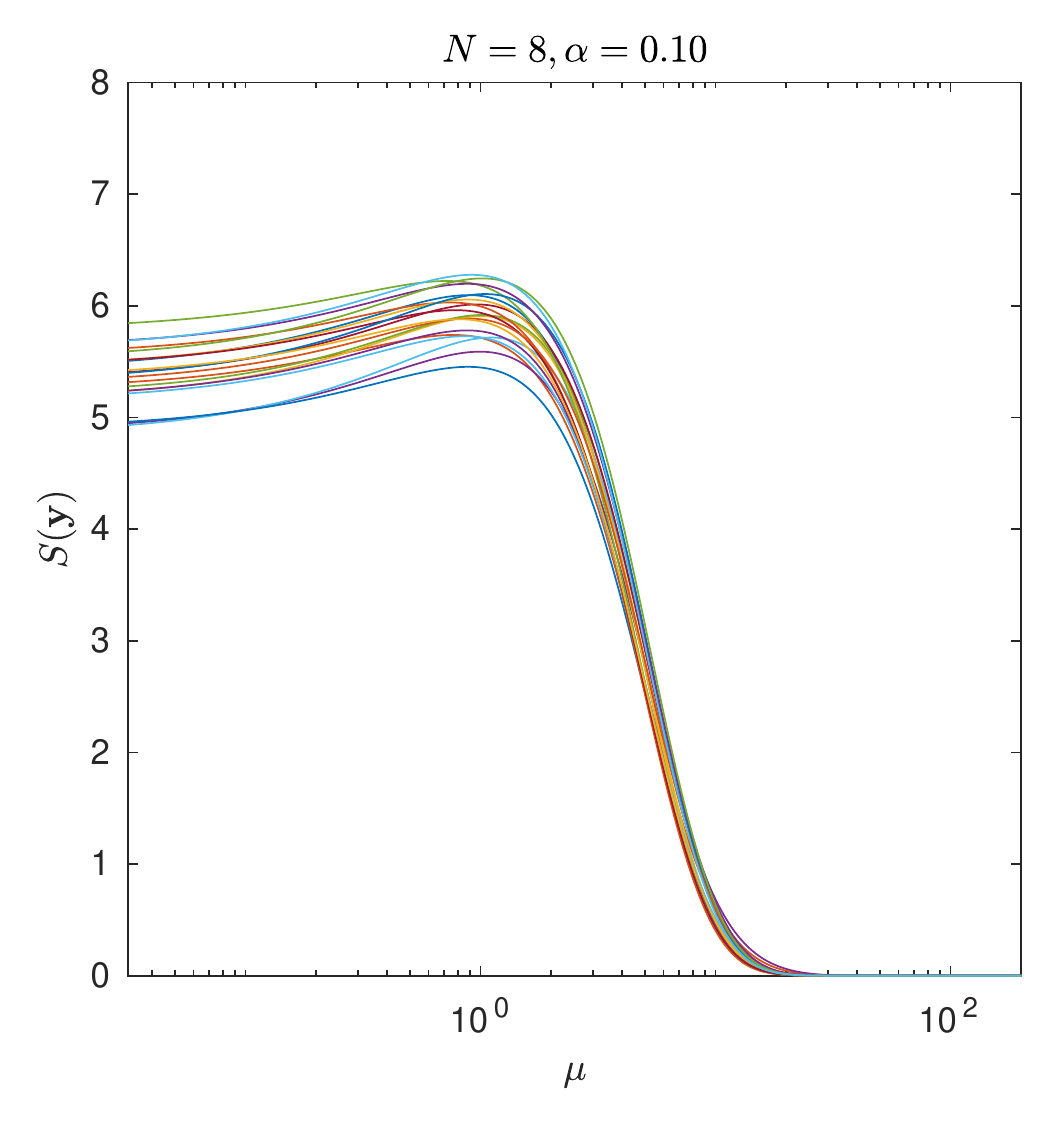} & \includegraphics[width=70mm]{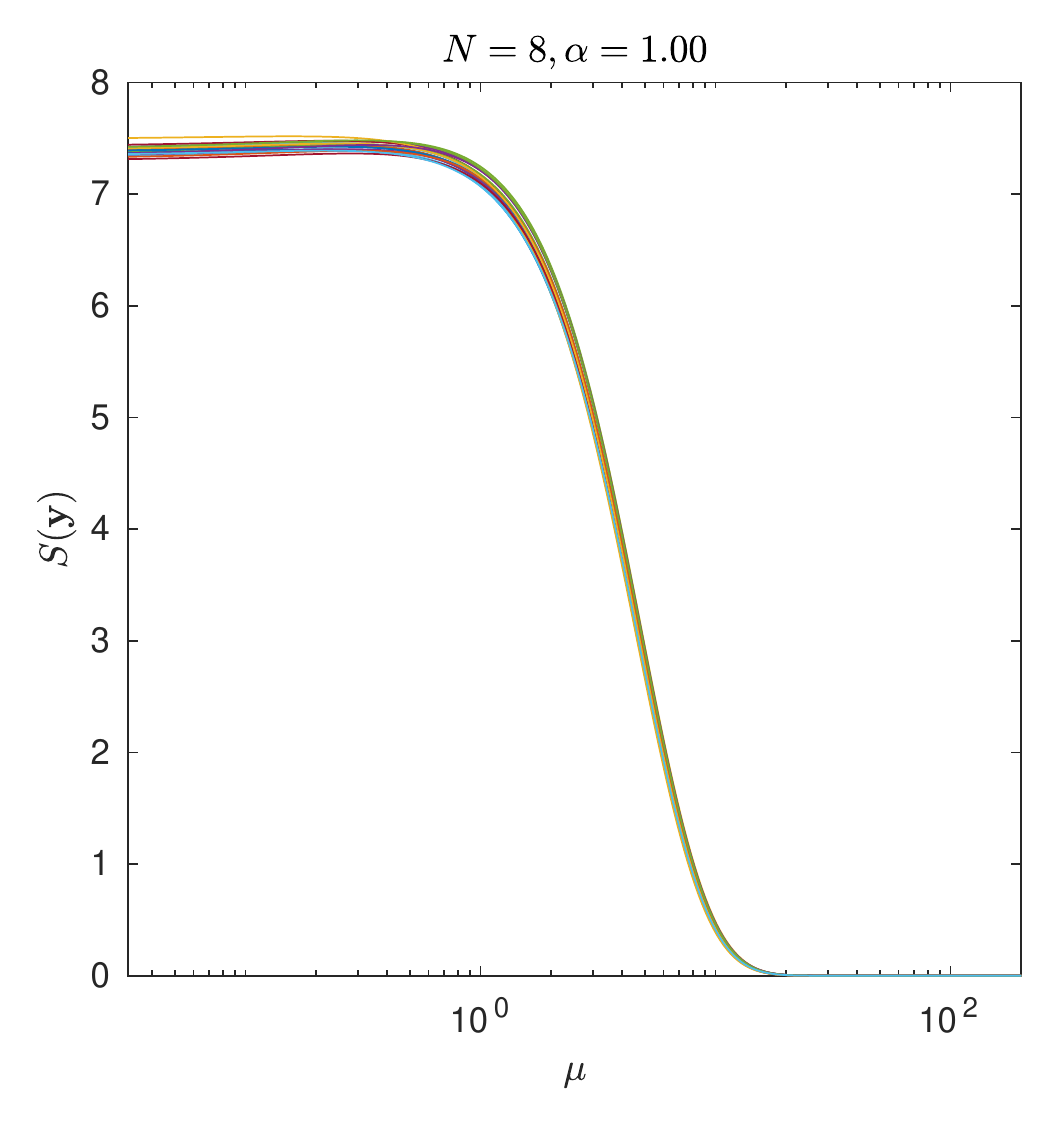}
\end{array}$
\caption{Shannon entropy $S(\mathbf{y})$ of the Dirichlet simulation realizations running as a function of $\mu$, where individual random realizations are visualized with different colors. }
\label{fig:dirichlet_realizations}
\end{figure}

To evaluate the performance of the inverse method, simulations are necessary. Standard numerical uncertainty methods such as bootstrapping may be used to quantify the uncertainties in-situ. We do the simulation in a fully controlled way and use the multidimensional Dirichlet distribution
\begin{equation}
f_D(x_1,\dots,x_n; \alpha_1,\dots,\alpha_n) = \frac{1}{B(\mathbf{\bm{\alpha}})} \prod_{i = 1}^n x_i^{\alpha_i - 1}, \text{ where } B(\mathbf{\bm{\alpha}}) = \frac{\prod_{i = 1}^n \Gamma(\alpha_i)}{\Gamma \left( \sum_{i = 1}^n \alpha_i \right)}
\end{equation}
as a random distribution for probabilities of $\mathbf{p}$ with constraint $\sum_c p_c = 1$. This constraint is by construction included in the Dirichlet distribution. The normalization factor is the multivariate Beta function, seemingly closely related to the famous Veneziano amplitudes of early string theory. We illustrate the Dirichlet distribution drawn realizations in Figure \ref{fig:dirichlet_realizations}, where we observe how the Shannon entropy $S(\mathbf{y})$ of the superposition result is running as a function of the compounding Poisson process $\mu$-value. Interestingly, the entropy has often non-monotonic behavior. Picking up random distributions allows us to sample the spectrum of different probability distributions of $\mathbf{p}$ maximally. This means a wide range of possible applications can be effectively probed, in a model independent way. We set the parameter vector of Dirichlet distribution $\bm{\alpha} = \alpha \mathbf{1}$ with concentration parameter $\alpha \in \{0.1, 1\}$, of which $\alpha = 1$ corresponds to the maximally uniform case in the multidimensional space of probabilities.

As a measure of agreement between the estimated probabilities and the true ones, we use the Kolmogorov-Smirnov type test statistic
\begin{equation}
D = \max |F(i) - \hat{F}(i)| \in [0,1], \;\;\; i = 1,\dots,n
\end{equation}
where $F(i) = \sum_{j = 1}^i p_j$ is the cumulative distribution function (CDF) of the true components of $\mathbf{p}$ and $\hat{F}(i)$ is the CDF of the estimate $\hat{\mathbf{p}}$. We are now neglecting all the  technicalities related to how rigorously extend the KS test to discrete distributions. We calculate only the test statistic itself, and do not calculate the significance level or $p$-value. In principle this could be done using the Kolmogorov distribution.

\begin{figure}[H]
\centering
$\begin{array}{ccc}
\includegraphics[width=70mm]{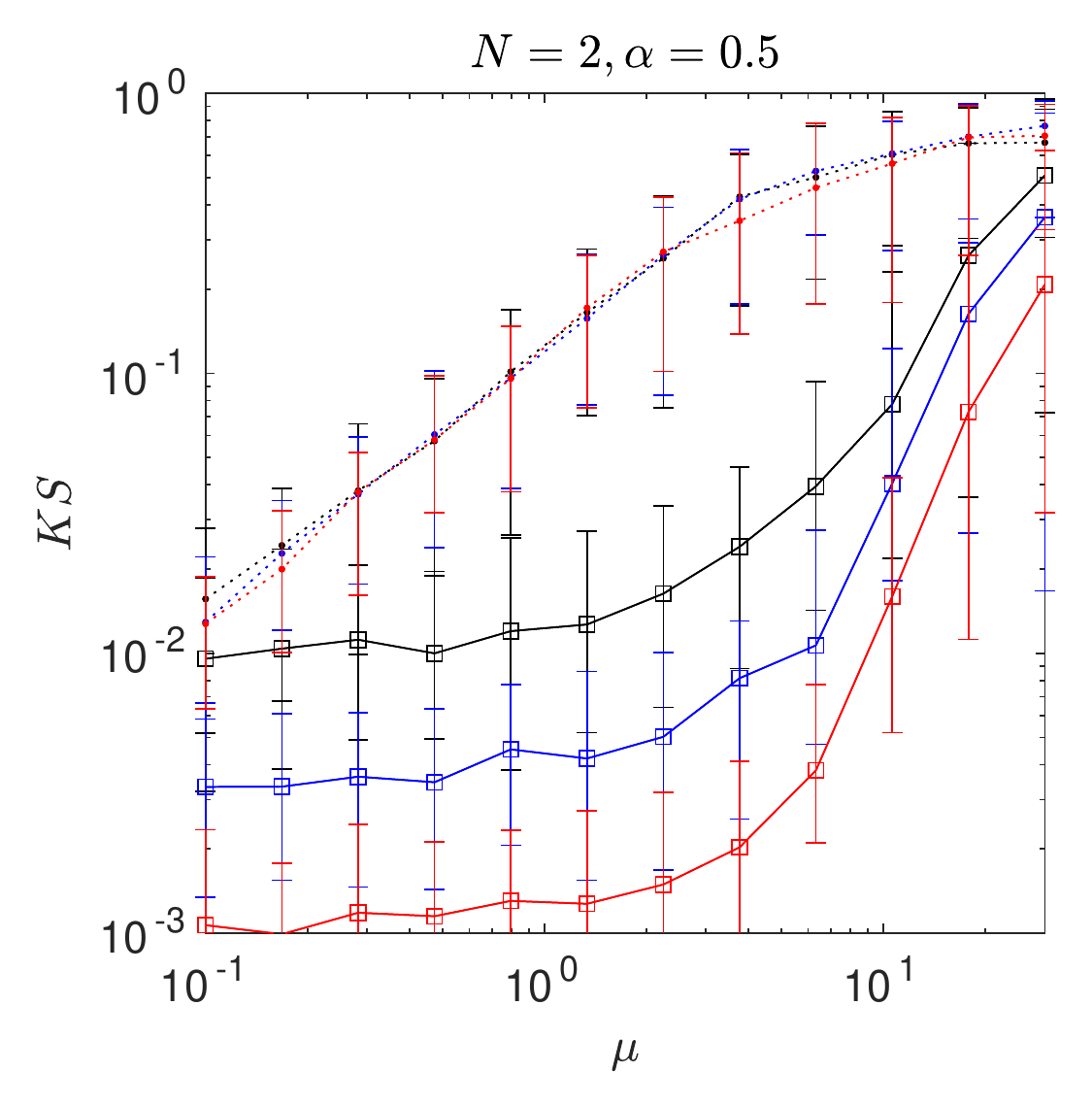} &
\includegraphics[width=70mm]{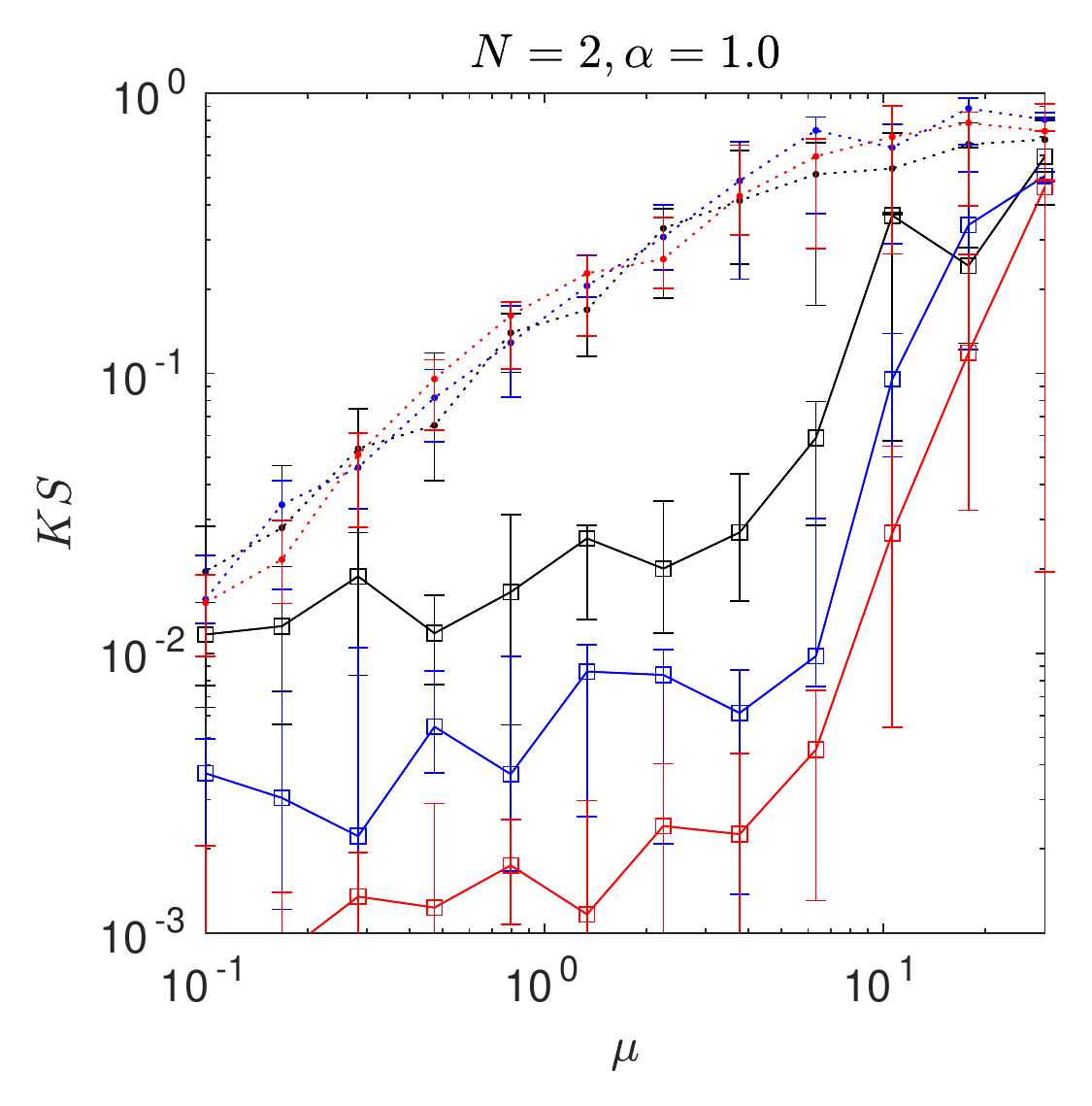} \\
\includegraphics[width=70mm]{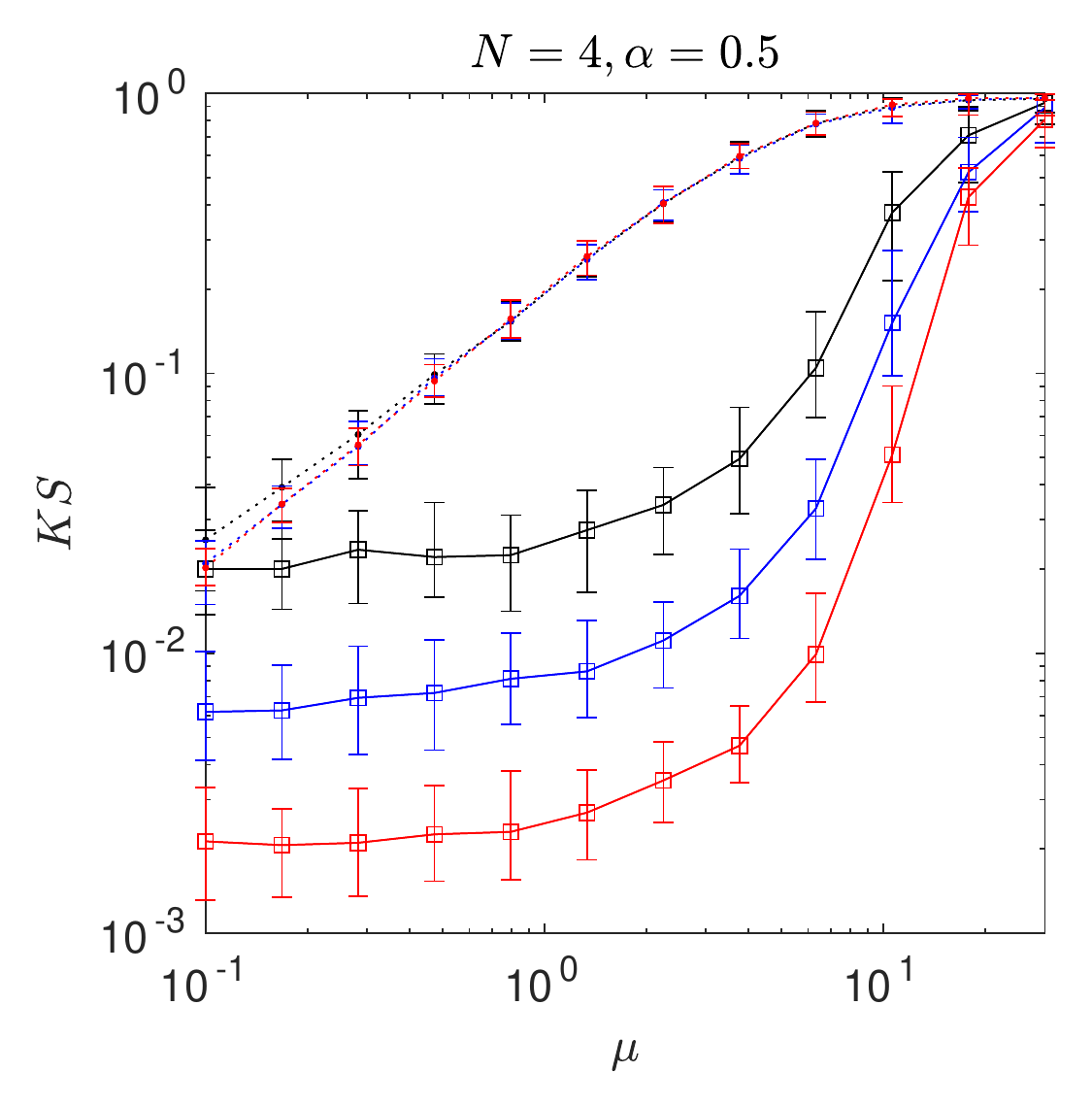} &
\includegraphics[width=70mm]{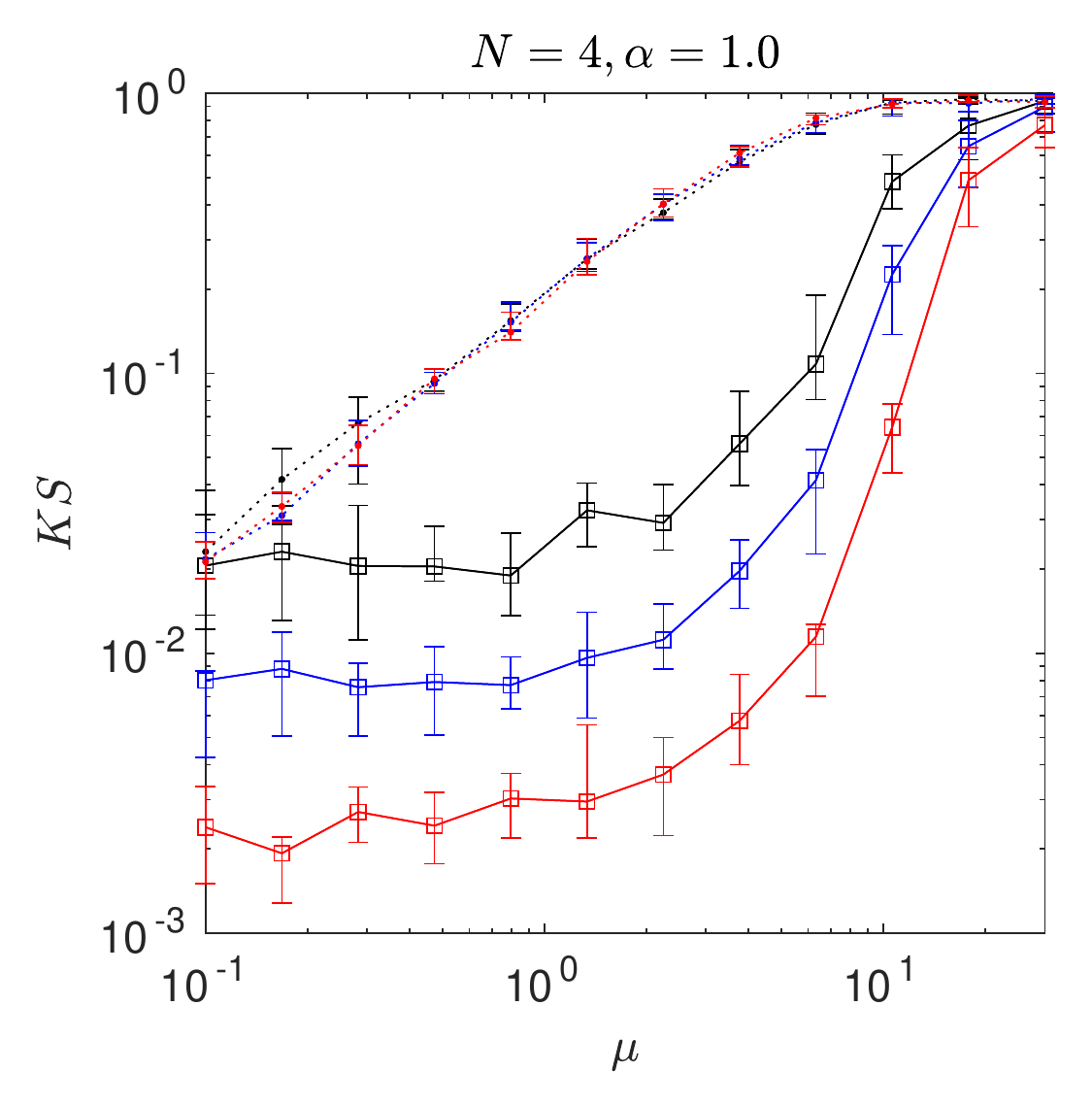} \\
\includegraphics[width=70mm]{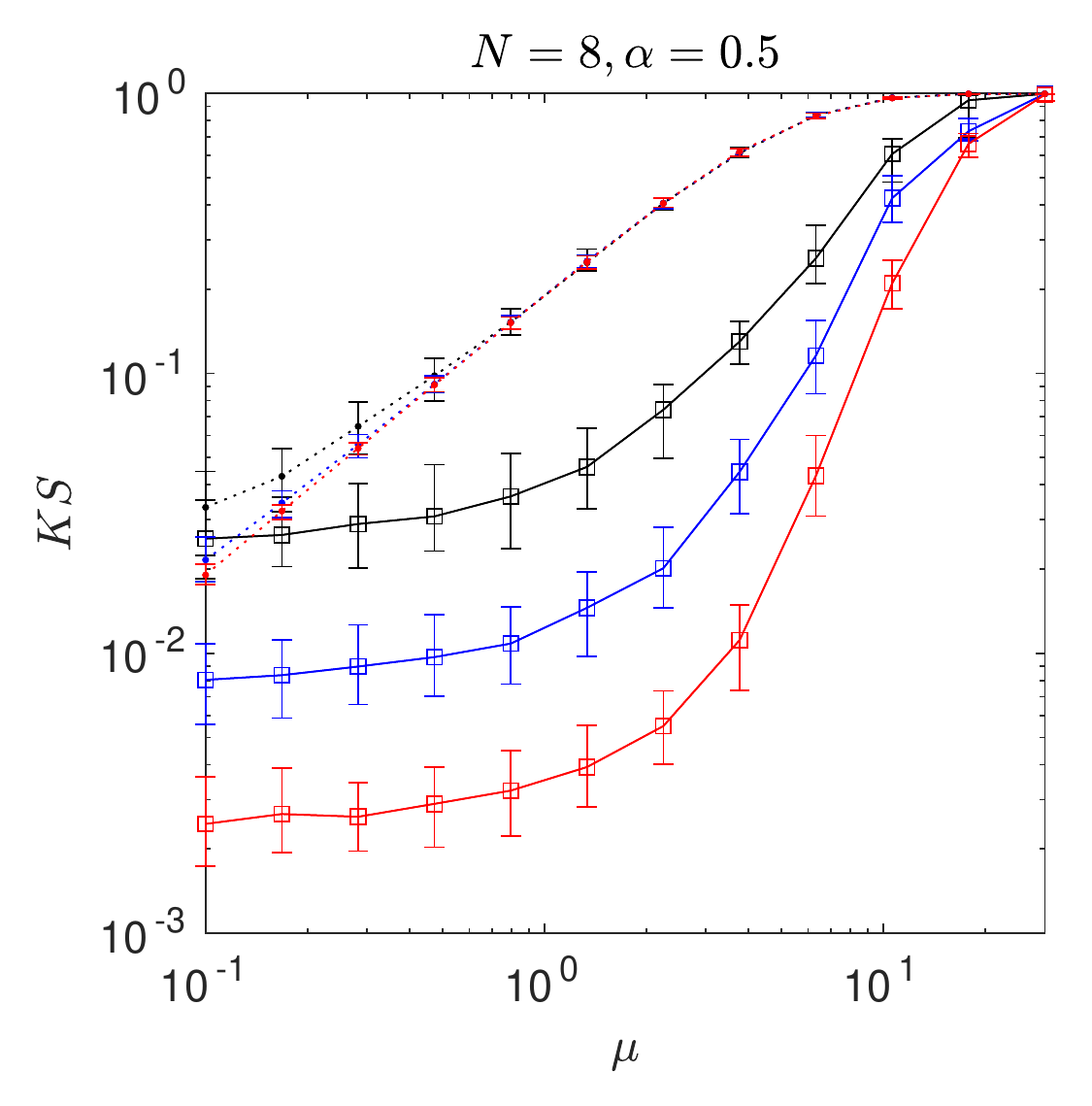} & \includegraphics[width=70mm]{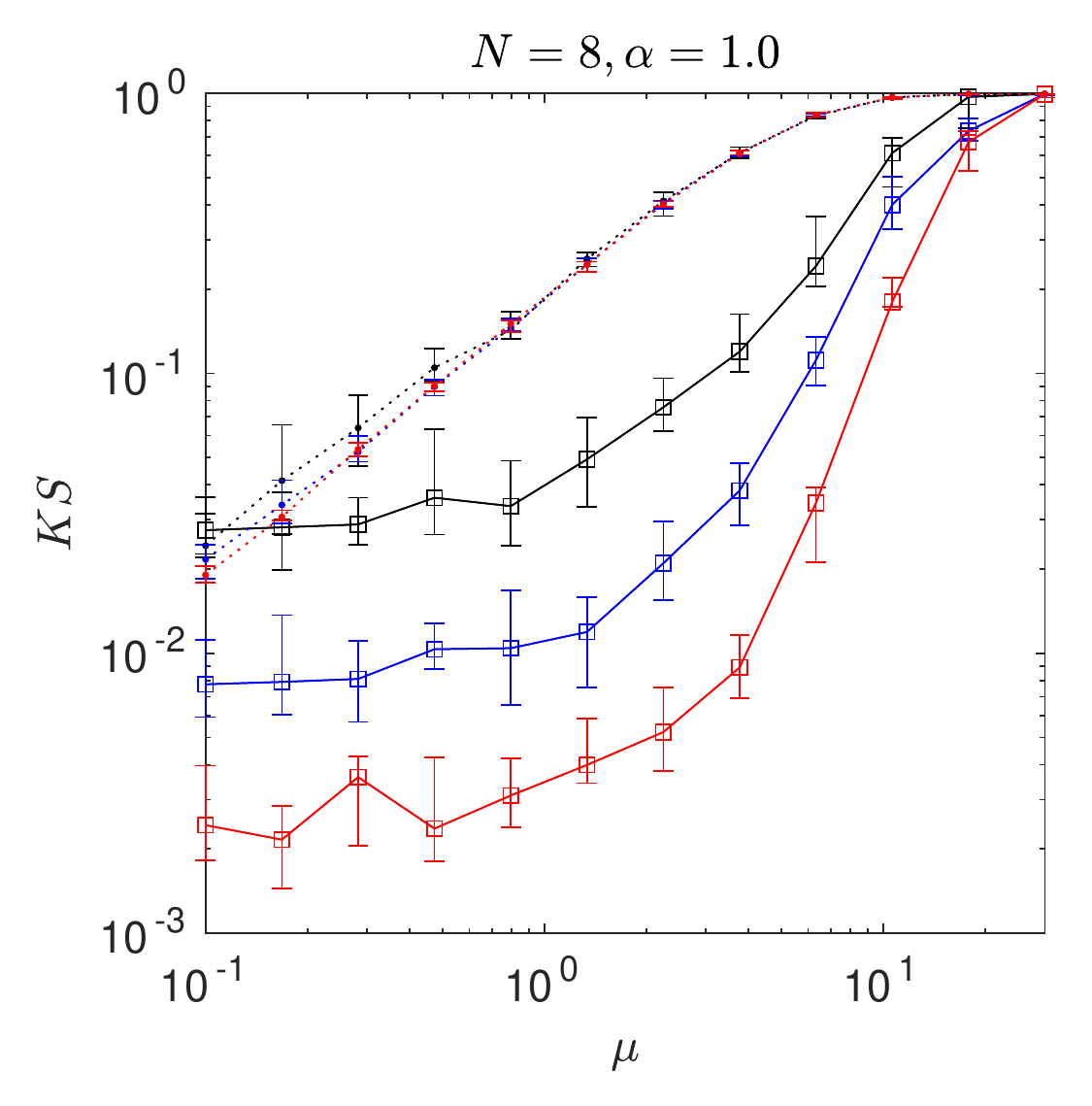}
\end{array}$
\caption{Kolmogorov-Smirnov error on vertical axis. Median values and $\pm 1\sigma$ intervals as a function of $\mu$, vector space dimension $N = 2,4,8$ and the number of simulated events $N_E = 10^3, 10^4, 10^5$ (black, blue, red). Solid lines are after, and dashed lines before the inversion.}
\label{fig:simulations}
\end{figure}

\begin{figure}[H]
\centering
$\begin{array}{ccc}
\includegraphics[width=70mm]{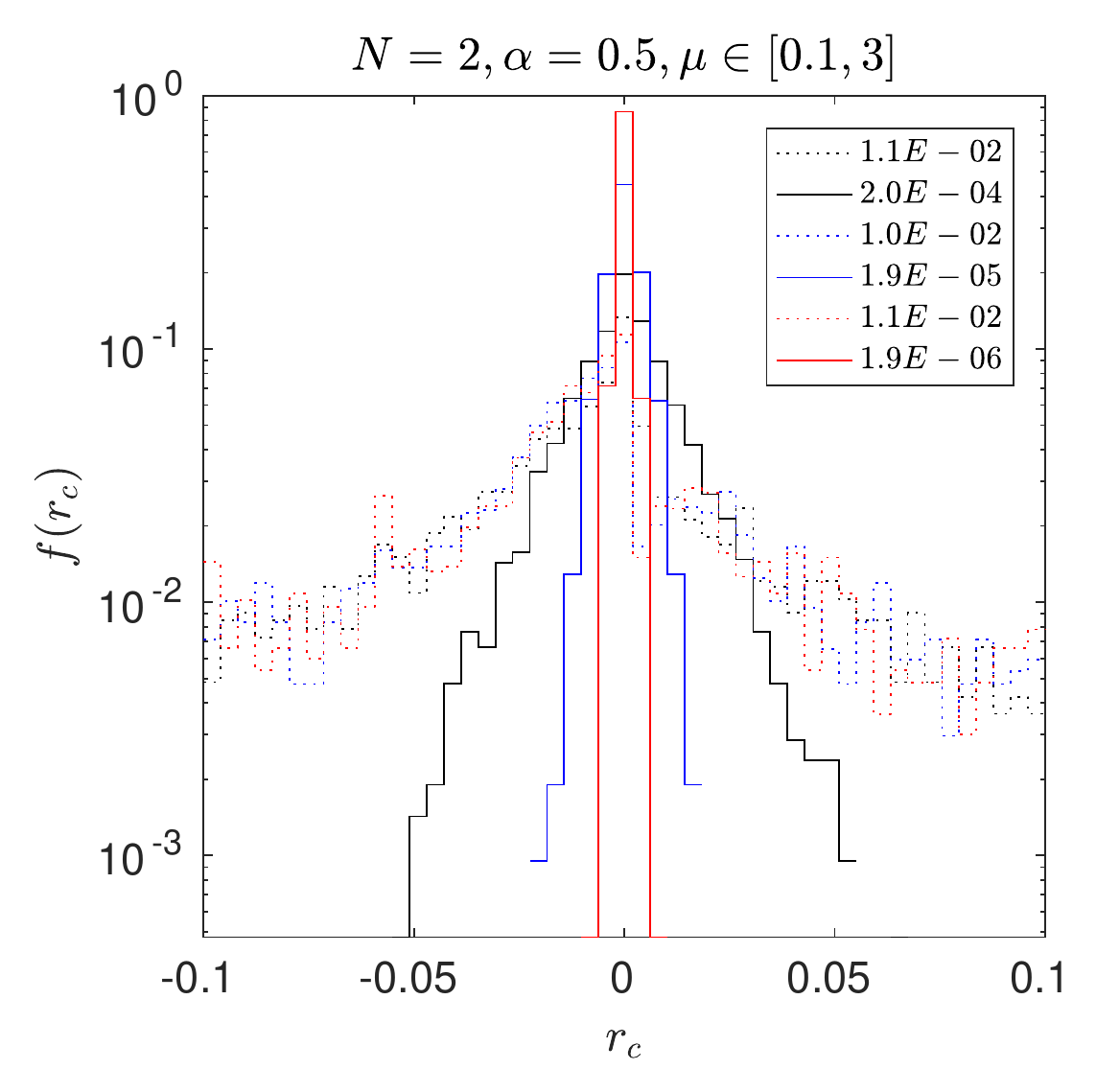} &
\includegraphics[width=70mm]{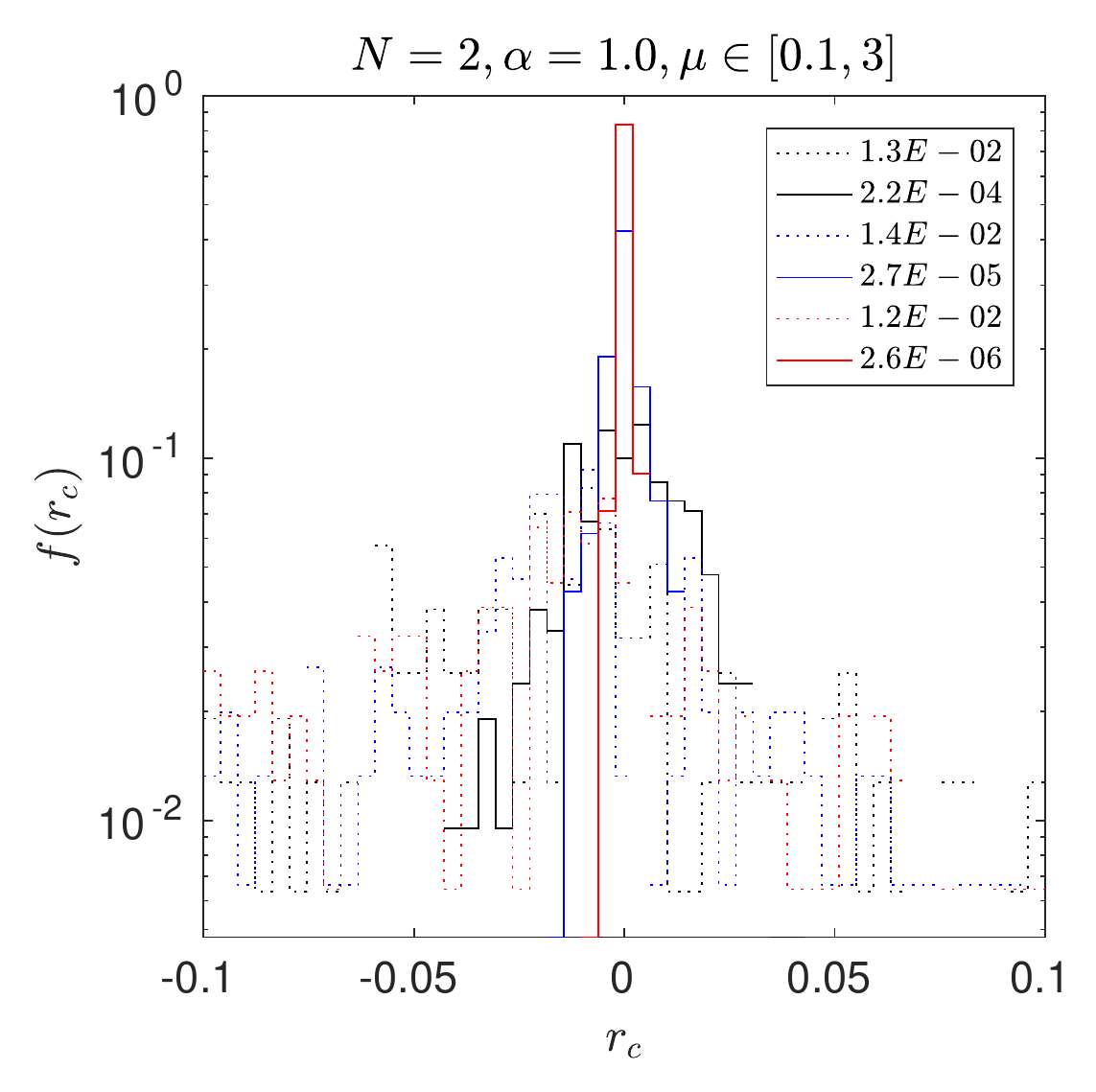} \\
\includegraphics[width=70mm]{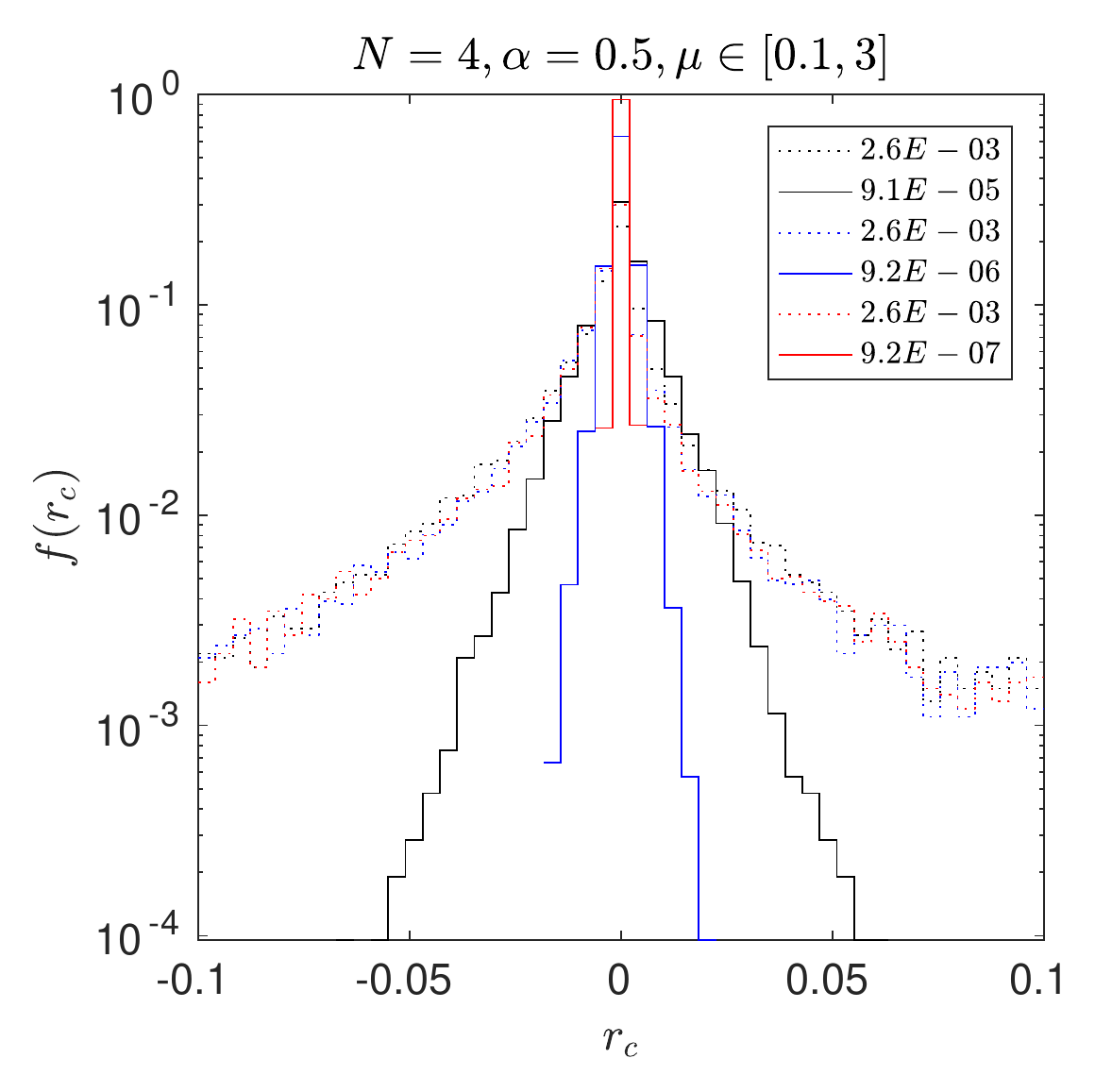} & \includegraphics[width=70mm]{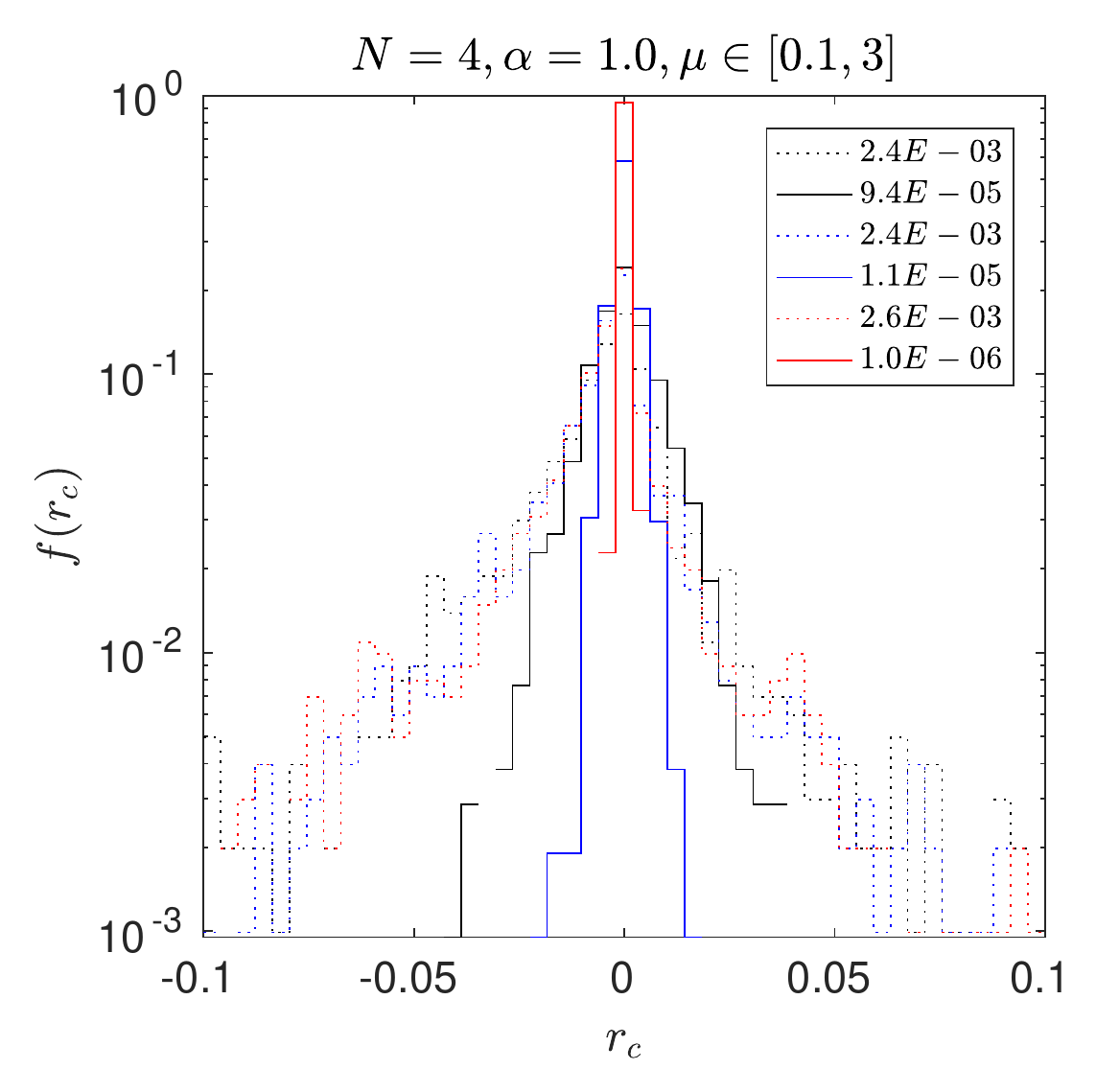} \\
\includegraphics[width=70mm]{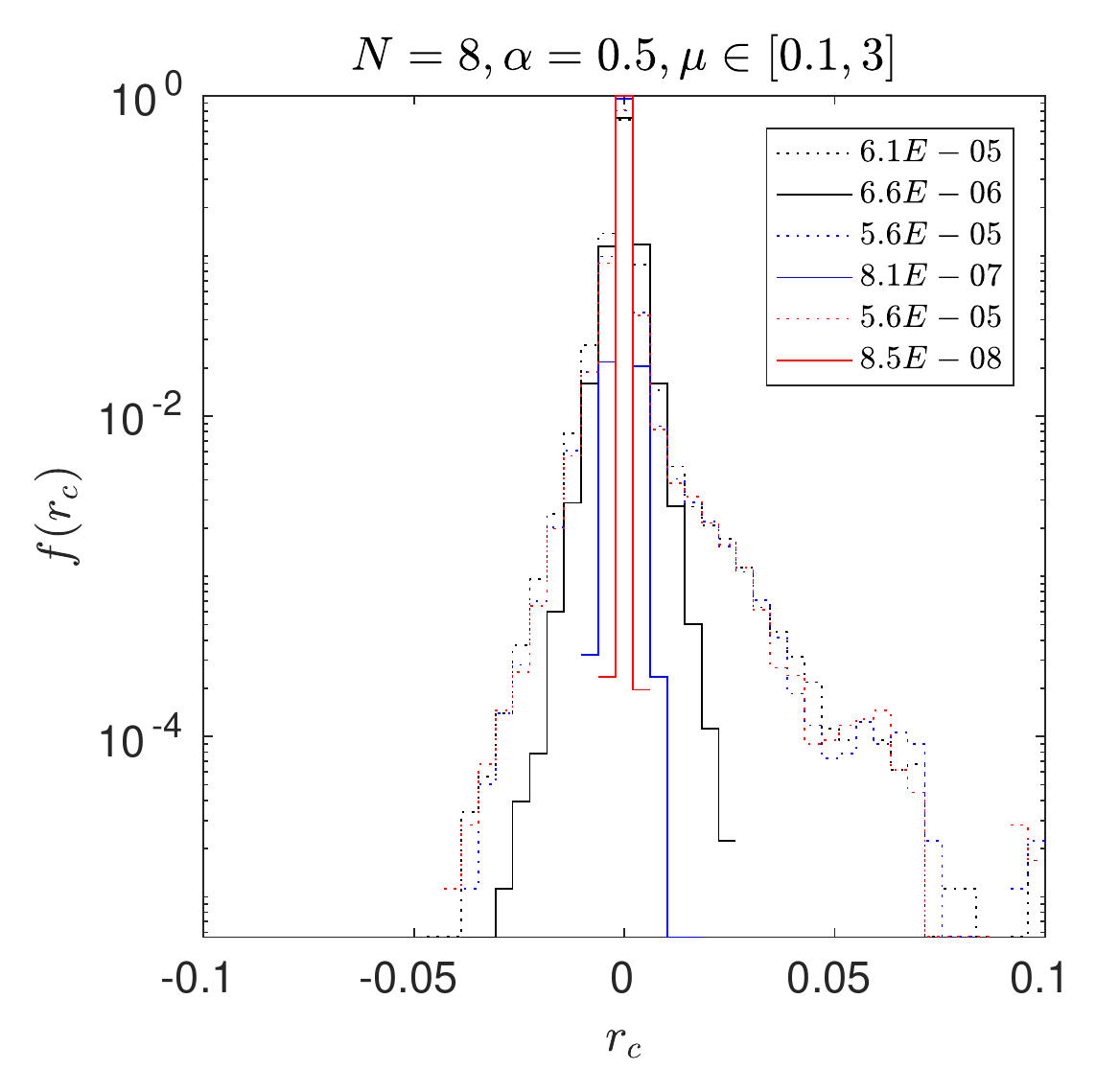} & \includegraphics[width=70mm]{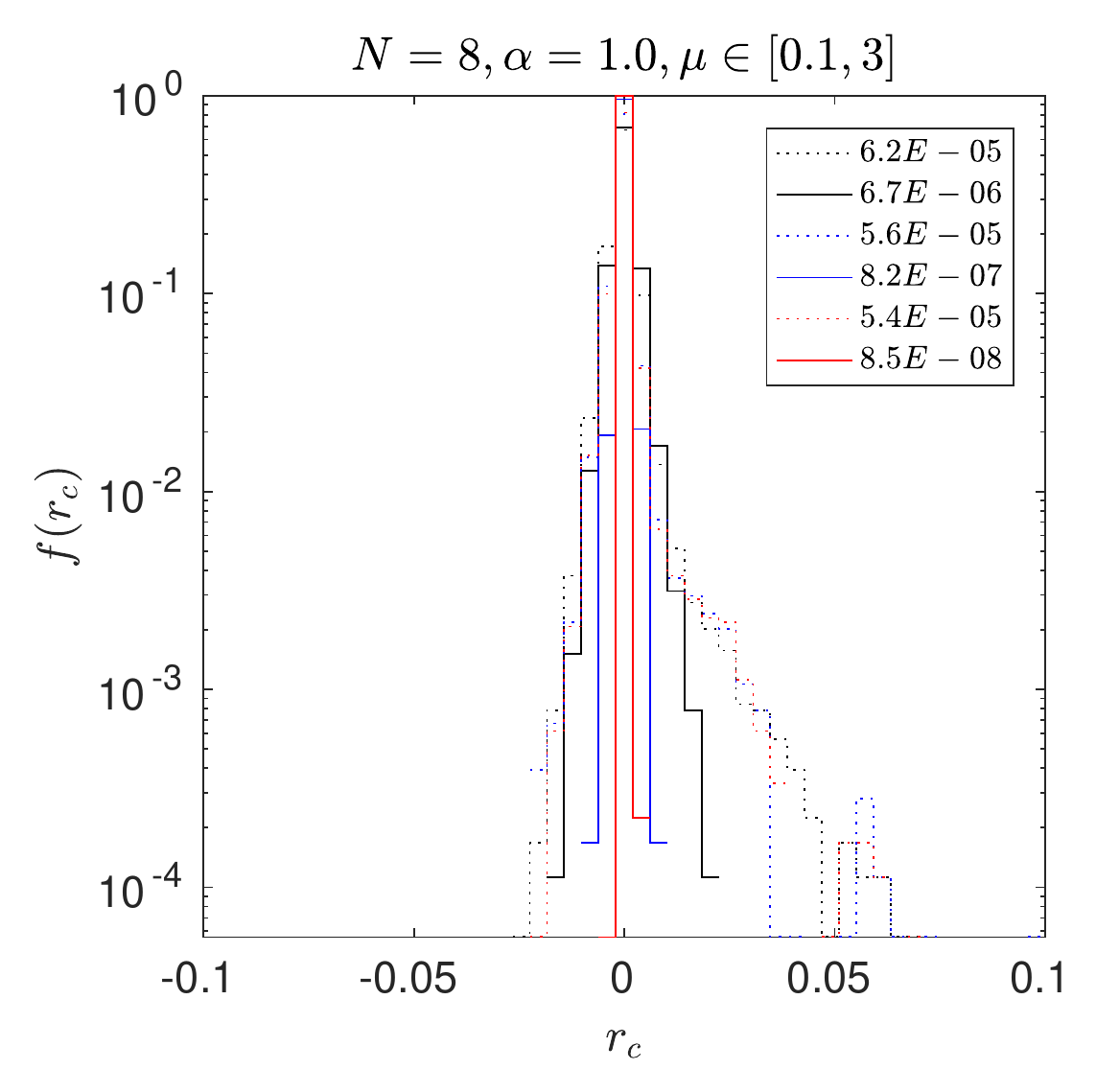}
\end{array}$
\caption{Error histograms $e_c = \hat{p}_c - p$ with dimensionality of binary vector space $N = 2,4,8$ and the simulated number of events is $N_E = 10^3, 10^4, 10^5$ with black, blue and red markers. Solid lines are after, and dashed lines before the inversion.}
\label{fig:simulations2}
\end{figure}

The results of several simulations are shown in Figures \ref{fig:simulations} and \ref{fig:simulations2}. The dashed lines correspond to uncorrected distributions and the solid ones for inverted results and error bars are 1 sigma (68 CL) uncertainties around the median value, obtained via Monte Carlo runs. The run $\mu$-value is varied between $10^{-3}$ and 30. To demonstrate the $\sqrt{N_E}$ scaling or asymptotic efficiency of the algorithm, we vary the number of simulated events $N_E$. We observe a clear saturation in the inversion process at high $\mu$-values, because at that point all event signatures are almost identical. That is, all binary observables are $B_i \equiv 1$ for all $i = 1,\dots,N$. The error residual $r_c = \hat{p}_c - p_c$ distribution histograms are shown in Figure \ref{fig:simulations2} sampled over different $\mu$-value runs. In the histogram legend, the variance of the residuals are shown $\text{var}[\hat{p}]=\text{E}[( \hat{p} - \mathbb{E}[\hat{p}] )^2]$. The estimates are seen to be unbiased, as expected from the construction, thus the mean square error (MSE) is equal to the variance of the estimates.

\newpage

\section{Summary}
\label{sec:conclusions}

We built explicitly the following new constructions:
\vspace{1em}

\begin{itemize}
\item A novel algebraic formulation for $N$-point correlations and complex event `topologies' over rapidity, transverse momentum and multiplicity designed as new observables of high energy soft QCD diffraction. The framework includes vector fiducial partial cross sections, fractal marginal distributions, transverse momentum or multiplicity dependent gap trajectories and multidimensional diffraction fit and projection algorithms.

\item A novel description of combinatorial compound processes based on the incidence algebras and inversion of the corresponding superposition Poisson problem based on the M\"obius inversion theorem. The invertability with finite statistics was simulated.
\end{itemize}

As a related technical remark regarding the superposition problem, the current event by event pileup correction methods at the LHC are essentially non-linear filtering operations (such as the median filter) applied on the observables event by event. A combination of the statistical approaches developed here and the event by event methods of \cite{bertolini2014pileup, larkoski2014soft, soyez2013pileup, cacciari2008pileup}could be possible.

\paragraph{Code} The code packages to reproduce the work here and beyond are available under the MIT license at \href{https://github.com/mieskolainen}{github.com/mieskolainen}.

% -----------------------------------------------

\acknowledgments

Risto Orava is acknowledged for discussions on the topic, and Abraham Villatoro Tello and Arseniy Shabanov for exploring the experimental issues.  I thank also Krzysztof Golec-Biernat for expressing interest in the work and pointing out misprints in formulas.

\newpage
% natbib
\bibliography{mainlatex}

% BIBLATEX
%\printbibliography

\newpage

\appendix
\section{Appendix}
\label{sec:appendix}

% -----------------------------------------------

This appendix holds some basic algebraic and other mathematical properties of the framework.

\subsection{Vector space subspaces over finite fields}

The $q$-binomial coefficient counts the number of subspaces of dimension $r$ in a vector space of dimension $N$, when the vector space is over a finite field with prime power $q$
\begin{equation}
\label{eq:qbinomial}
\begin{bmatrix}
N \\
r
\end{bmatrix}_q
= 
\begin{cases}
    \frac{(1-q^N)(1-q^{N-1})\cdots(1-q^{N-r+1})}{(1-q)(1-q^2)\cdots(1-q^r)}, & \text{if $r \leq N$}.\\
    0, & r > N.
\end{cases}
\end{equation}
These subspaces are encapsulated in the Grasmannian manifold Gr($r,N,\mathbb{F}_q$) with $\dim(\text{Gr}) = r(N-r)$.

\begin{table}[h]
\center
\begin{tabular}{c|lllllllll|c}
$N \backslash r$ & 0 & 1 & 2 & 3 & 4 & 5 & 6 & 7 & 8 & $\Sigma$ \\
\hline
1 & 1 & 1 & 0 & 0 & 0 & 0 & 0 & 0 & 0 & 2 \\
2 & 1 & 3 & 1 & 0 & 0 & 0 & 0 & 0 & 0 & 5 \\
3 & 1 & 7 & 7 & 1 & 0 & 0 & 0 & 0 & 0 & 16 \\
4 & 1 & 15 & 35 & 15 & 1 & 0 & 0 & 0 & 0 & 67 \\
5 & 1 & 31 & 155 & 155 & 31 & 1 & 0 & 0 & 0 & 374 \\
6 & 1 & 63 & 651 & 1395 & 651 & 63 & 1 & 0 & 0 & 2825 \\
7 & 1 & 127 & 2667 & 11811 & 11811 & 2667 & 127 & 1 & 0 & 29212 \\
8 & 1 & 255 & 10795 & 97155 & 200787 & 97155 & 10795 & 255 & 1 & 417199 \\
\end{tabular}
\caption[Table caption text]{Total number of $r$-dimensional subspaces of $\mathbb{F}_2^N$. }
% Mathematica: GaloisNumber[N\_, q\_] := Sum[QBinomial[N, r, q], {r, 0, N}]
\label{table: subspacemultiplicities}
\end{table}
The total number of subspaces is obtained with $\sum_{r=0}^N \begin{bmatrix} N \\ r \end{bmatrix}_q$. This count gives also the so-called simplexity of the $N$-cube in a minimal corner-cut triangulation of the cube, which is geometrically intuitive. For $q = 2$ and $N = 1,\dots,8$, these are listed in Table \ref{table: subspacemultiplicities}. Note that $r$ starts from zero.

\subsection{Pileup combinatorics}

\begin{figure}[h]
\centering
\vspace{1em}
\includegraphics[scale=0.7]{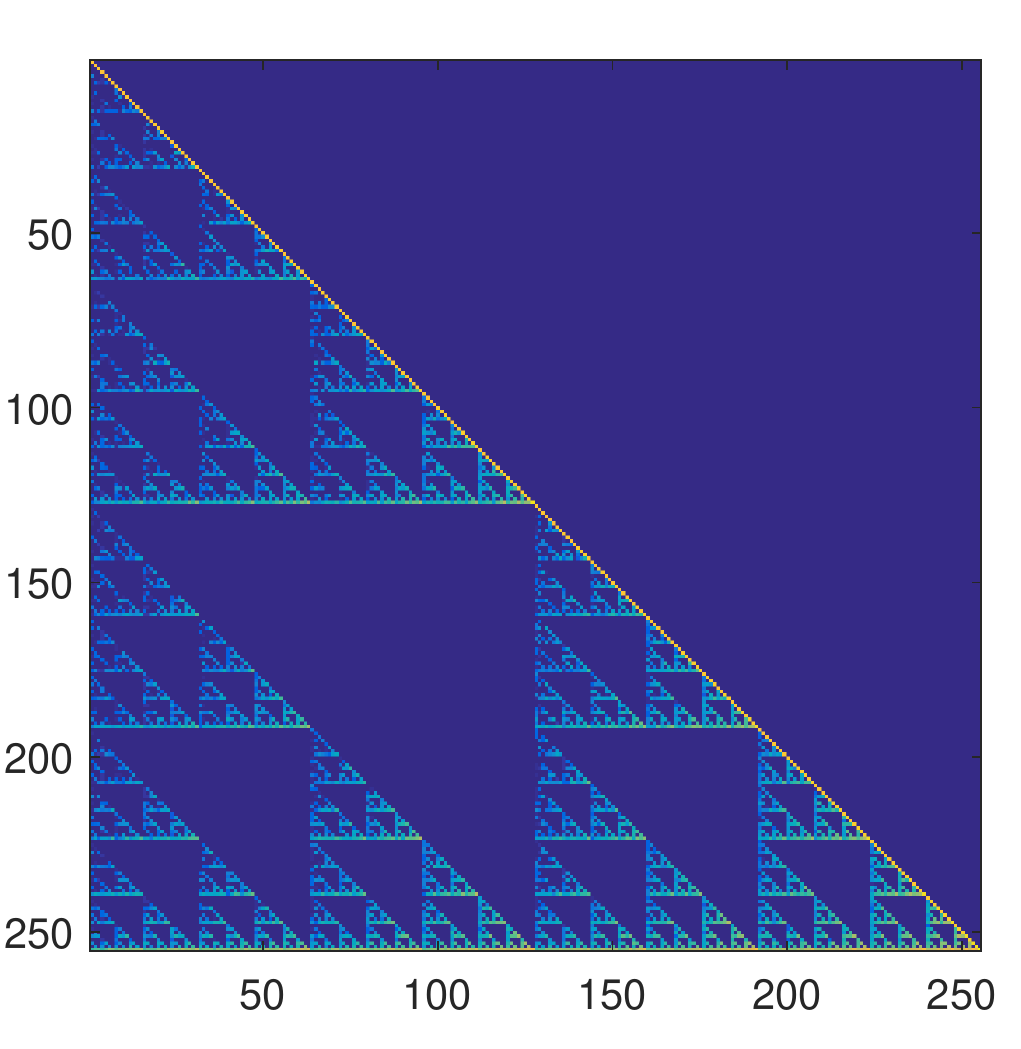}
\caption{Simulated pileup mixing matrix $M_{255 \times 255}$ for $N = 8$ and $\mu = 1.5$.}
\label{fig: mixing}
\end{figure}

Let $M:\Theta \rightarrow Y$ be a probabilistic or stochastic mixing matrix 
\begin{equation}
\mathbf{y} = M(\mathbf{p},\mu) \mathbf{p}, \; \; \; \sum_i M_{ij} = 1 \;\; \forall j
\end{equation}
which has each column with unit sum and size $2^N-1$. Each matrix element
\begin{equation}
M_{ij} \equiv P(c = i | c = j)
\end{equation}
gives the conditional probability of an event vector originating from the $j$-th final state class $c$ propagating to the $i$-th class, due to pileup. The matrix is thus interpreted vertically, which makes it a left stochastic matrix. This matrix is illustrated in Figure \ref{fig: mixing}.

The matrix was obtained by a toy Monte Carlo pileup simulation using Poisson random numbers with mean $\mu$ and by generating different final states according to uniform probabilities $p_c \sim 1/n$. The mixing matrix was obtained by counting the mixed final states event by event. Due to permutation ambiguity, the originating class $c$ was selected always to be the first of the random list, thus randomly. Because the matrix $M$ here is a function of (unknown) $\mathbf{p}$, and also $\mu$, it is not directly useful as an algorithmic inverse solution. Its function is to demonstrate the algebraic properties. In general the combinatorial enumerations can be approached by the `twelve-fold way', which is a systematic approach developed by G.C. Rota. Here we list some combinatorial aspects of this problem in order to make it more transparent.

The number of non-zero elements in the pile-up mixing matrix per row are given by
\begin{equation}
\label{eq:goulds}
a(i) = \sum_{r = 1}^i
\begin{pmatrix}
i \\
r
\end{pmatrix}
\text{mod}\,2 ,
\end{equation}
where mod denotes here the remainder after division and $i$ is the index of the row. The sequence for the first 15 rows is
\begin{equation} 1,1,3,1,3,3,7,1,3,3,7,3,7,7,15, \;\;\; i = 1,\dots,15.
\end{equation}
Similarly, the number of 1 in the binary vector representation, or the so-called Hamming weight, is given by adding one and taking base-2 logarithm of Equation \ref{eq:goulds}
\begin{equation}
a(i) = \log_2 \left( \sum_{r = 1}^i
\begin{pmatrix}
i \\
r
\end{pmatrix}
\text{mod}\,2 + 1\right).
\end{equation}
As an example
\begin{equation} 1,1,2,1,2,2,3,1,2,2,3,2,3,3,4, \;\;\; i = 1,\dots,15
\end{equation}
This sequence, namely the position of 1 in this sequence, give us directly the binary vectors which undergo only autocompound. That is, their signature cannot be imitated by a linear combination of other vectors. An example of this is given in Table \ref{table: multiplicities_per_N}, for example when $N = 3$, these are $c =$ 1,2 and 4. These are the unit basis vectors of $\mathbb{F}_2^N$. We also see the `fractal' structure, that is, increasing the dimension of binary space keeps the multiplicity structure from lower dimensions which is evident from Table \ref{table: multiplicities_per_N}. This is clear given the Hamming weight sequence.

The number of different terms for $c = 3$ (or $5,6,9,\dots$) follows from a sequence
\begin{equation}
a(k) = 
\begin{pmatrix}
k+1 \\
2
\end{pmatrix}
+k-1
=
\stirling{k+1}{k} + k - 1
,
\end{equation}
which generates a multiplicity sequence $1,4,8,13,19,26,\dots$ and the curly brackets on right give the Stirling partition number (Stirling number of the second kind). This is the base of all higher dimensional binary vector space sequences. All the higher order multiplicity matrices are self repeating results of this sequence or identity sequence $1,1,\dots$, which follows the algebraic binary expansion ordering. An example of this is given in Table \ref{table: multiplicities_per_N}.

\begin{table}[h]
\center
\begin{tabular}{c|llllll}
$\#_{(N,k)}$ & 1 & 2 & 3 & 4 & 5 & 6 \\
\hline
2 & 3 & 6 & 10 & 15 & 21 & 28  \\ 
3 & 7 & 28 & 84 & 210 & 462 & 924  \\ 
4 & 15 & 120 & 680 & 3060 & 11628 & 38760  \\ 
5 & 31 & 496 & 5456 & 46376 & 324632 & 1947792  \\ 
6 & 63 & 2016 & 43680 & 720720 & 9657648 & 109453344  \\ 
7 & 127 & 8128 & 349504 & 11358880 & 297602656 & 6547258432  \\ 
8 & 255 & 32640 & 2796160 & 180352320 & 9342250176 & 404830840960  \\ 
9 & 511 & 130816 & 22369536 & 2874485376 & 296071993728 & 25462191460608  \\ 
10 & 1023 & 523776 & 178956800 & 45902419200 & 9428356903680 & 1615391816163840  \\
\end{tabular}
\caption[Table caption text]{Combinatorial total multiplicities for $N \in [2,10]$ and $k \in [1,6]$.}
\label{table: multiplicities}
\end{table}

\begin{table}[h]
\center
\begin{tabular}{c|llllll}
$N = 2$ & $k$\\
$c$ & 1 & 2 & 3 & 4 & 5 & 6 \\
\hline
1 & 1 & 1 & 1 & 1 & 1 & 1 \\
2 & 1 & 1 & 1 & 1 & 1 & 1 \\
3 & 1 & 4 & 8 & 13 & 19 & 26 \\
\hline
$\Sigma$ & 3 & 6 & 10 & 15 & 21 & 28 \\
\hline
\\
$N = 3$ & $k$ \\
$c$ & 1 & 2 & 3 & 4 & 5 & 6 \\
\hline
1 & 1 & 1 & 1 & 1 & 1 & 1 \\
2 & 1 & 1 & 1 & 1 & 1 & 1 \\
3 & 1 & 4 & 8 & 13 & 19 & 26 \\
4 & 1 & 1 & 1 & 1 & 1 & 1 \\
5 & 1 & 4 & 8 & 13 & 19 & 26 \\
6 & 1 & 4 & 8 & 13 & 19 & 26 \\
7 & 1 & 13 & 57 & 168 & 402 & 843 \\
\hline
$\Sigma$ & 7 & 28 & 84 & 210 & 462 & 924  \\
\hline
\\
$N = 4$ & $k$ \\
$c$ & 1 & 2 & 3 & 4 & 5 & 6 \\
\hline
1  & 1 & 1  & 1   & 1    & 1    & 1 \\
2  & 1 & 1  & 1   & 1    & 1    & 1 \\
3  & 1 & 4  & 8   & 13   & 19   & 26 \\
4  & 1 & 1  & 1   & 1    & 1    & 1 \\
5  & 1 & 4  & 8   & 13   & 19   & 26 \\
6  & 1 & 4  & 8   & 13   & 19   & 26 \\
7  & 1 & 13 & 57  & 168  & 402  & 843 \\
8  & 1 & 1  & 1   & 1    & 1    & 1 \\
9  & 1 & 4  & 8   & 13   & 19   & 26 \\
10 & 1 & 4  & 8   & 13   & 19   & 26 \\
11 & 1 & 13 & 57  & 168  & 402  & 843 \\
12 & 1 & 4  & 8   & 13   & 19   & 26 \\
13 & 1 & 13 & 57  & 168  & 402  & 843 \\
14 & 1 & 13 & 57  & 168  & 402  & 843 \\
15 & 1 & 40 & 400 & 2306 & 9902 & 35228 \\
\hline
$\Sigma$ & 15 & 120 & 680 & 3060 & 11628 & 38760  \\
\hline
\end{tabular}
\caption[Table caption text]{Combinatorial multiplicities for each individual binary vector (unit-hypercube vertex) with $N \in [2,4]$ and $k \in [1,6]$.}
\label{table: multiplicities_per_N}
\end{table}

The total combinatorial or multinomial multiplicity (multiset multiplicity) is
\begin{equation}
\#_{(N,k)} =
\begin{pmatrix}
(2^N-1)+k-1 \\
(2^N-1)-1
\end{pmatrix}
\end{equation}
which is tabulated in Table \ref{table: multiplicities} for a finite number of $N$ and $k$ values. The factorial growth is obvious for high Poisson orders and binary dimensions.

\clearpage

\subsection{Exclusive efficiency}

A term \textit{exclusive efficiency} is often used in the context of pileup and large pseudorapidity gap veto based measurements of central diffraction. The probability for an event to be exclusive in a bunch cross is simply obtained from the zero-truncated Poisson distribution
\begin{equation}
P_P^{>0}(k;\mu) = P(X = k | X > 0) = \frac{P_P(k=1;\mu)}{1 - P_P(k=0;\mu)} = \frac{ \mu^{k} }{(e^{\mu} - 1)k! }.
\end{equation}
Then setting $k=1$, which is the probability of one visible interaction, gives us the value of interest $P_P^{>0}(k = 1;\mu) = \mu/(e^{\mu} - 1)$. This runs asymptotically to zero when $\mu \rightarrow \infty$. The exclusive efficiency should not be mixed with the fundamental probability of rapidity gap survival, such as random re-scattering or production of secondaries filling the rapidity gap in the elementary $pp$-interaction. Neither it should be mixed with signal selection efficiency or direct veto \textit{in}efficiency, these are either simulation based or control sample based corrections. The signal selection efficiency is driven by tracking and trigger efficiency where as veto inefficiency is driven by efficiency of (forward) detectors to see all soft particles. A signal efficiency loss is simply of type multiplicative efficiency correction, the veto inefficiency on the other hand can be either multiplicative or additive (subtractive background) correction.

One `obvious' idea would be to extrapolate the visible distributions of veto detectors down to zero multiplicity or zero energy deposit by assuming a certain shape for the distributions (such as negative binomial distribution or similar), to estimate the veto inefficiency from data. However, this approach relies on several assumptions.

\subsection{Poisson pileup problem}
\label{sec:Poisson}

Let us assume for now that the Poissonian fluctuations are fluctuations in the number of simultaneous $pp$-interactions. That is, purely experimental accelerator luminosity conditions driven. The parameter $\mu$ describes the mean of visible simultaneous $pp$-interactions. We highlight the word visible, because $k = 0$ includes not just `empty' or `non-interacting' bunch crossings but also experimentally non-visible inelastic interactions, the so-called low mass diffraction processes. These are the non-perturbative Regge domain QCD processes. Also, the elastic interactions belong to this category with cross section $\sim$ 25 $\%$ of the total $pp$-cross section, which is $\sigma_{tot}^{pp} \sim \mathcal{O}(100)$ mb at the LHC. The Poisson distribution comes from the law of \textit{rare events} and is an approximation of the binomial distribution Bin($n,p$), when the number of trials $n \rightarrow \infty$ and $np = \mu$, then Bin($n,p) \rightarrow $ Poi($\mu$). For proton bunches circulating the LHC, $n$ approximately $\sim 10^{11}$ -- the number of protons per bunch, and $p$ is order of $\sim 1/n$.

This gives us a finite probability of the triggered bunch cross event to originate from different Poisson terms as
\begin{equation}
P_P(k \geq 1 ; \mu) = 1 - P_P(0 ; \mu) = P_P(1 ; \mu) + P_P(2 ; \mu) + P_P(3 ; \mu) + \dots
\end{equation}
That is the sum of probabilities to have one visible $pp$-interaction, two $pp$-interactions (first order pileup) and so on, with the `zero-suppressed' or truncated mean value: $\mathbb{E}[K|K>0] = \mu/(1-e^{-\mu}) \geq 1$ and the variance $\text{Var}[K|K>0] = (\mu+\mu^2)/(1-e^{-\mu}) - \mu^2/(1-e^{-\mu})^2$, which both converge to $\mu$ at high values of $\mu$. Experimentally, the Poisson distribution $\mu$-value can be obtained from the mean trigger normalized rate $ R  \in [0,1)$ by
\begin{align}
\label{eq: mu}
\boxed{ R \equiv  1 - P_P(0 ; \mu) = 1 - e^{-\mu} \Leftrightarrow
\mu = -\ln (1 - R ) }.
\end{align}
Clearly when $R \ll 1$, then $\mu \simeq R$. The mean normalized trigger rate $R$ itself is obtained from the measured trigger frequency $f_R$ (Hz) as a function of time $t$. This we get by using the accelerator orbital frequency $f_{O} = 11.2455$ kHz at the LHC and the number of colliding bunch pairs $N_B$ circulating
\begin{equation}
\langle R \rangle_{t,N_B} = \frac{\frac{1}{\Delta t} \int_{\Delta t} f_R(t)\,dt}{N_B f_O} \;\; \text{or} \;\; \langle R \rangle_{N_{B}} = \frac{N_E}{N_{BC}}.
\end{equation}
The interval $\Delta t$ corresponds to the given data run span over time. The alternative version on right is just a reformulation using the number of triggered events $N_E$ and the number of bunch crosses $N_{BC}$. For the formalism it is important to emphasize that the trigger corresponds to Boolean OR $(\vee)$ operation between different (sub)detectors, which is the minimum bias trigger. The reason is that this operator is the most inclusive Boolean operator.

In general, care is required in determining the $\mu$-value experimentally because it is effectively a factor of both on-time and off-time pileup due to detector time integration windows which can span over several bunch crosses. It must be emphasized that the formalism here requires only an \textit{effective} $\mu$-value of a Poisson distribution. For the rest of the work, we assume that the effective modeling and measurement of the $\mu$-value is under control. The $\mu$ value determination at large $\mu$ is problematic directly from the $\mu = -\ln(1-R)$ due to $R \simeq 1$. However, in high luminosity conditions the $\mu$ value is measured with less saturated signals, by using for example track or primary vertex counting.

We assume that the pileup is linear with respect to the detector responses. Thus, we can take a linear incoherent superposition of the final states propagating from different simultaneous $pp$-interactions. Also to point out explicitly: when we `sum' binary vectors, we use the component wise OR, for example: $[1,0] \vee [1,1] = [1,1] \in \mathbb{F}_2^2$.

\subsection{Instantaneous luminosity and total inelastic cross section}
\label{luminosity}

The $\mu$-value can be understood also naturally in the context of instantaneous luminosity $L$ (cm$^{-2}\cdot\,$s$^{-1}$) measurement by using a simplified definition as in \cite{atlas2010luminosity}
\begin{equation}
\label{eq:muluminosity}
\mu_{abs} \equiv \frac{\sigma_{inel}L}{N_B f_O},
\end{equation}
where $\sigma_{inel}$ is the total inelastic cross section ($\sim 80$ mb in $pp$ at $\sqrt{s} = 13$ TeV), dictated by non-perturbative strong interactions, $N_B$ is the number of colliding bunch pairs and $f_O$ is the LHC orbital frequency (Hz). The $\mu_{abs}$ is by definition the absolute mean number of inelastic interactions per colliding bunch cross.

However, because the complete $\sigma_{inel}$ is essentially unknown due to limited low-mass diffraction acceptance at the LHC, also the $\mu_{\text{abs}}$ is unknown. The absolute mean number of inelastic interactions per bunch crossing $\mu_{abs}$, is reduced to the measured $\mu$ by the total integrated detector multiplicative acceptance $\times$ efficiency factor $0 \leq \epsilon \leq 1$ defining\footnote{Technically, we may have also `leakage' of events from outside the geometric acceptance, which do not exactly obey this definition, but nevertheless the effective $\epsilon$ is never larger than 1.} the visible inelastic cross section as 
\begin{equation}
\label{eq:visibletotalinelastic}
\sigma_{vis} \equiv \epsilon \sigma_{inel}.
\end{equation}
Then using linearity of Eq. \ref{eq:muluminosity} with respect to the cross section, we can write
\begin{equation}
\frac{\mu}{\mu_{abs}} = \frac{\sigma_{vis}}{\sigma_{inel}}
\end{equation}
and reformulating
\begin{equation}
L = \mu_{abs} \frac{N_B f_O}{\sigma_{inel}} = \mu \frac{N_B f_O}{\sigma_{vis}}.
\end{equation}

Now $L$ or equivalently $\sigma_{vis}$, can be measured by doing a van der Meer (vdM) scan, which is an absolute luminosity measurement calibration technique. However, the `total inelastic measurement' is based on an extrapolation inverting Eq. \ref{eq:visibletotalinelastic}. The efficiency $\times$ acceptance factor $\epsilon$ here has an uncertainty at least order of $\delta\epsilon / \epsilon \sim 0.1$. Also a crucial thing is to factorize the acceptance and efficiency, but experimentally one cannot always distinguish between these two and this creates the problem of defining the fiducial acceptance domain. This is because not all particles and their momentum of multibody decays of systems are measured, thus, the efficiency and acceptance have intrinsic detector simulation and minimum bias soft QCD Monte Carlo model dependence.

It is a significant challenge for experiments to carefully define the final state observables in forward domain. Fiducial definitions may be in terms of single particle acceptance, described by $(\eta,p_t)$ or $|\vec{p}|$, or in terms of the (diffractive) system invariant mass. Intuitively, the fiducial definition in terms of single particle kinematics should be less model dependent than definitions relying on the invariant mass, at least if momentum or energy measurements are available. A special case to discussion here is an indirect inference based on the elastic scattering, \textit{extrapolation} of $d\sigma_{el}/dt$ down to $t \rightarrow 0$ and using the optical theorem (unitarity) relating the total cross section and the imaginary part of the elastic forward amplitude.

\subsection{$F^*$-projection technique}
\label{sec:Fstar_projection}

We shall here shortly outline a data-MC hybrid interpolation and projection technique, what we may call the $F^*$-projection. The idea is simple: observables which cannot be directly reconstructed with the detector, may be multidimensionally interpolated using the Monte Carlo event generator based simulations, given the measured partial cross sections. This can be done at the generator level once the measured partial cross sections have been unfolded. This is closely related to non-orthogonal and overcomplete basis projection techniques known in the Wavelet and Compressed Sensing literature, for those see \cite{mallat1999wavelet}. In Monte Carlo, for each $j = 1, \dots, n = 2^N-1$ combinatorially selected sub-sample, construct the probability distribution $f^{\text{MC}}(\mathbf{O})_{|j}$ of the observable $\mathcal{O}$. We may use normalized histograms to represent the distributions, for example. This gives us a set 
\begin{equation}
\left \{ f^{\text{MC}}(\mathcal{O})_{|{j}} \right \}_{j=1}^{n}.
\end{equation}
Then we measure the set of combinatorial partial cross sections in data
\begin{equation}
\left \{ \sigma_{j}^{\text{DATA}} \right \}_{j=1}^{n}.
\end{equation}
The $F^*$-projected estimate of the differential cross section is obtained with
\begin{equation}
\frac{d\sigma^{F^*}}{d\mathcal{O}} = \sum_j \sigma_{j}^{\text{DATA}}f^{\text{MC}}(\mathcal{O})_{|{j}},
\end{equation}
which shows clearly that the results is a sum over Monte Carlo driven `dictionary functions' weighted with data driven coefficients. It is possible to show that for certain observables, when $N \rightarrow \infty$, the dependence on Monte Carlo is minimized. In practice, one wants to make the projection using several different Monte Carlo samples to obtain estimate of the model dependence. Natural distributions to reconstruct using this technique are rapidity gap size distributions $d\sigma/d\Delta y$ with varying boundary conditions, for example.

\subsection{Diffraction fit algorithms}
\label{sec:fitalgorithms}

Let us have a number of $\mathcal{C}$ different scattering processes for which we have a Monte Carlo event generator based predictions of the multidimensional combinatorial partial cross sections, normalized to probability densities. We write these down as rows in a model likelihood matrix $\mathcal{L}$ of size $\mathcal{C} \times 2^N-1$. Now given the measured partial cross sections, we want to obtain the Maximum Marginal Likelihood solution: the best fit of the measurement as a weighted sum of the Monte Carlo model based  `multidimensional template distributions'. The solution to this is obtained via Expectation Maximization iteration given in Algorithm \ref{algo:EMiteration}. If in addition, we want to re-weight the event generator distributions in order to simultaneously fit e.g. the effective Pomeron intercept, we proceed with Algorithm \ref{algo:OuterAlgorithm}. Statistical fit uncertainties can be obtained via bootstrap re-sampling.

\begin{algorithm}
\textbf{INPUT:} The unfolded measurement vector $\bm{\sigma}$ (or $\tilde{\bm{\sigma}}$) $(2^N - 1 \times 1)$, a model likelihood matrix $\mathcal{L}$ $(\mathcal{C} \times 2^N - 1)$ with $\forall i : \sum_j [\mathcal{L}]_{i,j} = 1$ and the initial estimate $\Theta$ $(\mathcal{C} \times 1)$, otherwise $\Theta = \mathbf{1}_\mathcal{C}/|\mathcal{C}|$.
\\
\begin{algorithmic}
\REPEAT
\STATE \textbf{A. Inverse step using Bayes theorem:} \\
\STATE Diagonal priors matrix $\times$ Likelihood matrix: \\
$\mathcal{P} \leftarrow \text{diag}(\Theta_k) \mathcal{L}$
\STATE Normalize each column to obtain the probabilistic mixed density operator:
\FORALL{$j = 1,\dots,2^N-1$}
\STATE
$[\mathcal{P}]_{:,j} \leftarrow [\mathcal{P}]_{:,j} / \sum_{i=1}^\mathcal{C} [\mathcal{P}]_{i,j}$
\ENDFOR
\STATE \textbf{B. Forward step:} \\
\STATE Operate on data with the mixed density operator: \\
$\Theta_{k+1} \leftarrow \mathcal{P} \bm{\sigma}$
\UNTIL{Convergence $\|\Theta_{k+1} - \Theta_k\|_2 < \delta$}
\end{algorithmic}
\textbf{OUTPUT:} The process cross section estimates $\hat{\Theta} = (\sigma^1, \sigma^2,..., \sigma^\mathcal{C})^T (\mathcal{C} \times 1)$.
\caption{Maximum Marginal Likelihood estimator via Expectation Maximization.}
\label{algo:EMiteration}
\end{algorithm}

\begin{algorithm}
\textbf{INPUT:}
\begin{algorithmic}
\FORALL{$M$ parameters in $\Phi = [\phi^1_a, \phi^1_b] \times [\phi^2_a, \phi^2_b] \times \cdots \times [\phi^M_a, \phi^M_b]$}
\FORALL{MC events}
\STATE Construct the MC observables functionally dependent on the parameters
\STATE Re-weight the event according to $(\phi^1, \phi^2, \dots, \phi^M)$
\ENDFOR
\STATE A. Construct re-weighted MC process Likelihood matrix with weighted event selection
\STATE B. Estimate the process cross sections (mixing weights) using Algorithm A$\ref{algo:EMiteration}$ \\
$\hat{\Theta} \leftarrow $A\ref{algo:EMiteration}$(\tilde{\bm{\sigma}}, \mathcal{L}_{RW})$
\STATE C. Construct the new Monte Carlo estimate of $2^N-1$ visible partial cross sections \\
$\tilde{\bm{\sigma}}^{MC} \leftarrow \mathcal{L}_{RW}^T \hat{\Theta}$
\STATE D. Calculate Kullback-Leibler divergence $D(\text{data}|\text{model})$ with \\
$D \leftarrow D(\tilde{\bm{\sigma}} | \bm{\sigma}^{MC})$
\ENDFOR
\end{algorithmic}
\textbf{OUTPUT:} Kullback-Leibler divergences for the parameter (hyper)grid $\Phi \subseteq \mathbb{R}^M$. Use these to infer optimal values and parameter sensitivity/uncertainty.
\caption{Event-by-event MC re-weighting via $M$ parameter brute force grid scan.}
\label{algo:OuterAlgorithm}
\end{algorithm}

%\end{fmffile} % Feynman diagrams, must be here at the end
\end{document}